\documentclass[acmtog]{acmart}
\nonstopmode

\usepackage[ruled]{algorithm2e} 
\usepackage{graphicx,multirow}

\SetAlFnt{\small}
\SetAlCapFnt{\small}
\SetAlCapNameFnt{\small}
\SetAlCapHSkip{0pt}
\IncMargin{-\parindent}
\numberwithin{equation}{section}

\acmJournal{TOG}
\acmVolume{9}
\acmNumber{4}
\acmArticle{39}
\acmYear{2017}
\acmMonth{3}

\setcopyright{none}

\acmDOI{0000001.0000001_2}

\setcitestyle{square}

\newcommand{\ignore}[1]{}

\newlength{\w}
\newlength{\h}

\newcommand{\ignorethis} [1] {}

\newcommand{\sectnum    } [1] {\ref{#1}}

\newcommand{\sect       } [1] {Section~\sectnum{#1}}

\newcommand{\tbl}[1]{Table~\ref{#1}}

\newcommand{\fignum     } [1] {\ref{#1}}
\newcommand{\fig        } [1] {Figure~\fignum{#1}}

\newcommand{\eqnnum     } [1] {\mbox{(\ref{#1})}}
\newcommand{\eqn        } [1] {equation~\eqnnum{#1}}

\definecolor{verydarkgreen}{rgb}{0,0.2,0}
\definecolor{verydarkorange}{rgb}{0.4,0.2,0}
\definecolor{verydarkyellow}{rgb}{0.55,0.55,0}

\ifdefined\ShowNotes
\newcommand{\Connelly}[1]{\colornote{darkgreen}{Connelly}{#1}}

\newcommand{\Yuting}[1]{\colornote{maroon}{Yuting}{#1}}
\newcommand{\todo}[1]{\colornote{darkgreen}{Todo}{#1}}
\else
\newcommand{\Connelly}[1]{}
\newcommand{\Quanquan}[1]{}
\newcommand{\Sam}[1]{}
\newcommand{\Yuting}[1]{}
\newcommand{\todo}[1]{}
\fi

\newcommand{\tablegap}[0]{\\[-0.05cm]}
\newcommand{\rottext}[2]{\raisebox{2.5\normalbaselineskip}[0pt][0pt]{\rotatebox[origin=c]{90}{#1}}\hspace{0.03cm}\raisebox{2.5\normalbaselineskip}[0pt][0pt]{\rotatebox[origin=c]{90}{#2}}}

\newcommand{\dorn}[0]{Dorn~et~al.~\shortcite{dorn2015}}

\ifdefined\ShowNotes
  \newcommand{\colornote}[3]{{\color{#1}\bf{#2: #3}\normalfont}}
\else
  \newcommand{\colornote}[3]{}
\fi

\definecolor{darkred}{rgb}{0.7,0.1,0.1}
\definecolor{darkgreen}{rgb}{0.1,0.7,0.1}
\definecolor{verydarkgreen}{rgb}{0.0,0.5,0.0}
\definecolor{verydarkblue}{rgb}{0.0,0.0,0.7}
\definecolor{cyan}{rgb}{0.7,0.0,0.7}
\definecolor{dblue}{rgb}{0.2,0.2,0.8}
\definecolor{maroon}{rgb}{0.76,.13,.28}
\definecolor{burntorange}{rgb}{0.81,.33,0}

\definecolor{uhref_color}{HTML}{000066}

\newcommand\ulinec{\bgroup\markoverwith
      {\textcolor{uhref_color}{\rule[-0.5ex]{2pt}{0.4pt}}}\ULon}

\newcommand{\nsfurl}[1]{\ulinec{\small{\color{uhref_color}{\url{#1}}}}}

\newcommand{\beginintegraltable}[2]{
\begin{tabular}{|l|l|}
	\hline
	Function $f(x)$ & Bandlimited with #1 kernel: $\hat{f}^{(#2)}(x)$ \\
	\hline
}
\newcommand{\finishintegraltable}{
	\hline
	\end{tabular}\\ \\
}

\DeclareMathOperator{\sinc}{sinc}
\DeclareMathOperator{\fract}{fract}
\DeclareMathOperator{\erf}{erf}

\newcommand\missing{} 

\newcommand{\floor}[1]{\left\lfloor #1 \right\rfloor}
\newcommand{\ceil}[1]{\lceil #1 \rceil}

\newcommand\tbltab{\hspace{0.25cm}}

\begin{document}
\title{Approximate Program Smoothing Using Mean-Variance Statistics, with Application to Procedural Shader Bandlimiting}

\author{Yuting Yang}
\affiliation{
    \institution{University of Virginia}
}
\email{yyuting@virginia.edu}

\author{Connelly Barnes}
\affiliation{
    \institution{University of Virginia}
}
\email{connelly@cs.virginia.edu}



\begin{abstract}
This paper introduces a general method to approximate the convolution of an arbitrary program with a Gaussian kernel. This process has the effect of smoothing out a program. Our compiler framework models intermediate values in the program as random variables, by using mean and variance statistics. Our approach breaks the input program into parts and relates the statistics of the different parts, under the smoothing process. We give several approximations that can be used for the different parts of the program. These include the approximation of Dorn et al.~\shortcite{dorn2015}, a novel adaptive Gaussian approximation, Monte Carlo sampling, and compactly supported kernels. Our adaptive Gaussian approximation is accurate up to the second order in the standard deviation of the smoothing kernel, and mathematically smooth. We show how to construct a compiler that applies chosen approximations to given parts of the input program. Because each expression can have multiple approximation choices, we use a genetic search to automatically select the best approximations. We apply this framework to the problem of automatically bandlimiting procedural shader programs. We evaluate our method on a variety of complex shaders, including shaders with parallax mapping, animation, and spatially varying statistics. The resulting smoothed shader programs outperform previous approaches both numerically, and aesthetically, due to the smoothing properties of our approximations.


\end{abstract}

%
%


%
%



%

\begin{teaserfigure}
  \setlength{\tabcolsep}{1pt}
  \setlength{\w}{1.37in}
  \begin{tabular}{cccccc}
(a) Ground Truth & (b) No Antialiasing & (c) Our Result & (d) Dorn~et~al.~2015 & (e) MSAA & \\
\includegraphics[width=\w]{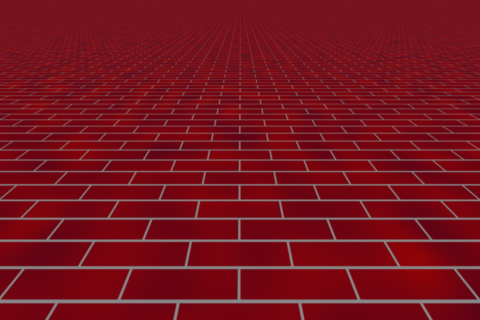} & \includegraphics[width=\w]{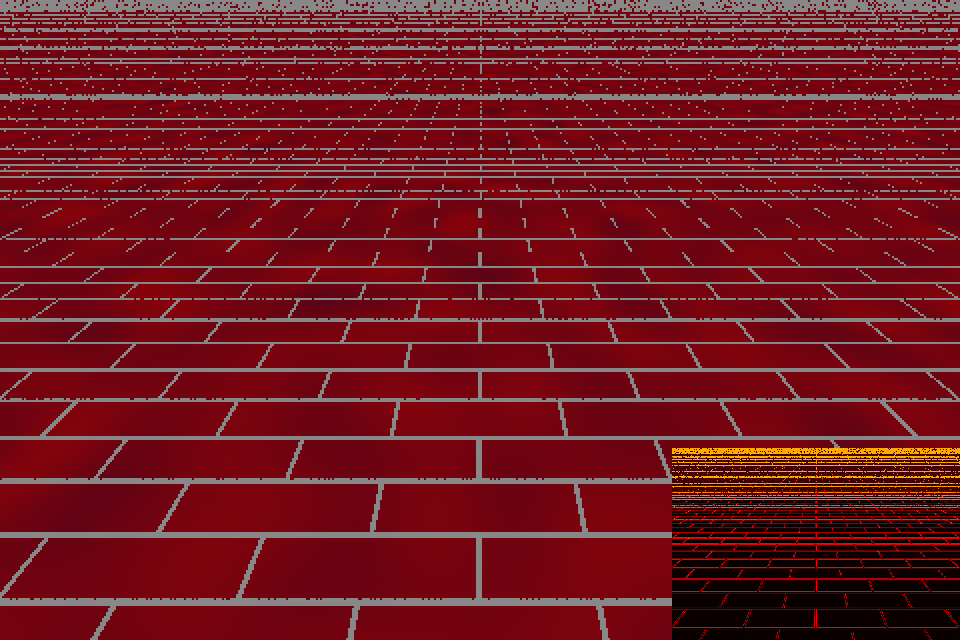} & \includegraphics[width=\w]{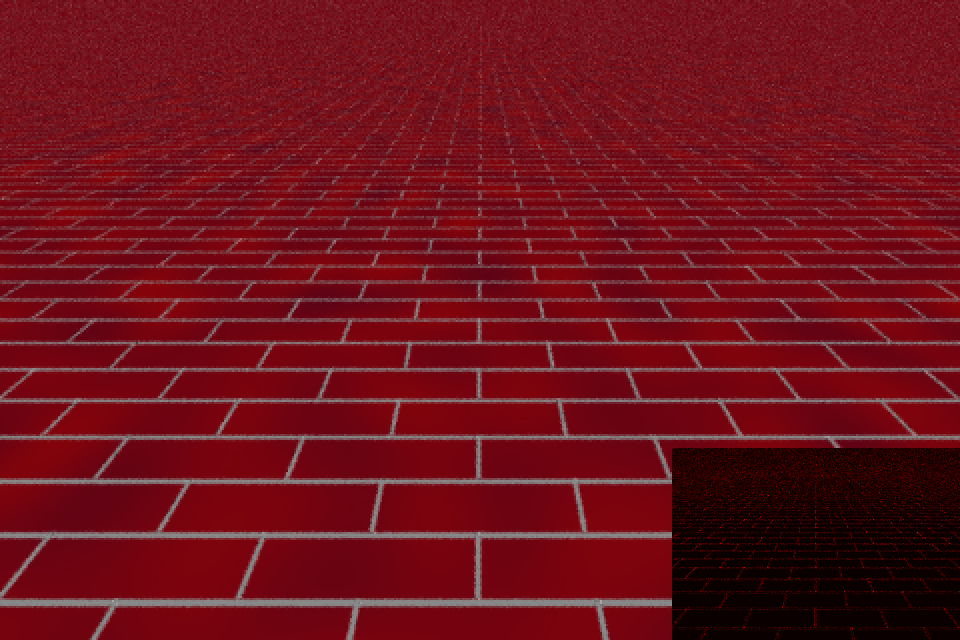} & \includegraphics[width=\w]{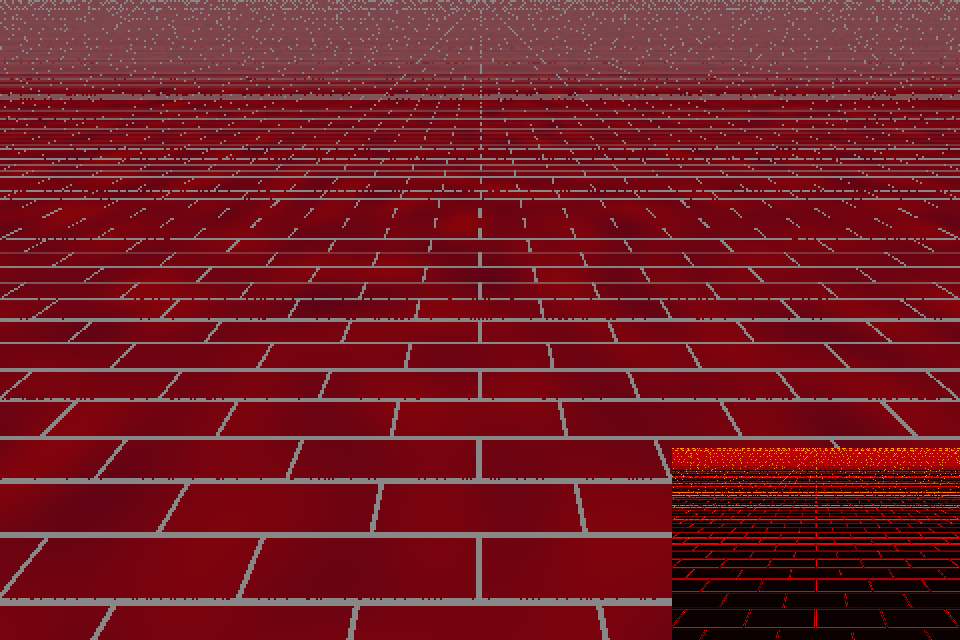} & \includegraphics[width=\w]{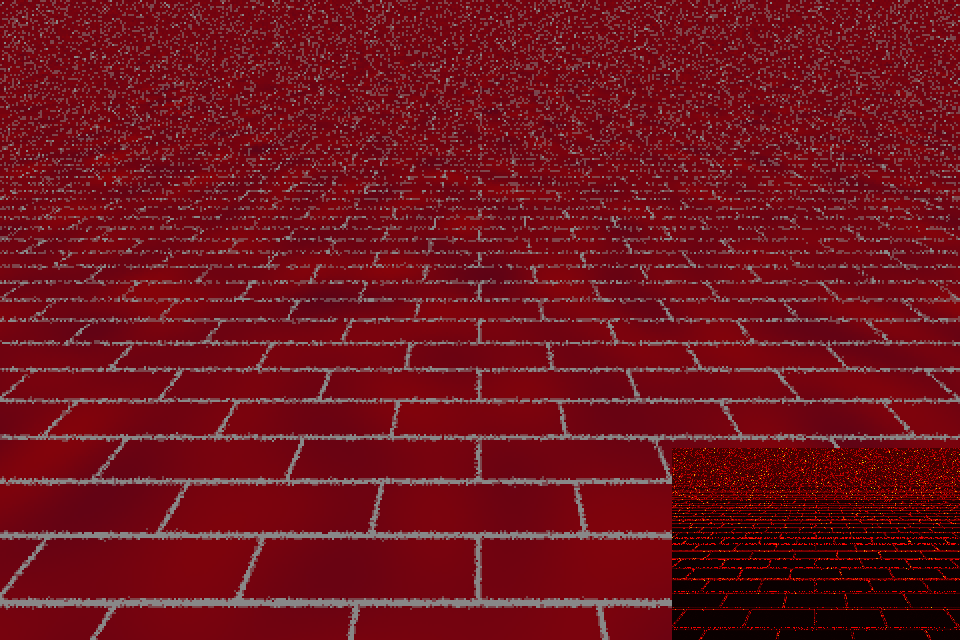} & \includegraphics[height=0.93in]{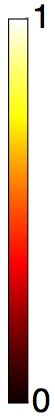} \tablegap
& {\small 2115 ms, $L^2$ error: 0.117} & {\small 4041 ms (2x), $L^2$ error: 0.021} & {\small 1892 ms (1x), $L^2$ error: 0.095} & {\small 3659 ms (2x), $L^2$ error: 0.072} & \\
  \end{tabular}
  \vspace{-2ex}
  \caption{Our paper gives a novel compiler framework for smoothing out programs in an approximate manner. Here we show how our smoothing framework can be applied to bandlimiting (antialiasing) procedural shader programs. In (a) is the ground truth result for a brick shader, estimated by using 1000 samples; (b) is the aliased result due to naively evaluating the original shader program; (c) is our result; (d) is the result of previous work; and (e) is multi-sample antialiasing (MSAA), chosen to use comparable run-time as our result. The $L^2$ errors are reported in sRGB color space, with the inset heatmap depicting per-pixel $L^2$ error. Our result has significantly less error, noise, and aliasing than other approaches.}
  \label{fig:teaser}
\end{teaserfigure}

\maketitle
\thispagestyle{empty}

\section{Introduction}

In many contexts, functions that have aliasing or noise could be viewed as undesirable. In this paper, we develop a general compiler-driven machinery to approximately smooth out arbitrary programs, and thereby suppress aliasing or noise. We then apply this machinery to bandlimit procedural shader programs. In order to motivate our approach concretely by an application, we first discuss how procedural shaders may be bandlimited, and then return to our smoothing compiler.

Procedural shaders are important in rendering systems, because they can be used to flexibly specify material appearance in virtual scenes~\cite{akenine2008real}. One visual error that can appear in procedural shaders is \textit{aliasing}. \textit{Aliasing} artifacts occur when the sampling rate is below the Nyquist limit \cite{crow1977aliasing}. There are two more conventional approaches used to reduce such aliasing: supersampling and prefiltering. We discuss these before discussing our smoothing compiler.



Supersampling increases the spatial sampling rate, so that the output value for each pixel is based on multiple samples. The sampling rate can be uniform across the image. The sampling rate can also be chosen adaptively based on measurements such as local contrast \cite{dippe1985antialiasing,mitchell1987generating,hachisuka2008multidimensional,mitchell1991spectrally}. This approach in the limit recovers the ground truth image, but can be time-consuming due to requiring multiple samples per pixel. 

Prefiltering typically stores precomputed integrals in mipmaps \cite{williams1983pyramidal} or summed area tables \cite{crow1984summed}. This approach offers the benefit of exact solutions with a constant number of operations, provided that the shading function can be tiled or otherwise represented on a compact domain. However, in practice many interesting shaders do not tile, so this limits the applicability of this method. Further, prefiltering increases storage requirements and may replace inexpensive computations with more expensive memory accesses. This approach is not practical for functions of more than two or three variables because memory costs scale exponentially.

An alternative strategy is to construct a bandlimited variant of the shading function by symbolic integration. This can be expressed by convolving the shading function with a low-pass filter~\cite{norton1982clamping}. Exact analytic band-limited formulas are known for some specialized functions such as noise functions~\cite{lagae2009procedural}. In most cases, however, the shader developer must manually calculate the convolution integral. But frequently the integrals cannot be solved in closed form, which limits this strategy.

Our framework takes a different approach from most previous work. Our goal is to smooth out an arbitrary input function represented as a program, by approximately convolving it with a Gaussian filter. We take the program as input, break it into different parts, and relate the statistics of the different parts, under the desired smoothing process. Specifically, we treat each intermediate value in the computation as a random variable with a certain probability distribution. We use mean and variance statistics to model these random variables. In this manner, we can smooth out arbitrary programs that operate over floating-point numbers. Our approach can be applied to bandlimit shader programs, because we take as input an original shader that may have aliasing, and produce as output bandlimited approximations that have been convolved with the Gaussian kernel.  



In our framework, we explore a number of approximation rules. We first improve the approximations of Dorn et al.~\shortcite{dorn2015} (\sect{sec:dorn}) and relate them to the mean and variance statistics in our framework. We then develop a novel adaptive Gaussian approximation (\sect{sec:gaussian}). For a class of programs that are real analytic, this approximation if used at all nodes in the computation results in a smoothed program that is accurate to the second power of the standard deviation. We next relate Monte Carlo sampling (\sect{sec:mc}) to our framework. This can give good approximations for non-analytic functions, because it converges for large numbers of samples to the ground truth. Finally, we discuss how compactly supported kernels (\sect{sec:undef}) can be used for parts of the computation that would otherwise be undefined or interact with infinite values. 

Because each computation node can choose from a variety of approximation methods, the search space for the optimal approximations is combinatoric. We use genetic search to find the \textit{Pareto frontier} of  approximation choices that optimally trade off the running time and error of the program. A programmer can then simply select a program variant from the Pareto frontier according to the desired running time or error level. 

To evaluate our framework, we developed a variety of complex shaders, including shaders with parallax mapping, animation, and spatially varying statistics, and compare the performance with \dorn{} and commonly used multisample antialiasing (MSAA). Our framework gives a wider selection of band-limited programs with less error than \dorn. Our shaders are frequently an order of magnitude faster than MSAA for comparable errors.

\section{Related work}

\textbf{Mathematics and smoothing}. Smoothing a function is beneficial in domains such as optimizing non-convex, or non-differentiable problems \cite{nesterov2005smooth,chen1999global2,chen1999global}. For our purposes, smoothing can be understood as convolving a function with smooth kernels. When used in numerical optimization, this approach is sometimes known as the continuation method or mollification  \cite{ermoliev1995minimization,ermoliev1997nonsmooth,wu1996effective}. The amount of smoothing can be controlled simply by changing the width of the kernel. In our framework, we model the input program by relating the statistics of each variable in the program, and apply a variety of approximations to smooth the program. Our idea of associating a range with each intermediate value of a program is conceptually similar to interval analysis~\cite{moore1979methods}. Chaudhuri and Solar-Lezama~\shortcite{chaudhuri2011smoothing} developed a smoothing interpreter that uses intervals to reason about smoothed semantics of programs. The homogeneous heat equation with initial conditions given by a nonsmoothed function results in a smoothing process, via convolution with the Gaussian that is the Green's function of the heat equation. Thus, connections can be made between convolution with a Gaussian and results for the heat equation, such as \L{}ysik~\shortcite{lysik2012mean}.

\textbf{Procedural shader antialiasing}. The use of \textit{antialiasing} to remove sampling artifacts is important and well studied in computer graphics. The most general and common approach is to numerically approach the band-limited signal using supersampling \cite{apodaca2000advanced}. Stochastic sampling \cite{dippe1985antialiasing,crow1977aliasing} is one effective way to achieve this. The sampling rate can be effectively lowered if it is adaptively chosen according to the contrast of the pixel~ \cite{dippe1985antialiasing,mitchell1987generating,hachisuka2008multidimensional,mitchell1991spectrally}. In video rendering, samples from previous frames can also be reused for computation efficiency \cite{yang2009amortized}. An alternative to sample-based \textit{antialiasing} is to create a band-limited version of a procedural shader. This can be a difficult task because analytically integrating the function is often infeasible. There are several practical approaches \cite{ebert2003texturing} that approximate the band-limited shader functions by sampling. This includes clamping the high-frequency components in the frequency domain \cite{norton1982clamping}, and producing lookup tables for static textures using mipmapping \cite{williams1983pyramidal} and summed area tables \cite{crow1984summed}. Like our work, and unlike most other work in this area, Dorn et al.~\shortcite{dorn2015} uses a compiler-driven technique to apply closed-form integral approximations to compute nodes of an arbitrary program. Unlike Dorn et al.~\shortcite{dorn2015}, our framework flexibly incorporates both mean and variance statistics, and we use several approximations that have higher accuracy. Our approach is general and can apply to arbitrary programs: we simply explore shaders as an example application.

\textbf{Genetic algorithms}. Genetic algorithms and genetic programming (GP) are general machine learning strategies that use an evolutionary methodology to search for a set of programs that optimize some fitness criterion \cite{koza1992genetic}. In computer graphics, recent work by Sitthi-Amorn et al.~\shortcite{sitthi2011genetic} describes a GP approach to the problem of automatic procedural shader simplification. Brady and colleagues \cite{brady2014genbrdf} showed how to use GP to discover new analytic reflectance functions. We use a similar approach as \cite{sitthi2011genetic} to automatically generate the Pareto frontier of approximated smoothed functions.

%
%
%
%
%

\section{Overview}

This section gives an overview of our system. We first discuss the goal of our smoothing process. Next, we give an overview of the key assumptions and components in our framework.

The goal for our framework is to take an arbitrary program and produce a smoothed output program which closely approximates the convolution of the input program with a Gaussian kernel. This convolution could be multidimensional: for shader programs, the dimension is typically 2D for spatial coordinates. When producing such approximations, we would also like to keep high computation efficiency.

In our compiler-based framework, we assume the input program has a compute graph, where each node represents a floating-point computation, and the graph is a directed acyclic graph (DAG). We label each node in the DAG as a random variable and compute its mean and variance statistics. Note we use these random variables as a helpful conceptual device to determine statistics, but in most cases, we never actually sample from these random variables.\footnote{Except for Monte Carlo sampling (\sect{sec:mc}), which of course is sampled.} We assume the input variables have specified mean and variance, and for simplicity assume they are independent. For example, in a shader, the input variables might be the $(x, y)$ pixel coordinate of the shader, and the output might be the color. We carry mean and variance computations forward through the compute graph, and collect the output by taking the mean value of the output variable. 

We now give an brief conceptual example of how random variables can be used to collect statistics associated with values in a program. Suppose that a part of an input program contains a function application $y = f(x)$, where both $x$ and $y$ are intermediate values, and $f$ is a built-in mathematical function (e.g. cosine). In our framework, we model this computation as $Y = f(X)$. Here, $X$ and $Y$ are random variables with mean and variance statistics. In our framework, we simplify the problem of smoothing out the overall function by breaking the program into sub-parts, which can each be separately smoothed.\ignore{So the problem of convolution integration on a complicated function is simplified into approximation of the sub-part computations that are relatively simple.} In this paper, we use lower-case letters such as $x$ and $y$ to represent real values (scalars) in the original non-smoothed input function. These can be either input, output, or intermediate values. We use corresponding capital letters such as $X$ and $Y$ to represent random variables for the corresponding smoothed computation node.

%

As we just mentioned, the random variables in our framework have two key values associated with each individual: mean and variance. We denote these as $\mu_X$ and $\sigma^2_X$ respectively.

\begin{figure}
	\centering
	\includegraphics[width=3.4in]{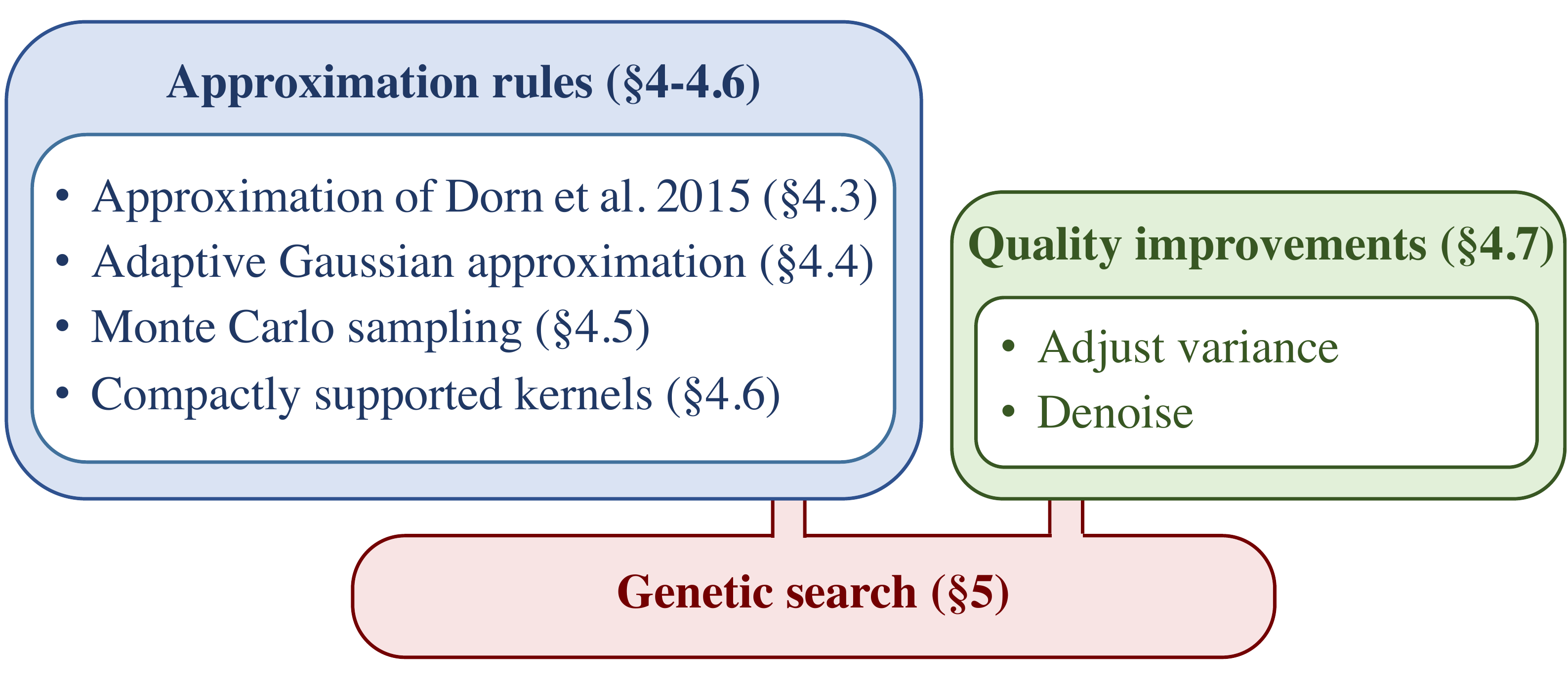}
	\caption{An overview of our system showing its components. The key components are: approximation rules, quality improvements, and genetic search. In approximation rules (\sect{sec:approximation} - \ref{sec:undef}), a variety of approximation methods are implemented to smooth the input function. We can optionally improve the quality of the output program by using the simplex method to adjust variances and denoising the output (\sect{sec:fine_tuning}). All smoothed program variants are selected through a genetic search (\sect{sec:autotuner}), which finds a Pareto frontier that optimally trades off program running time and error.}
	\label{fig_overview}
\end{figure}

An overview of the components of this framework is shown in \fig{fig_overview}. Our framework has three main parts: approximation rules, quality improvements, and genetic search. Under the random variable assumption, the compiler first approximates the smoothed function by determining the mean and variances of each compute node (discussed in ~\sect{sec:approximation} - \ref{sec:undef}). Next, additional quality improvements can be optionally made. These include heuristically adjusting the variance of each node using simplex search, and denoising (discussed in ~\sect{sec:fine_tuning}).\ignore{choices including adjusting the variance and denoising are optionally added to the search space.} Finally, we use genetic search to select the best-performing program variants based on the criterion of fast run-time and low error (discussed in ~\sect{sec:autotuner}).

\begin{figure*}
	\centering
\setlength\tabcolsep{0.01in}

\setlength{\h}{1.2in}

\begin{tabular}{ccccc}
	\includegraphics[height=\h]{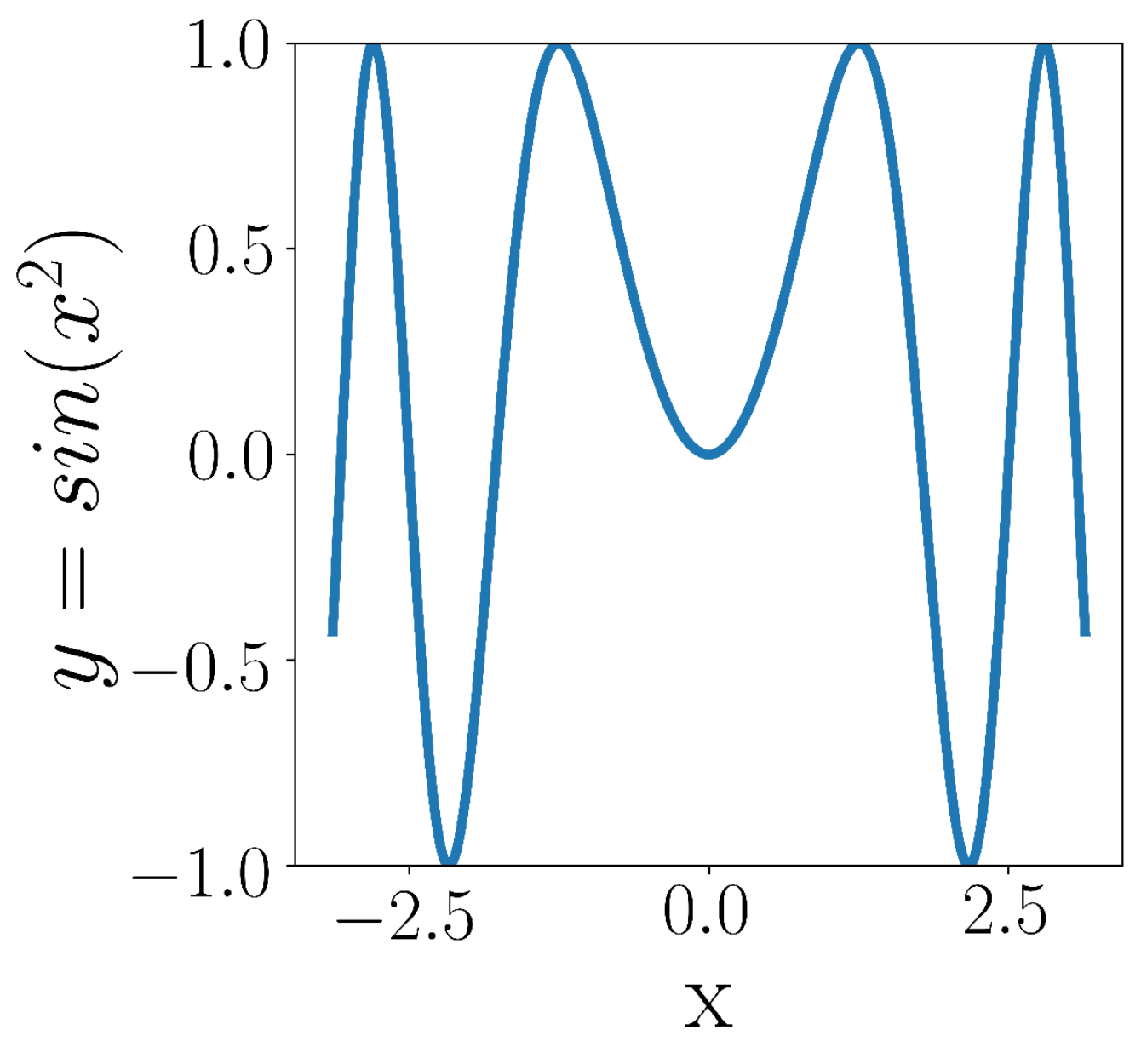} & \includegraphics[height=\h]{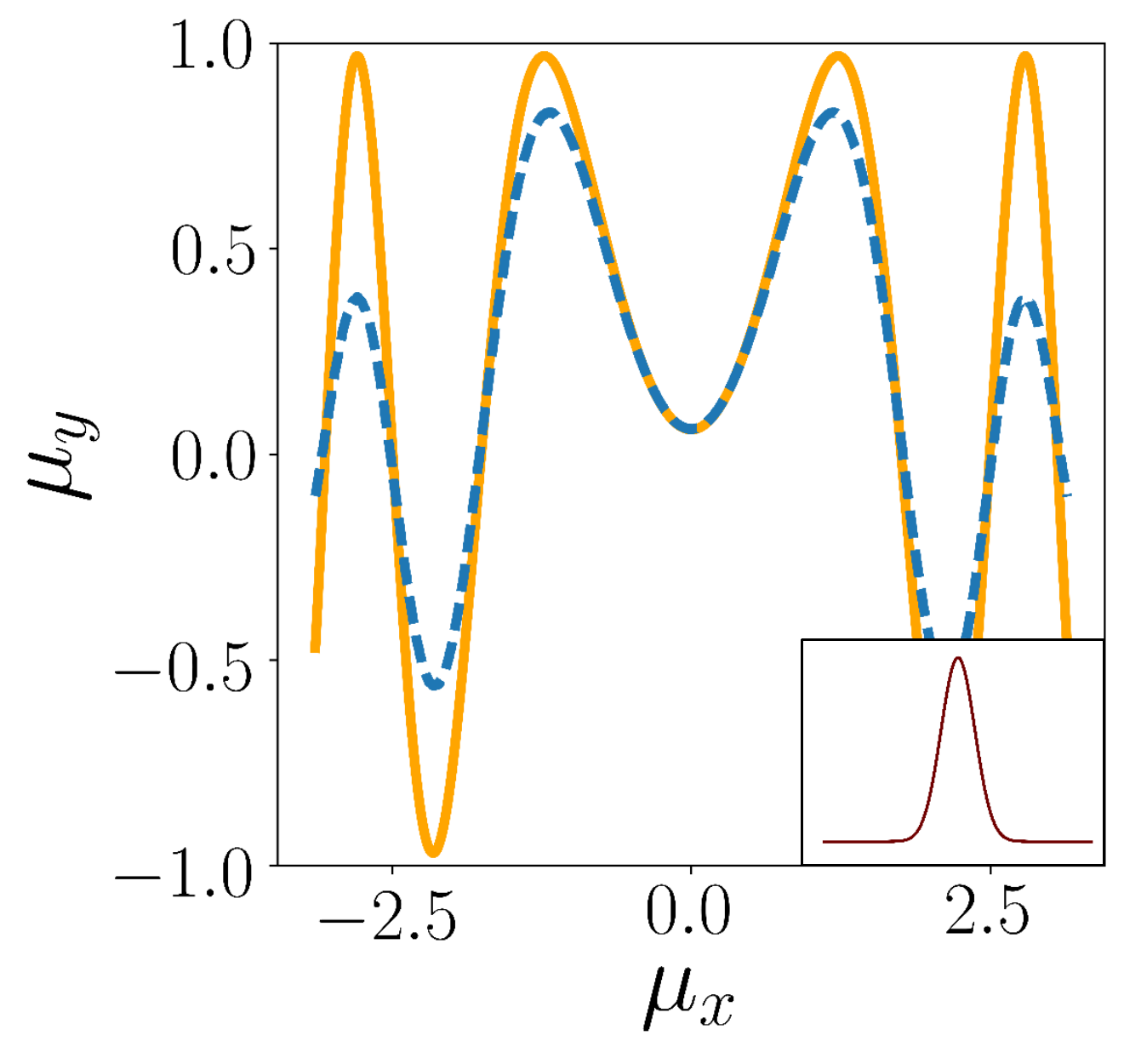} & \includegraphics[height=\h]{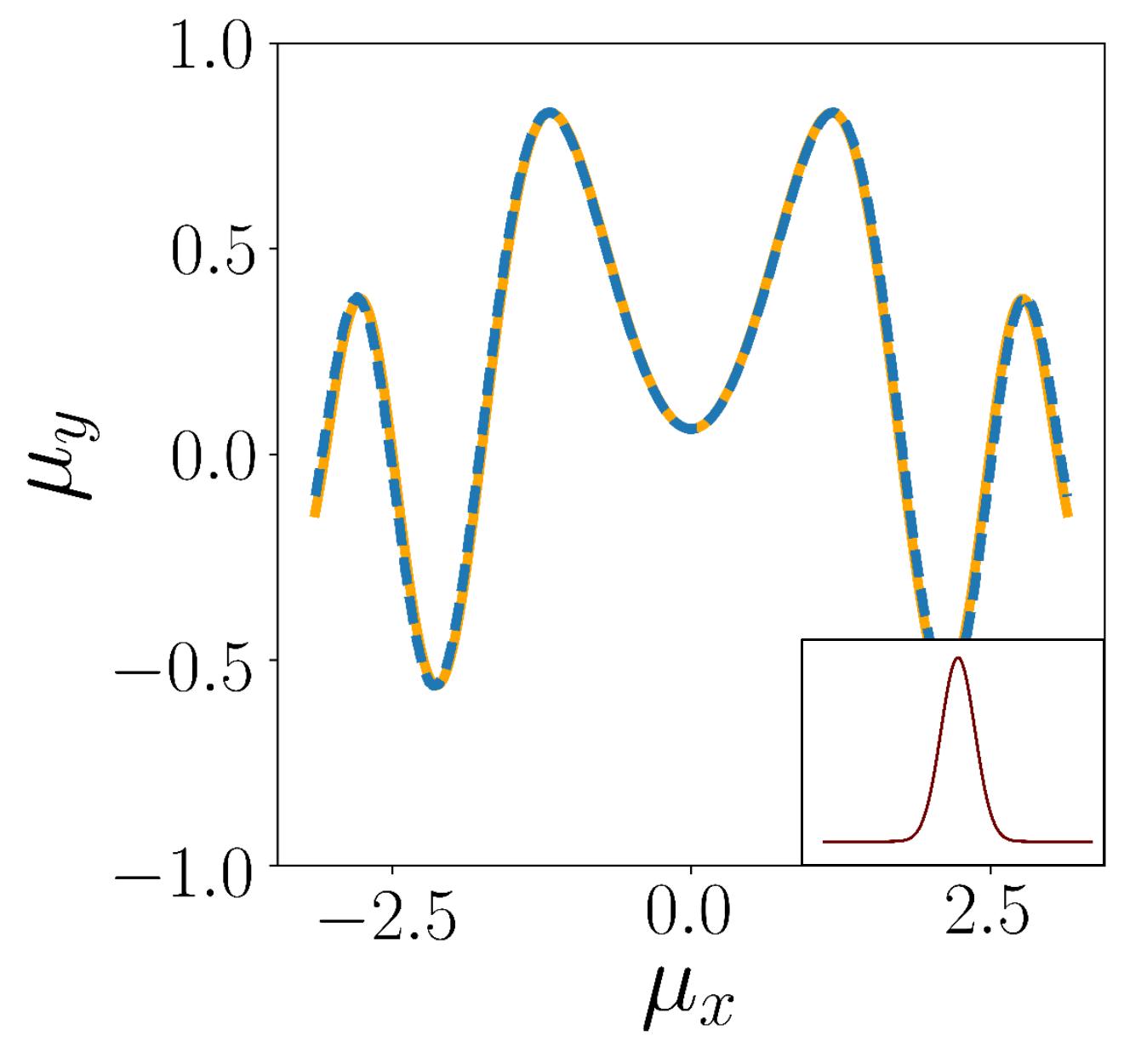} & \includegraphics[height=\h]{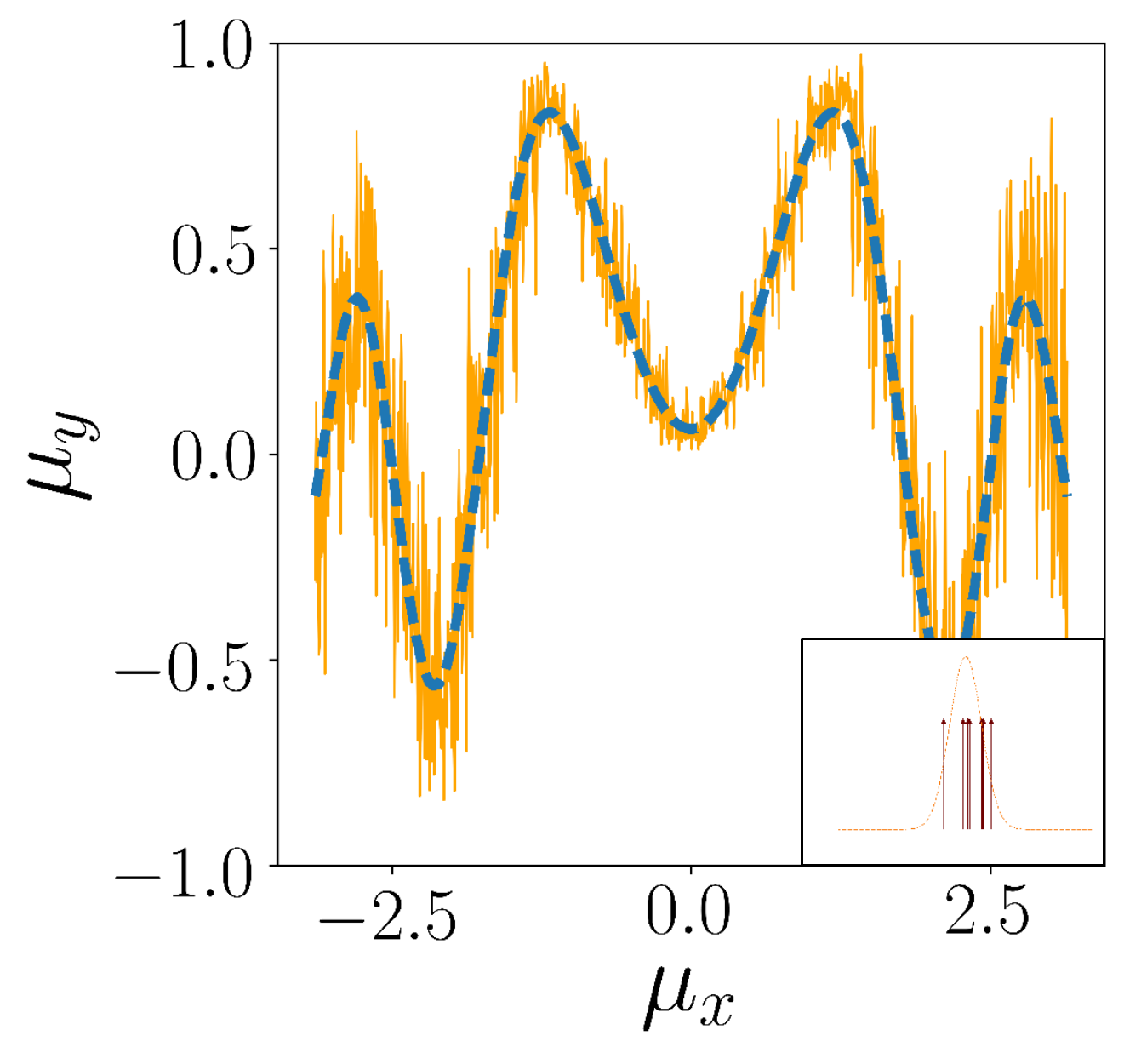}  & \includegraphics[height=\h]{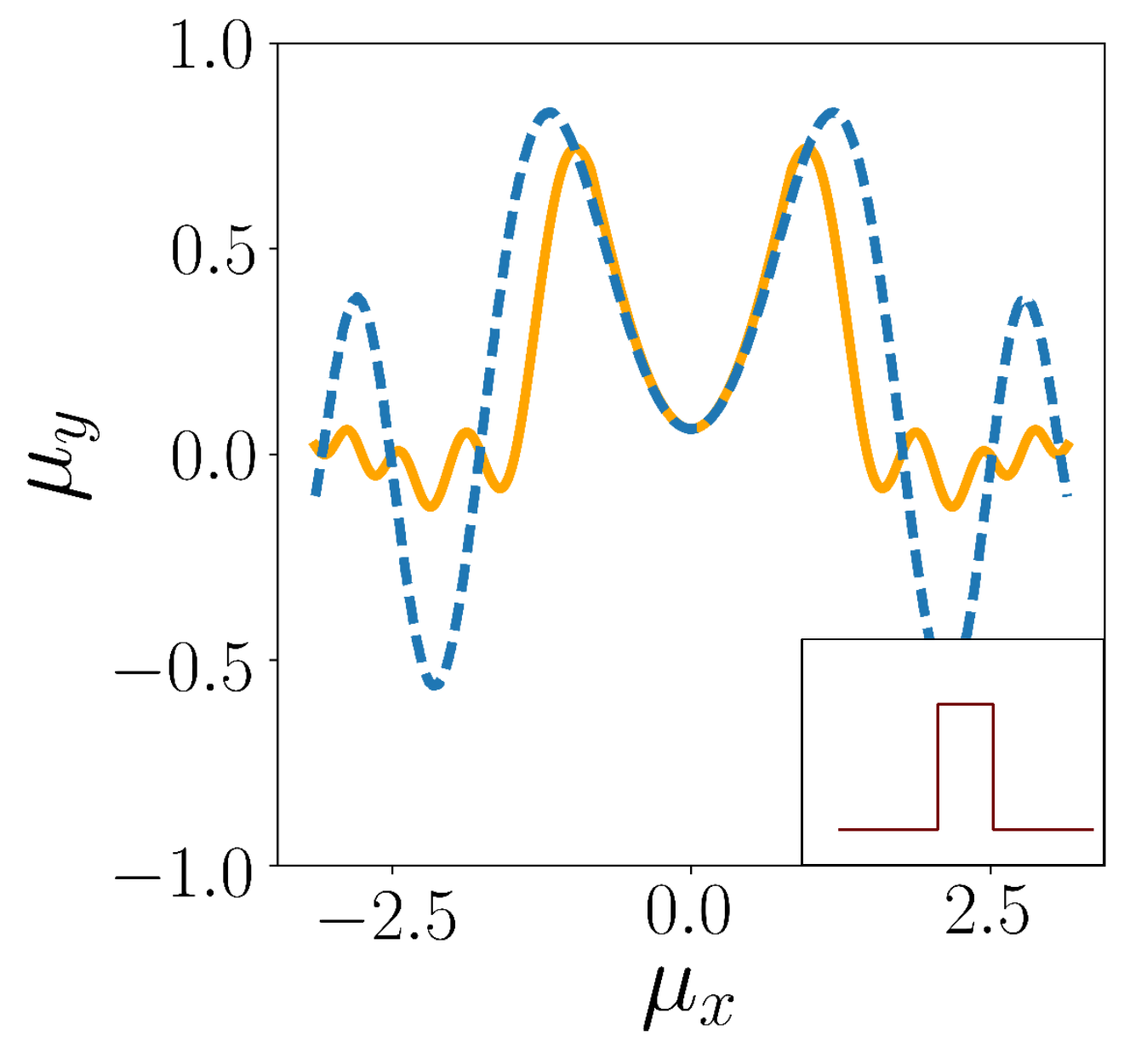}	\vspace{-1ex} \\
	{\small (a) Input function} & {\small (b) ~\cite{dorn2015}}, & {\small (c) Adaptive Gaussian}, & {\small (d) Monte Carlo sampling} & {\small (e) Compactly supported kernels}\\
	& {\small (\sect{sec:dorn})} & {\small (\sect{sec:gaussian})} & {\small (\sect{sec:mc})} & {\small (\sect{sec:undef})} \\
\end{tabular} \vspace{-1.2ex}
	\caption{An overview of different approximation techniques. (a) The input function $y = f(x) = \sin(x^2)$. This function is evaluated in our framework as the composition of two ``atomic" functions that we do know how to smooth: $\sin()$ and $x^2$. The ``ground truth" correctly smoothed function $\hat{f}(x)$ is shown in blue dashed curves in subfigures (b-e). This is determined by sampling at a very high sample rate. The orange lines in subfigures (b-e) are the mean of the output random variable calculated by different approximations. The dark red subplots in (b-e) give an abstract illustration of the kernels that were actually used to evaluate these. (b) The approximation by Dorn et al. \shortcite{dorn2015} (\sect{sec:dorn}); (c) Our adaptive Gaussian approximation (\sect{sec:gaussian}); (d) Monte Carlo sampling approximation with 8 samples (\sect{sec:mc}); (e) Compactly supported kernels approximation: here we use a box kernel (\sect{sec:undef}). We use a standard deviation of $\sigma = 0.25$ for all input distributions.}\vspace{-4ex}
	\label{fig_approx_overview}
\end{figure*}

As shown in \fig{fig_approx_overview}, we implemented several different approximation methods to calculate the mean and variance of each random variable. These approximations are:

\begin{itemize}
    \item Dorn et al.~\shortcite{dorn2015} (\sect{sec:dorn}): we improve the approximations described in \cite{dorn2015} and relate them to the mean and variance of our framework. The variance calculation here is naive yet fast.
    \item Adaptive Gaussian approximations (\sect{sec:gaussian}): these use convolution to estimate both the mean and variance for a given compute stage assuming Gaussian distributed inputs and outputs. When applied to a whole program that is in a certain real analytic class, this approximation rule gives accuracy that is accurate to the second power of $\sigma$, the standard deviation of the input variables.
    \item Monte Carlo sampling: we integrate this widely used method into our framework. The mean and variance are given by sampled estimators. For large number of samples, this converges to the ground truth. 
    \item Compactly supported kernels: for a sub-part of the computation that contains undefined or infinite values, (e.g. $y = f(x) = 1/x$ for $x = 0$), the corresponding integrals with the Gaussian for mean and variance may not exist. However, the full program may have a well-defined result, so smoothing should still be possible through such a sub-part. To handle this case, we use compactly supported kernels such as box or tent. The kernel size is limited based on the distance to the nearest undefined point.
\end{itemize}

In \fig{fig_approx_overview}(b-e), we show an example of using these different approximations to smooth the function $y = \sin(x^2)$. The approximations are shown as orange curves and the ground truth as blue dashed curves. Specifically, we have considered this function as the composition of two primitive functions $\sin()$ and $x^2$, which have each been atomically smoothed. The adaptive Gaussian rule in \fig{fig_approx_overview}(c) gives a close approximation to the ground truth for small $\sigma$. For demonstration purposes, in \fig{fig_approx_overview}(e) we also use $y = \sin(x^2)$ to show the approximation for compactly supported kernels. But in our implementation, only functions with undefined regions, or those that do not have a closed form Gaussian convolution will be approximated using this method.





\section{Approximation Rules}
\label{sec:approximation}

Suppose that the programmer writes down a given input program that we wish to smooth. In our framework,  we carry out smoothing by conceptually replacing each intermediate float value in the computation with a random variable having a known probability density function. We thus represent each compute node by having instead of a specific value, having specified mean and standard deviation statistics. For each node X in the computation, we use $\mu_X$ to denote its mean, and $\sigma_X$ for its standard deviation. The output of the program is then the mean value for the output variable.

As a concrete example, for shader bandlimiting, each input pixel coordinate $(x, y)$ could be regarded as two independent random variables, $X$ and $Y$. The means of these random variables could represent pixel positions, $\mu_X = x$, and $\mu_Y = y$. Because we wish to evaluate an antialiased variant of the shader, we can model the standard deviations of the inputs as $\sigma_X = \sigma_Y = 0.5$, i.e. half a pixel, to suppress aliasing beyond the Nyquist rate. Then the mean of the output variables in the computation is simply used as the rendered color.

In this section, we will first use an example to illustrate how we use this approach to smooth functions. Then, we will describe how our composition rules can be used to combine different smoothed sub-parts of a program. After that, we will talk about different approximation rules to estimate the mean and standard deviation statistics.

\subsection{Motivating example}
\label{sec:motivation}

Assume we are smoothing a function $y = f(x)$. The smoothed function can be denoted as $\hat{f}$, and the output value can be computed as $y = \hat{f}(x, \sigma^2)$. Here, $\sigma$ controls the smoothness. Higher $\sigma$ means $\hat{f}$ is more smooth. We will use the $\hat{}$ operator throughout this paper to denote that a function is being smoothed. We use convolution to define $\hat{f}(x, \sigma^2)$ as follows.
\begin{equation} \label{eq_convolution_integration}
    \begin{split}
        \hat{f}(x, \sigma^2) & = f(x) * G(x, \sigma^2)\\
		& = \int_{-\infty}^{\infty} f(x - u) \cdot G(u, \sigma^2) du
    \end{split}
\end{equation}

In \eqn{eq_convolution_integration}, $G(x, \sigma^2)$ is the smoothing kernel that is used to smooth out the original function $f(x)$. To more explicitly identify the kernel as being $G$, we can also use the notation $\hat{f}^G(x, \sigma^2)$. The convolution kernel $G(x, \sigma^2)$ can be any non-negative kernel that has an integral over $x$ from $-\infty$ to $\infty$ of 1. This allows us to interpret the kernel also as a probability density function. For example, we might use the Gaussian kernel:
\begin{equation} \label{eq_gaussian_kernel}
    G(x, \sigma^2) = \frac{1}{\sqrt{2 \pi} \sigma} \exp\left(-\frac{x^2}{2\sigma^2}\right)
\end{equation}

If $f(x)$ happens to be a shader program, then as is discussed in \cite{dorn2015}, $\hat{f}(x, \sigma^2)$ is simply a band-limited version of the same procedural shader function.

We now show how the convolution from \eqn{eq_convolution_integration} fits into our framework. We assume that in the input program, an intermediate value $Y$ is computed by applying a function to a previous value $X$, i.e., $Y = f(X)$. But in our framework, both $X$ and $Y$ are random variables. If the probability density function of $X$ is $f_X$, then $\mu_Y$ can be computed from $\mu_X$ and $\sigma_X$ as:
\begin{equation} \label{eq_mean_pdf}
    \mu_Y = E[f(X)] = \int_{-\infty}^{\infty} f(u) \cdot f_X(u) du
\end{equation}

As an example, if we assume X is normally distributed, $X \sim \mathcal{N}(\mu_X,\,\sigma_X^{2})$, then \eqn{eq_mean_pdf} can be rewritten as: 
%
%
\begin{equation} \label{eq_mean_normal}
    \begin{split}
        \mu_Y & = \int_{-\infty}^{\infty} f(u) \cdot \frac{1}{\sqrt{2 \pi} \sigma_X} \exp\left(-\frac{(u-\mu_X)^2}{2\sigma_X^2}\right) du \\
        & = f(\mu_X) * G(\mu_X, \sigma_X^2) \\
        & = \hat{f}(\mu_X, \sigma_X^2)
    \end{split}
\end{equation}

In \eqn{eq_mean_normal}, $*$ indicates convolution. If $X$ has a different probability density function, then $G(x, \sigma^2)$ will be a different kernel. Thus, $\mu_Y$ is the smoothed value for $f(X)$. This gives some intuition for how our model is used for smoothing functions. Our framework provides different methods to estimate $\mu$ and $\sigma$. We will describe them in the following subsections.

\subsection{Composition rules}
\label{sec:composition}

In principle \eqn{eq_convolution_integration} can be used to determine the correct smoothed function $\hat{f}$ from any input function $f$. However, in practice, the associated integrals often do not have a closed-form solution. Therefore, a key observation used in our paper is that we can break up the computation into different sub-parts, and compute approximate mean and standard deviation statistics for each sub-part. We do this by simply substituting the smoothed mean and standard deviation that are output from one compute node as the inputs for any subsequent compute nodes.

As an example, suppose we have the computation of \fig{fig_approx_overview}. From the input $x_0$, we compute $x_1 = x_0^2$, and then the output $x_2 = \sin(x_1)$, with associated random variables $X_0, X_1, X_2$, respectively. We are given $\mu_{X_0}, \sigma_{X_0}$, the mean and standard deviation of the input. Using the approximation rule chosen for the node $X_1$, we compute $\mu_{X_1}, \sigma_{X_1}$ from $\mu_{X_0}, \sigma_{X_0}$. Then from $\mu_{X_1}, \sigma_{X_1}$, using the approximation rule for the node $X_2$, we compute $\mu_{X_2}, \sigma_{X_2}$: $\mu_{X_2}$ is the smoothed output of the program ($\sigma_{X_2}$ can be discarded).

\subsection{Approximation of Dorn et al. 2015}
\label{sec:dorn}

We integrate the approximation methods described in Dorn~et~al. \shortcite{dorn2015} as one of our approximation options. Dorn's method involves computing the mean for a smoothed function by convolving with a Gaussian kernel. Suppose an intermediate variable $y$ is computed from another variable $x$, and the associated random variables are $Y$ and $X$, respectively, where $Y = f(X)$. Then $\mu_Y$ is:
\begin{equation} \label{eq_dorn}
    \mu_Y = \hat{f}(\mu_X, \sigma_{X}^2)
\end{equation}

This is the same as the result we derived in \eqn{eq_mean_normal}. Here $\hat{f}(x, \sigma_{X}^2)$ is computed from \eqn{eq_convolution_integration}. In \tbl{tbl:bandlimited}, we show commonly used functions $f$ and their corresponding smoothed functions $\hat{f}$. In Dorn's method, $\sigma_Y$ is determined based on the following simplifying assumption: output $\sigma$ is a linear combination of the axis-aligned input $\sigma$s in each dimension. Simple rules are used, such as $\sigma$ for addition and subtraction are the sum of input $\sigma$s, and $\sigma$ for multiplication or division are the product or quotient, respectively, of the input $\sigma$s. In all other cases, including function calls, the output $\sigma$ is the average of the non-zero $\sigma$s of all the inputs.

We make two improvements to Dorn et al.~\shortcite{dorn2015}, and use the improved variant of this method for all comparisons in our paper. The first improvement gives better standard deviation estimates, and the second collects a Pareto frontier. For the standard deviations (known as ``sample spacing" in Dorn et al.~\shortcite{dorn2015}), we detect the case of multiplication or division by a constant and adjust the standard deviation accordingly (i.e. $\sigma_{aX} = a\sigma_X$). This improvement helps give more accurate estimates of the standard deviations and thus reduces the problem seen in Dorn et al.'s Figure 5(c-d), where the initial program found by the genetic search is quite different from the final program after the variances have been adjusted by a simplex search. The adjustment process is described later in more detail in \sect{sec:fine_tuning}.

Our second improvement is to collect not just a single program variant with least error, but instead a Pareto frontier of program variants that optimally trade off running time and error. This process is described later in \sect{sec:autotuner}.

\subsection{Adaptive Gaussian Approximations}
\label{sec:gaussian}

In this novel approximation, we model the input variables to a compute node as being Gaussian distributed, and then fit a Gaussian to the produced output by collecting its mean and standard deviation. Thus, this rule more accurately and adaptively calculates standard deviations.

We first consider the case that we have a function of one variable. In the same manner as \sect{sec:dorn}, if a random variable $Y$ is computed from another random variable $X$ as $Y = f(X)$, then $\mu_Y$ can be determined from \eqn{eq_mean_normal}, \eqn{eq_convolution_integration} and \tbl{tbl:bandlimited}. However, the standard deviation $\sigma_Y$ is determined differently based on the definition of variance of $Y$:
\begin{equation} \label{eq_unary_var}
    \begin{split}
        \sigma_Y^2 & = E[Y^2] - E[Y]^{2} \\
        & = \widehat{f^2}(\mu_X, \sigma_{X}^{2}) - \hat{f}^2(\mu_X, \sigma_X^2)
    \end{split}
\end{equation}

The approximation of $\hat{f}$ can also be extended to multiple dimensions. Suppose $f(\mathbf{x})$ is a function applied to an $n$-dimensional vector. Then $\hat{f}$ can be computed by the convolution of  $f(\mathbf{x})$ and an $n$-dimensional multivariate Gaussian with zero mean and covariance matrix  $\mathbf{\Sigma}$. Suppose for simplicity that the input variables have zero correlation ($\rho = 0$) and equal standard deviation, so $\mathbf{\Sigma} = \mathbf{I}\sigma^2$, where $\mathbf{I}$ is the identity matrix. By using Green's function on this convolution \cite{baker2003green}, we can find a Taylor expansion for the function $\hat{f}(\mathbf{x}, \sigma^2)$ in terms of $f(\mathbf{x})$:
\begin{equation} \label{eq_taylor}
    \hat{f}(\mathbf{x}, \sigma^2) = f(\mathbf{x}) + \frac{1}{2} \sigma^2 \mathbf{\nabla}^2 f(\mathbf{x}) + \frac{1}{(2!) 2^2} \sigma^4 \mathbf{\nabla}^4 f(\mathbf{x}) + \frac{1}{(3!) 2^3} \sigma^6 \mathbf{\nabla}^6 f(\mathbf{x}) + \ldots
\end{equation}

The derivation of \eqn{eq_taylor} assumes that $f$ is \textit{real analytic} on $\mathbb{R}^n$, and can be extended to a holomorphic function on $\mathbb{C}^n$, so that all the derivatives exist, and the Taylor series has an infinite radius of convergence~\cite{wiki:entire_function}. This class of functions includes polynomials, sines, cosines, and compositions of these. It is also necessary to assume that the function is bounded by exponentials: the precise conditions are discussed by \L{}ysik~\shortcite{lysik2012mean}. These properties could hold for some shader programs, but even if they do not hold for an entire program, they could often hold for sub-parts of a program. We show in Appendix~\ref{sec:secondorder} that a single function composition gives a result accurate to second order in $\sigma$ for this rule. Similarly, this property can be proved via induction for multiple function compositions.


\begin{figure}
	\centering
\setlength\tabcolsep{0.01in}

\setlength{\h}{1.0in}

\begin{tabular}{ccc}
	\includegraphics[height=\h]{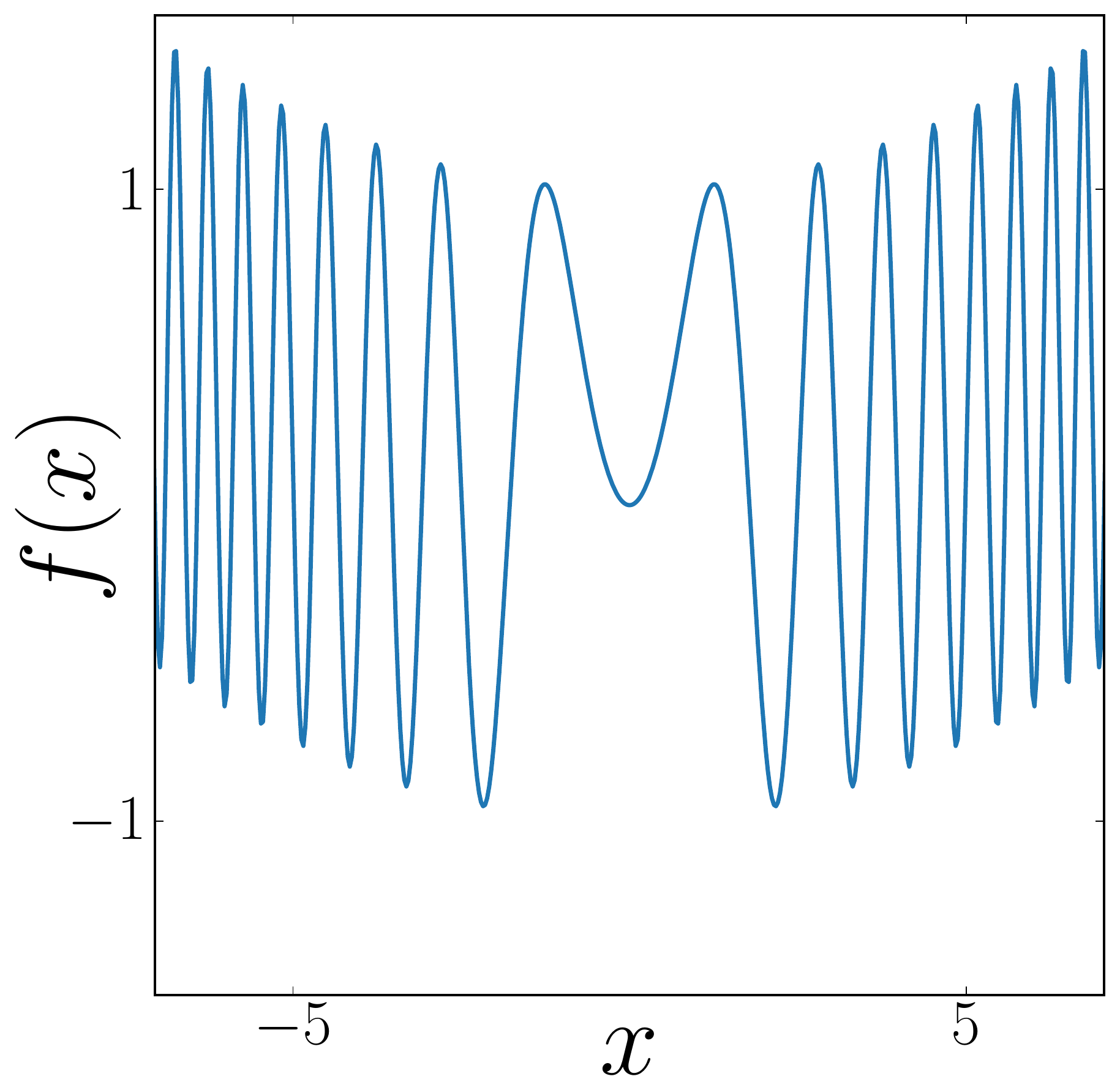} &  \includegraphics[height=\h]{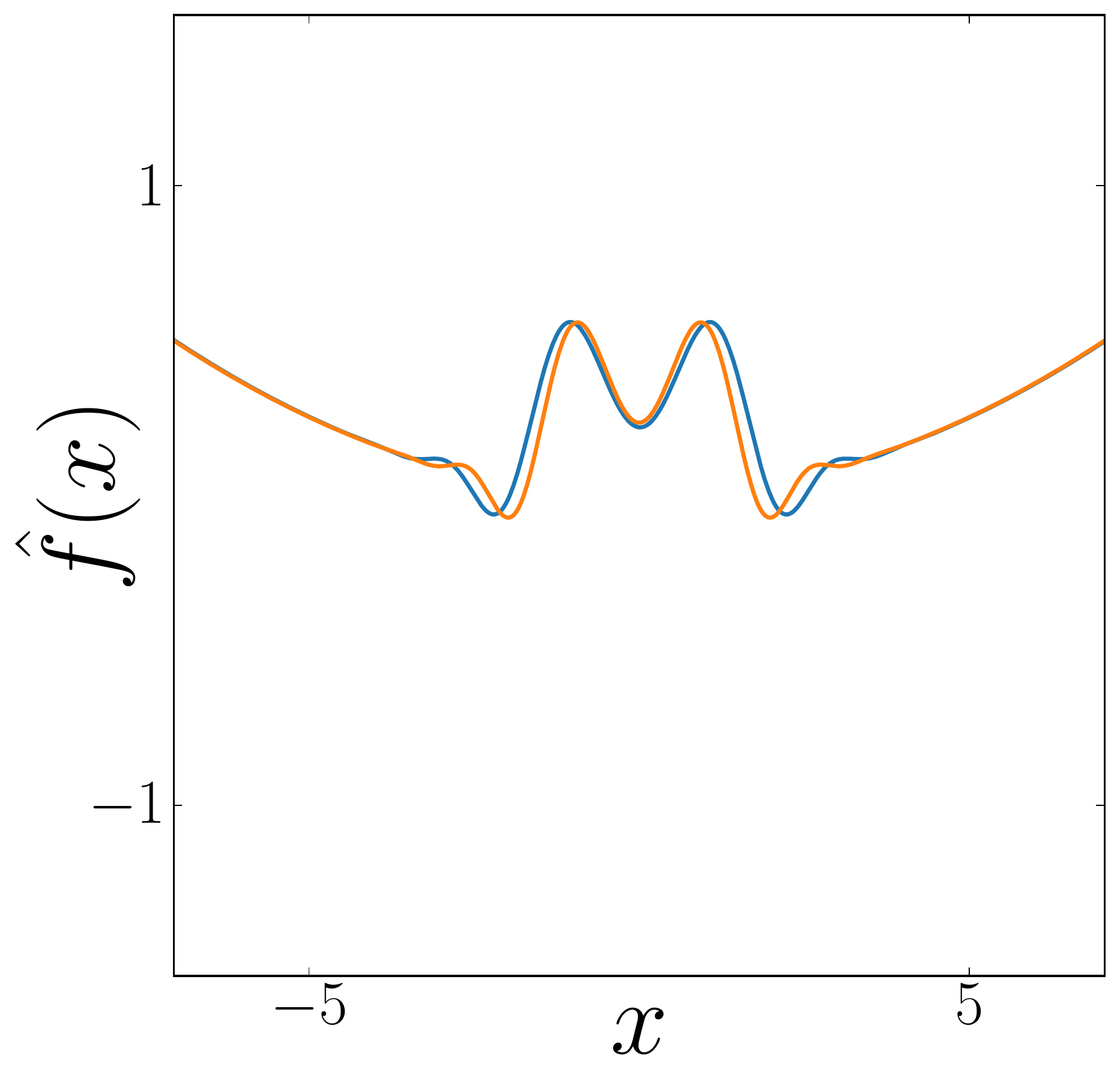} & \includegraphics[height=\h]{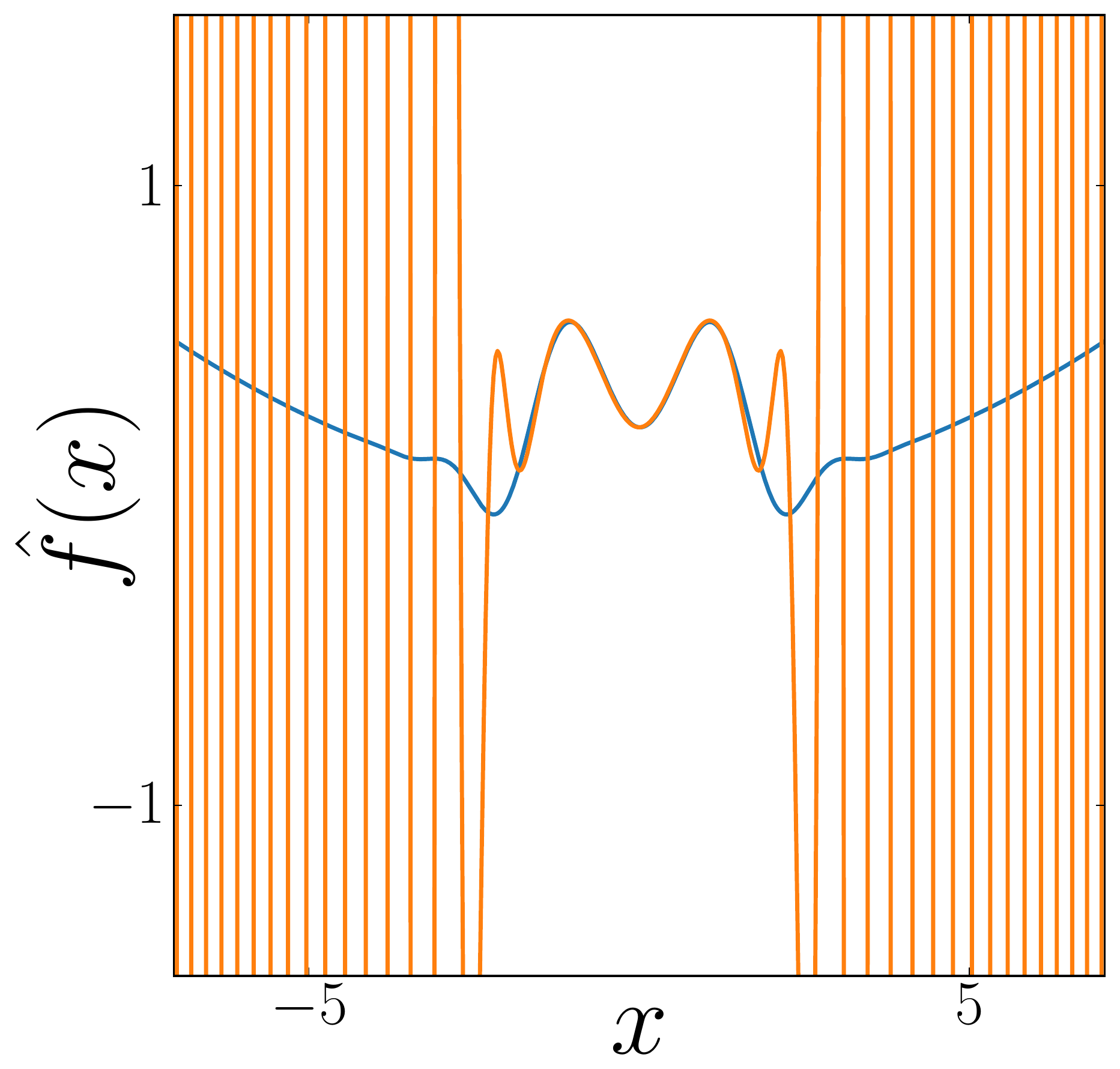}	\vspace{-1ex} \\
	{\small (a) Input function} & {\small (b) Adaptive Gaussian} & {\small (c) Truncated Taylor}\\
	&  {\small approximation~($\S$\sectnum{sec:gaussian})} & {\small expansion~(\eqn{eq_taylor})} \\
	& {\small $\sigma=0.5$} & {\small $\sigma=0.5$} \\ 
\end{tabular} \vspace{-1.2ex}
	\caption{A comparison of different approximation techniques. (a) The input function $f(x) = \sin(x^2) + \frac{1}{100}x^2$. The ground truth correctly band-limited functions $\hat{f}(x)$ are shown in blue in subfigures (b-c). These were determined by sampling at a high sample rate. (b) Our adaptive Gaussian approximation (\sect{sec:gaussian}) is shown in orange and compared against the ground truth in blue. The approximation is good. (c) A truncated Taylor expansion with 10 terms does not result in smoothing. \ignore{\Yuting{Make this a one column figure with (a) Input function, (b) Adaptive Gaussian approximation, (c) Truncated Taylor expansion.}}}\vspace{-4ex}
	\label{fig_compare_mollify}
\end{figure}

There are also other second order accurate approximations, such as simply truncating the Taylor expansion in \eqn{eq_taylor} to use only the first and second term. Why would we bother to propose an adaptive Gaussian approximation?

To illustrate why adaptive Gaussian gives a more accurate approximation, we show an example in \fig{fig_compare_mollify}. Here, we use the function $f(x) = \sin(x^2) + \frac{1}{100}x^2$. We show the approximation using adaptive Gaussian and simply truncating the Taylor expansion in \eqn{eq_taylor}. In \fig{fig_compare_mollify}, the blue lines represent the ground truth, and the orange lines represent different approximations. From \fig{fig_compare_mollify}(c), we can see that simply truncating the Taylor expansion results in large extrapolation errors and actually amplifies high frequencies, instead of smoothing them out. In \fig{fig_compare_mollify}(b), we show that the function is smoothed more accurately using the adaptive Gaussian approximation.

For binary functions, we can still show that adaptive Gaussian is accurate to second order if a correlation term is carefully chosen. But for brevity, we are only going to show the rules to compute the mean and standard deviation for binary functions we used in our compiler.

In general, the inputs of a binary function $f(a, b)$ can be considered as two random variables $A$ and $B$ (corresponding to $a$ and $b$). We make the assumption that $A$ and $B$ are distributed according to a bivariate Gaussian:
\begin{equation} \label{eq_bi_gaussian}
    A, B \sim \mathcal{N}\left(\begin{bmatrix} \mu_A \\ \mu_B \end{bmatrix}, \begin{bmatrix} \sigma_{A}^{2} & \rho \sigma_A \sigma_B\\ \rho \sigma_A \sigma_B & \sigma_{B}^{2} \end{bmatrix}\right)
\end{equation}

Here, $\sigma_{A}$ and $\sigma_{B}$ are standard deviations of $A$ and $B$. These can be determined directly from the approximation of previous computation. Here $\rho$ is the correlation term between $A$ and $B$. We will talk about how we choose $\rho$ later in this section. The mean and standard deviation for binary function plus $(+)$, minus $(-)$ and multiplication $(\cdot)$ can be derived from these assumptions based on properties of the Gaussian distribution~\cite{petersen2008matrix}:

\begin{equation} \label{eq_hat_plus_minus_mul}
    \begin{split}
        \mu_{\mathrm{plus}} = & \mu_A + \mu_B \\
        \sigma_{\mathrm{plus}}^2 = & \sigma_{A}^{2} + \sigma_{B}^{2} + 2 \rho \sigma_A \sigma_B \\ \\
        \mu_{\mathrm{minus}} = & \mu_A - \mu_B \\
        \sigma_{\mathrm{minus}}^2 = & \sigma_{A}^{2} + \sigma_{B}^{2} - 2 \rho \sigma_A \sigma_B \\ \\
        \mu_{\mathrm{mul}} = & \mu_A \mu_B + \rho \sigma_A \sigma_B \\
        \sigma_{\mathrm{mul}}^{2} = & \mu_{A}^{2} \sigma_B + \sigma_A \mu_{B}^{2} + \\
        & 2 \rho \mu_A \mu_B \sigma_A \sigma_B + \sigma_{A}^{2} \sigma_{B}^{2} (1 + \rho^2)
    \end{split}
\end{equation}

For the binary function divide $f_{\mathrm{div}}(a, b) = \frac{a}{b}$, we reduce this to multiplication by using $(\cdot)$ as $f_{\mathrm{div}}(a, b) = f_{\mathrm{mul}}(a, b^{-1})$. The mean and standard deviation for division can be calculated via the composition rules. Here, $g(b) = b^{-1}$ is an univariate function with singularity at $b = 0$. Technically, the mean and variance therefore do not exist if the Gaussian kernel is used. We work around this singularity by approximating using a compact kernel with finite support. This will be described in detail in \sect{sec:undef}. The modulo function, $f_{\mathrm{mod}}(a, b) = a\%b$, can be rewritten as $f_{\mathrm{mod}}(a, b) = b \cdot \text{fract}(\frac{a}{b})$. Here, fract($x$) is the fractional part of $x$. We make the simplifying assumption that the second argument $b$ of mod is an ordinary (non-random) variable (so $\sigma_B = 0$), to obtain: 
\begin{equation} \label{eq_hat_mod}
    \begin{split}
        \mu_{mod}^2 = & \mu_B \cdot \widehat{\text{fract}}(\frac{\mu_A}{\mu_B}, \frac{\sigma_{A}^{2}}{\mu_{B}^{2}}) \\
        \sigma_{mod} = & \mu_{B}^{2} \cdot \widehat{\text{fract}^2}(\frac{\mu_A}{\mu_B}, \frac{\sigma_{A}^{2}}{\mu_{B}^{2}}) - \mu_{mod}^2
    \end{split}
\end{equation}

Comparison functions $(>, \ge, <, \le)$ are approximated by converting them to univariate functions including the Heaviside step function $H(x)$. As an example, the function greater than $(>)$ can be rewritten as $f_{>}(a, b) = H(a - b)$. This can be approximated using rules described previously.


One other important multi-variate function we approximated is the \textit{select} function. We approximated this in the same manner as Dorn~et~al.~\shortcite{dorn2015} as a linear interpolation: $select(a, b, c) = a \cdot b + (1 - a) \cdot c$. But in our case we apply the above univariate and binary function approximations to this formula.

As we discussed before, for binary functions, we approximate the input random variables $A$ and $B$ as bivariate Gaussian with correlation coefficient $\rho$ (\eqn{eq_bi_gaussian}). In general, it is difficult to determine $\rho$, because determining $\rho$ exactly involves an integral over the entire subtrees of the computation. In our framework, we provide three options to approximate $\rho$: (1) Assume $\rho$ is zero; (2) Assume $\rho$ is a constant for each node. The constant value is estimated at training stage by sampling; (3) Estimate $\rho$ to accuracy that is second order in $\sigma$, based on a simplified assumption that the given nodes are affine functions of the inputs. For case (3) we simply take the gradients with respect to the input of the terms $a$ and $b$, normalize these gradients, and take the dot product, which recovers $\rho$. This can be done using reverse mode automatic differentiation.

We explored these different rules in our genetic search. In practice, we find that for shader programs, using only rule (1), $\rho = 0$ typically gives good results. If the other rules (2, 3) are also included, minor quality improvements are gained, but these rules are used relatively rarely by our genetic search process of \sect{sec:autotuner}. We include in Appendix~\ref{sec:rho} the details for the these other choices for correlation coefficients, because they may be more beneficial for other applications, and the second order accuracy property is interesting.

\subsection{Monte Carlo Sampling}
\label{sec:mc}

%

We relate the Monte Carlo stochastic sampling \cite{Cook:1986:SSC:7529.8927, dippe1985antialiasing} to our framework. Here $n$ is the number of samples. The computation for a node computing $f$ is modeled as $Y = f(X_1, ..., X_m)$, where the output mean and standard deviation are given by sampled estimators as follows:
\begin{equation} \label{eq_msaa}
    \begin{split}
        \mu_Y = & \frac{1}{n} \sum_{i=1}^{n} f(X_1 + \mathcal{N}_{i,1} \sigma_{X_1}, ..., X_m + \mathcal{N}_{i,m} \sigma_{X_m}) \\
        \sigma_{Y}^{2} = & \frac{1}{n} \sum_{i=1}^{n} f^2(X_1 + \mathcal{N}_{i,1} \sigma_{X_1}, ..., X_m + \mathcal{N}_{i,m} \sigma_{X_m}) - \mu_{Y}^{2}
    \end{split}
\end{equation}

Here, each $\mathcal{N}_{ij}$ are random numbers independently drawn from normal distribution $\mathcal{N}(0, 1)$. We experimented with applying the Bessel's correction \cite{wiki:bessel} to correct  the bias variance that occurs for small sample counts $n$. In practice, we found it does not have a significant improvement on the result for our system. This is mainly because the variance can also be adjusted in the ``quality improvement" phase (discussed in \sect{sec:fine_tuning}).

We choose sample numbers from ($2, 4, 6, 8, 16, 32$). The approximation converges to the ground truth for large sample numbers, and the output program simplifies to MSAA \cite{Cook:1986:SSC:7529.8927} when the entire input program is approximated under Monte Carlo sampling. 


The error of the Monte Carlo sampling $\sigma_M$ is estimated as follows \cite{mc_error}.
\begin{equation} \label{eq_error_mc}
    \sigma_M \approx \frac{\sigma}{\sqrt{n}}
\end{equation}

Here, $\sigma$ is the standard deviation computed from \eqn{eq_msaa} and $n$ is sample number. This error estimate becomes more accurate in the limit of large sample numbers.


\subsection{Compactly Supported Kernels Approximation}
\label{sec:undef}

Because the Gaussian kernel has infinite support, it cannot be used on functions with undefined regions. For example, $\sqrt{x}$ is only defined on non-negative $x$, and its convolution with Gaussian using \eqn{eq_convolution_integration} does not exist. Monte Carlo sampling may also encounter such problem. However, even if an input program contains such functions as sub-parts, the full program may have a well-defined result, so smoothing should still be possible for such programs. To handle this case, we use compactly supported kernels.

Results for certain compactly supported kernels can be obtained by using repeated convolution \cite{heckbert1986filtering} of boxcar functions. This is because such kernels approximate the Gaussian by the central limit theorem \cite{wells1986efficient}. In our framework, we use box and tent kernels to approximately smooth functions with undefined values. Because the convolution with a box kernel is easier to compute, this approximation can also be used when the Gaussian convolution does not have a closed-form solution. In \tbl{tbl:bandlimited}, we list the smoothed result using the box kernel for commonly used functions.

When integrating against a function that has an undefined region, it is important to make sure that the integral is not applied at any undefined regions. Our solution to this is to make the kernel size adapt to the location at which the integral is evaluated at. Thus, the integral is no longer technically a convolution, because it is not shift-invariant. We first measure the distance $r$ from value $x$ that we are determining the integral at to the function's nearest undefined point. If the kernel half-width was  $h$ before re-scaling, then we rescale the half-width to be $\min(h, \lambda r)$. Here $\lambda$ is a constant less than one, and in practice we use $\lambda = \frac{1}{2}$.

We can also use this truncation mechanism to better model functions such as $\fract(x)$ (the fractional part of $x$), which have discontinuities. We observe that $\fract()$ is discontinuous at integer $x$. If we input a distribution that spans a discontinuity, such as $X \sim \mathcal{N}(0, 0.1^2)$, into $\fract()$, then we find that the output $Y = \fract(X)$ may be bimodal, with some values close to zero (due to $x$ being a positive value), and other values close to one (due to $x$ being negative with small absolute value). If we fit a Gaussian to this resulting bimodal distribution, as our adaptive Gaussian rule proposes, then the mean would be  $\frac{1}{2}$, even though most of the values of $Y$ are either near $0$ or $1$. This may result in a poor approximation, which can show up in tiled pattern shaders (which use $\fract$) as an improper bias towards the center of the tile's texture. \Connelly{Add figure showing this.} One fix would be to randomly select either mode, based on the probability contained in each mode. However, this introduces sampling noise in the result, which we wish to avoid. Therefore, our solution in practice is to first identify whether the output distribution is bimodal: for $\fract()$ we can do this by simply checking if input distribution when represented as a uniform (box) distribution contains exactly one discontinuity, i.e. one integer. If so, we simply truncate the filter at the location of the discontinuity. 


\subsection{Quality Improvements}
\label{sec:fine_tuning}

At this point, we assume we have now applied the approximation rules described in Sections~\sectnum{sec:dorn} through \sectnum{sec:undef} to an input program. We can optionally improve the approximation quality in two ways: a) adjust the standard deviation made in the approximations, and b) apply denoising to program variants that use Monte Carlo sampling.

Because our approximations are not exact, the standard deviation of some nodes may be too high or too low. Following \cite{dorn2015}, we learn coefficients to refine the standard deviation estimates. By comparing with the ground truth image for the shader rendering, we use the Nelder-Mead simplex search \cite{nelder1965simplex} to learn multiplicative factors for standard deviations.

When Monte Carlo sampling is used as part of the approximation, noise is introduced because of the relatively small sample count. A variety of techniques have been developed to filter such noise \cite{kalantari2015machine, Bako17, rousselle2012adaptive}. We implement the non-local means denoising method \cite{buades2005non, buades2011non} with Laplacian pyramid \cite{liu2008robust}. We find that aesthetically appealing denoising results can be using a three level Laplacian pyramid, with a patch size of 5, search radius of 10, and denoising parameter $h$ is 10 for the lower resolutions, and searched over or set by the user for the finest resolution. In the genetic search process (\sect{sec:autotuner}), we experimented with allowing the algorithm to search from a variety of denoising parameters for the best result. However, because our denoising algorithm incurs some time overhead, it ends up being only rarely chosen. Thus, in our current setup, denoising is typically specified by the user manually choosing that he or she wants to denoise a result.

\section{Genetic Search}
\label{sec:autotuner}


In this section, we describe the genetic search algorithm. This automatically assigns approximation rules to each computation node. The algorithm finds the Pareto frontier of approximation choices that optimally trade off the running time and error of the program.

We developed this genetic search because it gives users the opportunity to explore the trade-off between efficiency and accuracy of the smoothed program. Although developers can manually assign approximation rules to each node, we found this to be a time-consuming process that can easily overlook beneficial approximation combinations. This is because each individual computation node may choose from several approximation rules, and the search space for this is combinatoric.

Our genetic search closely follows the method of Sitthi-Amron et al. \shortcite{sitthi2011genetic}. We adopt their fitness function and tournament selection rules, and we use the same method to compute the Pareto frontier of program variants that optimally trade-off running time and error with ground truth.

We start with ``decent initial guesses''. For each approximation method, we create a program where the rule is applied to all the expression nodes. For such initial guesses, we also apply cross-over with a probability of 0.5. Then we employ standard mutation and cross-over operations to explore the search space. The mutation step chooses a new approximation rule, and with equal probability, assigns this new rule to 1, 2, or 4 adjacent expression nodes in depth-first order. As an alternative, with equal probability, the new approximation rule can also be assigned to the whole subtree of an arbitrary node. In the genetic search algorithm, we choose our probability of crossover as 0.4, the probability of retaining elite individuals to be 0.25, and the mutation probability to be 0.35. Also, we use a tournament size of 4, and population size of 40, with 20 generations. Finally, we run 3 random restarts for the algorithm.

For the Monte Carlo sampling approximation, during initialization and mutate, we select sample counts with equal probability from the set $\{2, 4, 8, 16, 32\}$. For the determination of correlation coefficients described in \sect{sec:gaussian}, we pick with equal probability one of the three options.

%

%

\section{Evaluation}
\label{sec:evaluation}

The previous work of Dorn~et~al.~\shortcite{dorn2015} was evaluated primarily on relatively simple shaders. To provide a more challenging and realistic benchmark, we authored 21 shaders. Unlike the simple shaders of Dorn~et~al., these include shaders that have a Phong lighting model, animation, spatially varying statistics, and which include parallax mapping~\cite{szirmay2008displacement}. Our 21 shaders were produced by combining 7 \emph{base shaders} with 3 choices for parallax mapping: none, bumps, and ripples. In \tbl{tab:shaders}, we describe our base shaders, the choices for parallax mapping, and the associated code complexity.

\begin{table}
\caption {A table of our 21 shaders. At the top we list our 7 \emph{base shaders}, which are each combined with 3 different choices for parallax mapping, listed at the bottom. We also report the number of non-comment lines used to construct the shader program, and the number of scalar expressions in the program's compute graph.}
\setlength{\tabcolsep}{3pt}
\begin{tabular}{|l|l|l|l|}
	\hline
    Shader & Lines & Exprs & Description \\
	\hline
	\emph{Base shaders} & & & \\
	\tbltab{}Bricks & 38 & 192 & Bricks with noise pattern \\
	\tbltab{}Checkerboard & 20 & 103 & Greyscale checkerboard \\
	\tbltab{}Circles & 16 & 53 & Tiled greyscale circles \\
	\tbltab{}Color circles & 26 & 164 & Aperiodic colored circles \\
	\tbltab{}Fire & 49 & 589 & Animating faux fire\\
	\tbltab{}Quadratic sine & 26 & 166 & Animating sinewave\\
	& & & of quadratic (non-stationary) \\
	\tbltab{}Zigzag & 24 & 224 & Colorful zigzag pattern\\
	\hline
	\emph{Parallax mappings} & & & \\
	\tbltab{}None & 0 & 0 & No parallax mapping \\
	\tbltab{}Bumps & 21 & 203 & Spherical bumps \\
	\tbltab{}Ripples & 23 & 178 & Animating ripples \\
    \hline
\end{tabular}
\label{tab:shaders}
\end{table}

Results for 7 of our shaders are presented in \fig{fig:teaser} and \fig{fig:results}, including one result for each base shader. The result for our method was selected by a human choosing for each shader a program variant that has sufficiently low error. Dorn~et~al.~\shortcite{dorn2015} typically cannot reach sufficiently low errors to remove the aliasing, so we simply selected the program variant from Dorn~et~al. that reaches the lowest error. The MSAA result was selected based on evaluating MSAA program variants that use 2, 4, 8, 16, 32 samples, and selecting the one that has most similar time as ours. Please see our supplemental video for results with a rotating camera for all 21 shaders.

We also show in \fig{fig:pareto_results} time versus error plots for the Pareto frontiers associated with these 7 shaders. Note that Dorn~et~al. typically has significantly higher error, which manifests in noticeable aliasing. Also note that the MSAA method frequently takes an order of magnitude more time for equal error. Plots for all 21 of our shaders are included in the supplemental document.

Statistics for the approximations used are presented in \tbl{tab:stats}. Note that a rich variety of approximation strategies are used: all five choices for approximation are selected for different programs. For the correlation term discussed in \sect{sec:gaussian}, when aggregated across all 21 shaders, nearly all approximations for programs on the Pareto frontier prefer the simple choice of $\rho = 0$. We weight each shader's contribution equally, and find 87\% of program variants prefer $\rho = 0$, whereas only 4\% use $\rho$ a constant, and 6\% use $\rho$ estimated to second order accuracy. We conclude that for shader programs, the simple choice of $\rho = 0$ in most cases suffices.


Note that our brick shader gives poor results for the method of Dorn~et~al.~\shortcite{dorn2015}, while in that paper, a brick shader with similar appearance shows good results. This is because the brick shader in Dorn~et~al.~\shortcite{dorn2015} was implemented using floor() functions which can each be bandlimited independently, and then a good result is obtained by linearity of the integral. In our paper, we implemented a number of shaders using the fract() function to perform tilings that are exactly or appropriately periodic, including the brick shader. The fract() function ends up being more challenging to bandlimit for the framework of Dorn~et~al.~\shortcite{dorn2015}, but our method can handle such shaders.

\begin{table}
    \centering
    \caption{Statistics of which approximations were chosen for different shaders. We show statistics for the 7 program variants for the shaders presented in \fig{fig:teaser} and \fig{fig:results}. We also show aggregate statistics over all 21 shaders, with each shader's contribution weighted equally. For the aggregate statistics we  report statistics from the entire Pareto frontier, as well as for each shader choosing only the slowest, fastest, or median speed program variant. Our results show that a rich variety of our different approximation rules are needed for the best performance.}
    \setlength{\tabcolsep}{1pt}
    \begin{tabular}{|c|c|c|c|c|}
        \hline
         Shader & Dorn~et~al. & Adaptive & Monte Carlo & None  \\
          & ~\shortcite{dorn2015} & Gaussian & Sampling & \\
         \hline
         Bricks w/ None & 28\% & 0\% & 30\% & 29\% \\
         Checkerboard w/ Ripples & 66\% & 34\% & 0\% & 1\% \\
         Circles w/ None & 4\% & 21\% & 71\% & 4\% \\
         Color Circles w/ Bumps & 8\% & 47\% & 44\% & 0\% \\
         Fire w/ Bumps & 1\% & 7\% & 33\% & 60\% \\
         Quadratic sine w/ Ripples & 13\% & 80\% & 0\% & 8\% \\
         Zigzag w/ Ripples & 0\% & 91\% & 1\% & 8\% \\
         
         All shaders (Pareto frontier) & 29\% & 15\% & 25\% & 30\% \\
         All shaders (fastest time) & 13\% & 10\% & 0\% & 77\% \\
         All shaders (median time) & 20\% & 19\% & 49\% & 13\% \\
         All shaders (slowest time) & 10\% & 27\% & 49\% & 14\% \\
         \hline
    \end{tabular}
    \label{tab:stats}
\end{table}

We performed our evaluation on an Intel Core i7 6950X 3 GHz (Broadwell), with 10 physical cores (20 hyperthreaded), and 64 GB DDR4-2400 RAM. All shaders were evaluated on the CPU using parallelization. The tuning of each shader took between 1 and 3 hours of wall clock time. However, we note that good program variants are available after just a few minutes, and most of the remaining tuning time is spent making slight improvements to the best individuals. Also, our tuner is intentionally a research prototype that is not particularly optimized: it could be significantly faster if it was parallelized more effectively, cached more redundant computations, or targeted the GPU.

\begin{figure*}
\setlength{\tabcolsep}{1pt}
\setlength{\w}{1.3in}
\centering
\begin{tabular}{cccccc}
 & Ground Truth & No Antialiasing & Our Result & Dorn~et~al.~2015 & MSAA \\
 


\rottext{Checkerboard}{with Ripples} &
\includegraphics[width=\w]{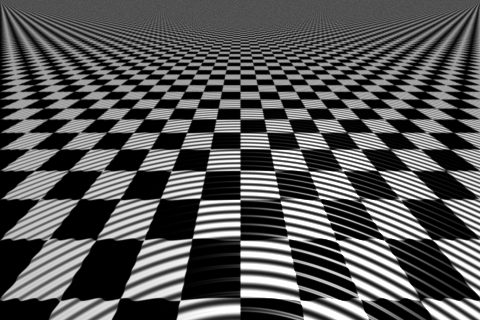} & \includegraphics[width=\w]{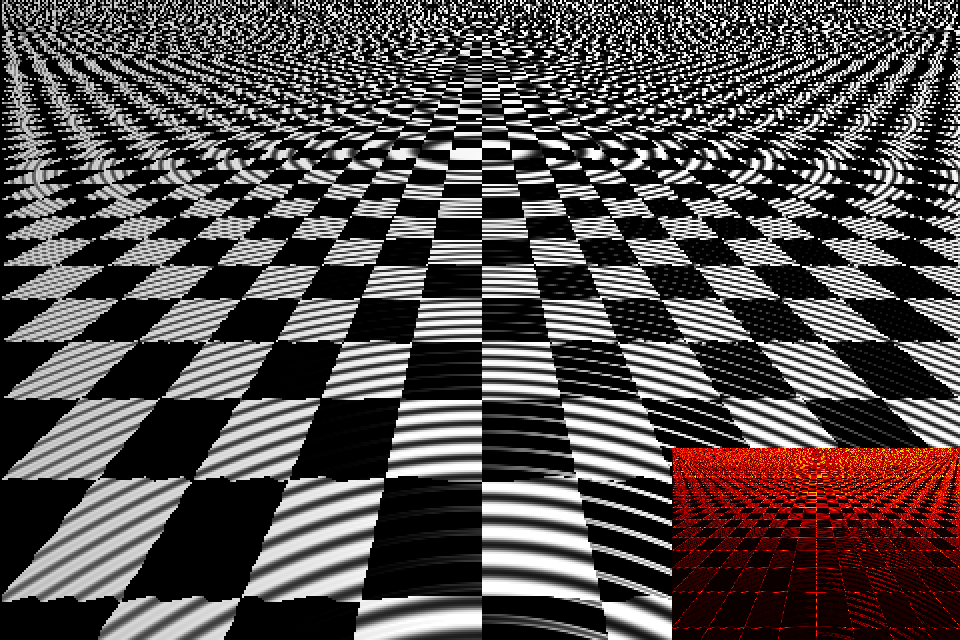} & \includegraphics[width=\w]{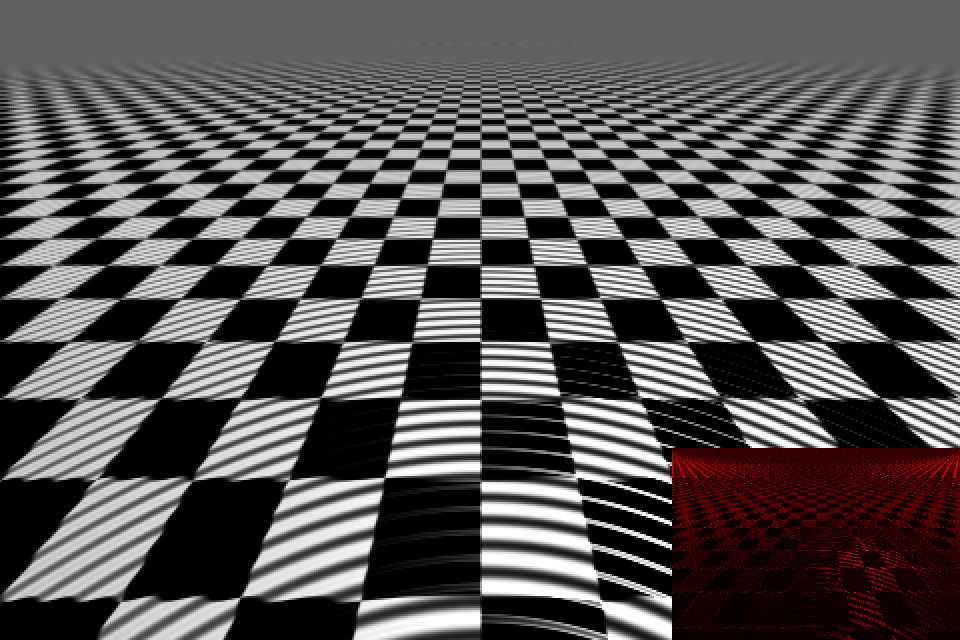} & \includegraphics[width=\w]{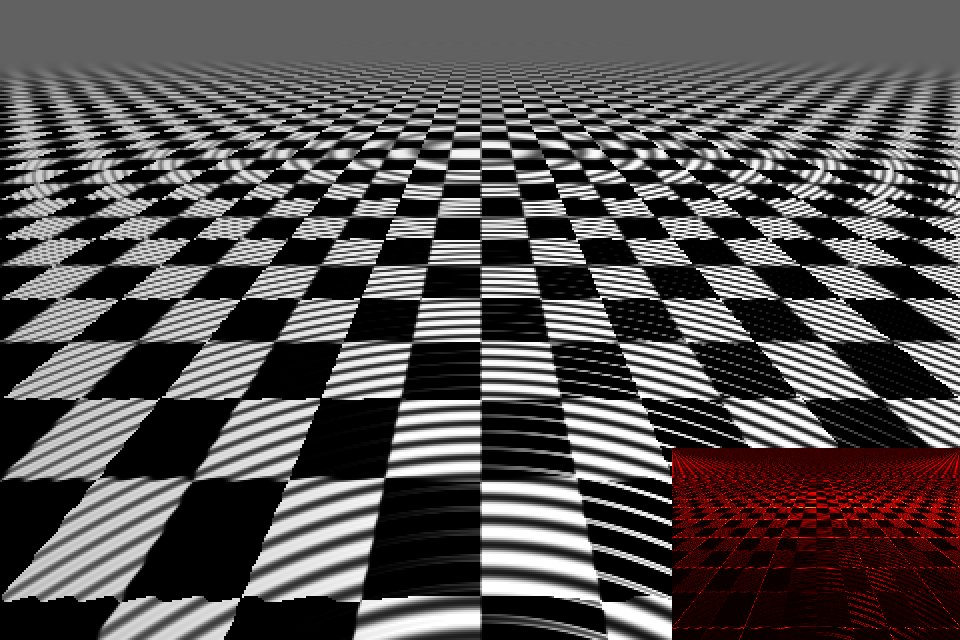} & \includegraphics[width=\w]{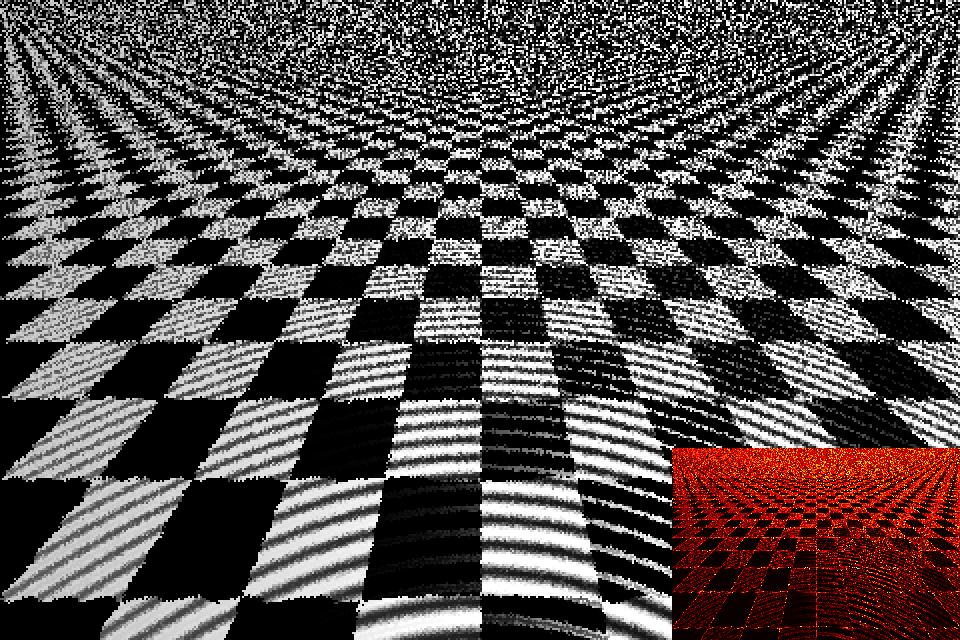}\tablegap

& & {\small 30 ms, $L^2$ error: 0.194} & {\small 54 ms (2x), $L^2$ error: 0.071} & {\small 50 ms (2x), $L^2$ error: 0.102} & {\small 47 ms (2x), $L^2$ error: 0.233} \\

\rottext{Circles}{with None} &
\includegraphics[width=\w]{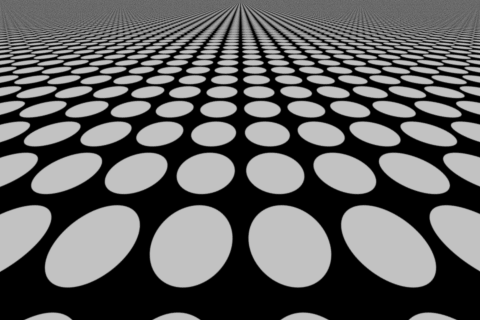} & \includegraphics[width=\w]{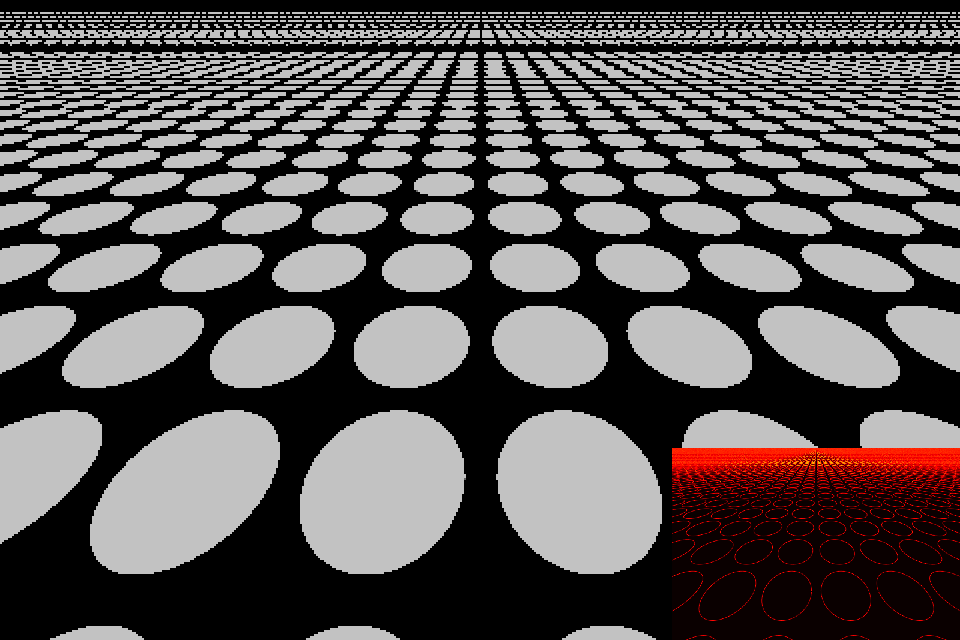} & \includegraphics[width=\w]{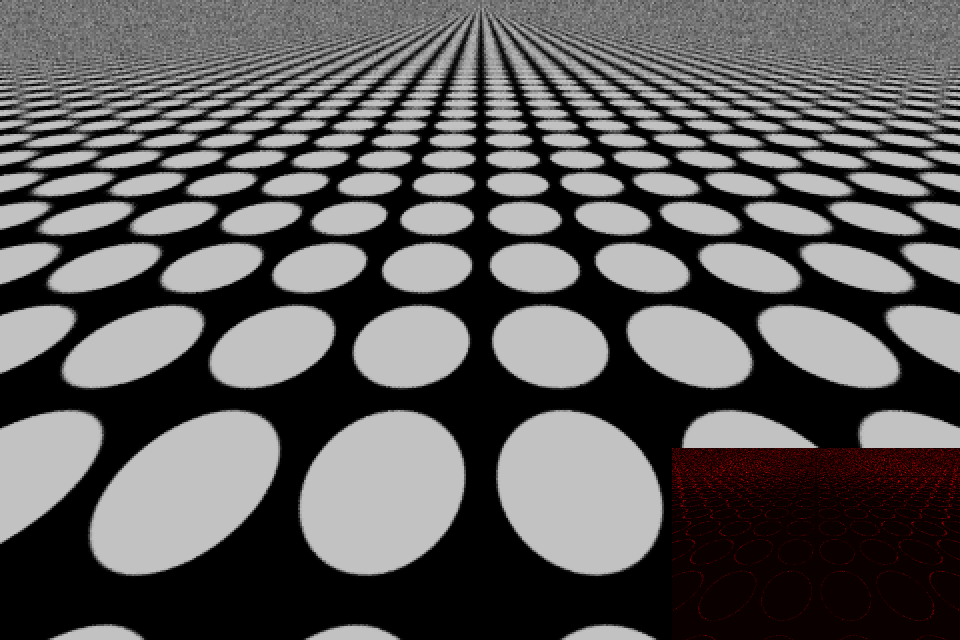} & \includegraphics[width=\w]{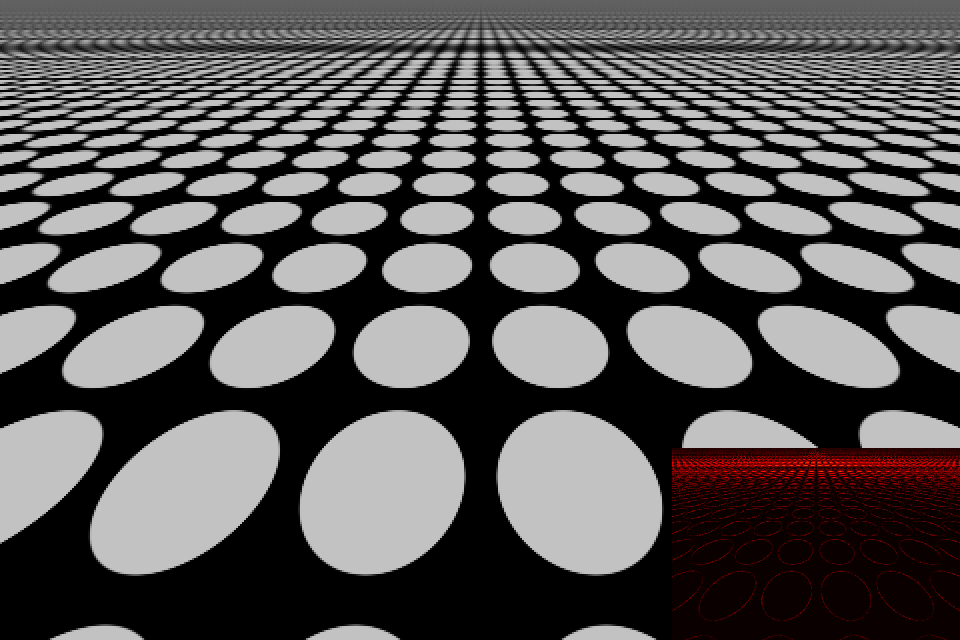} & \includegraphics[width=\w]{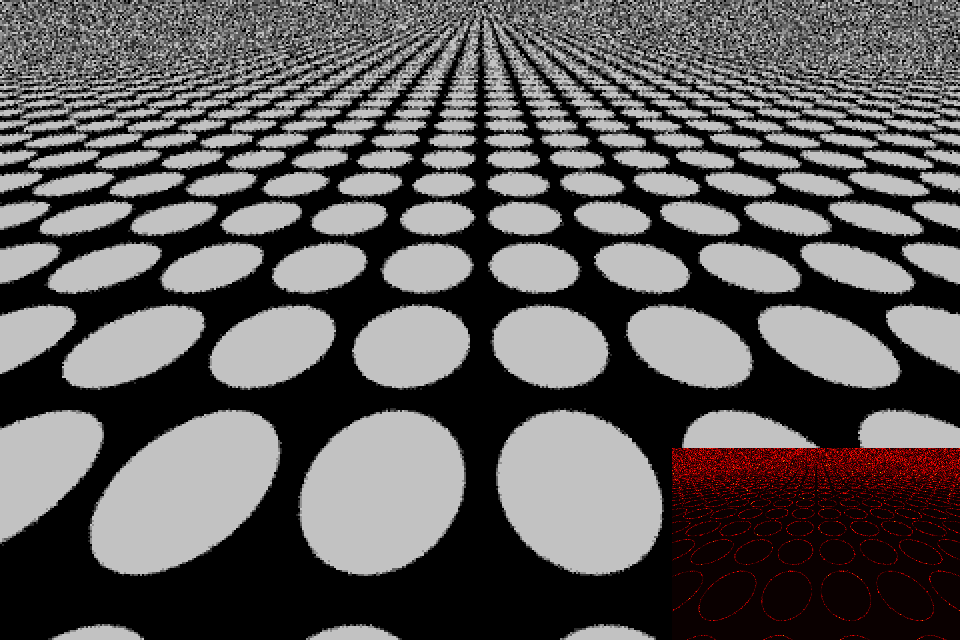}\tablegap

& & {\small 20 ms, $L^2$ error: 0.148} & {\small 71 ms (4x), $L^2$ error: 0.035} & {\small 39 ms (2x), $L^2$ error: 0.063} & {\small 67 ms (3x), $L^2$ error: 0.087} \\

\rottext{Color Circles}{with Bumps} &
\includegraphics[width=\w]{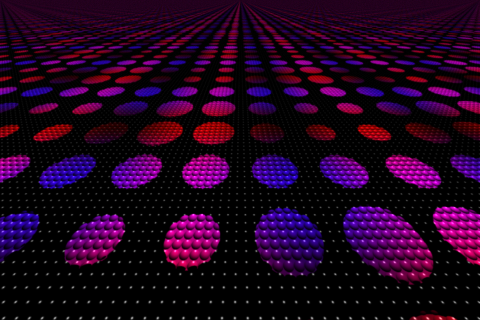} & \includegraphics[width=\w]{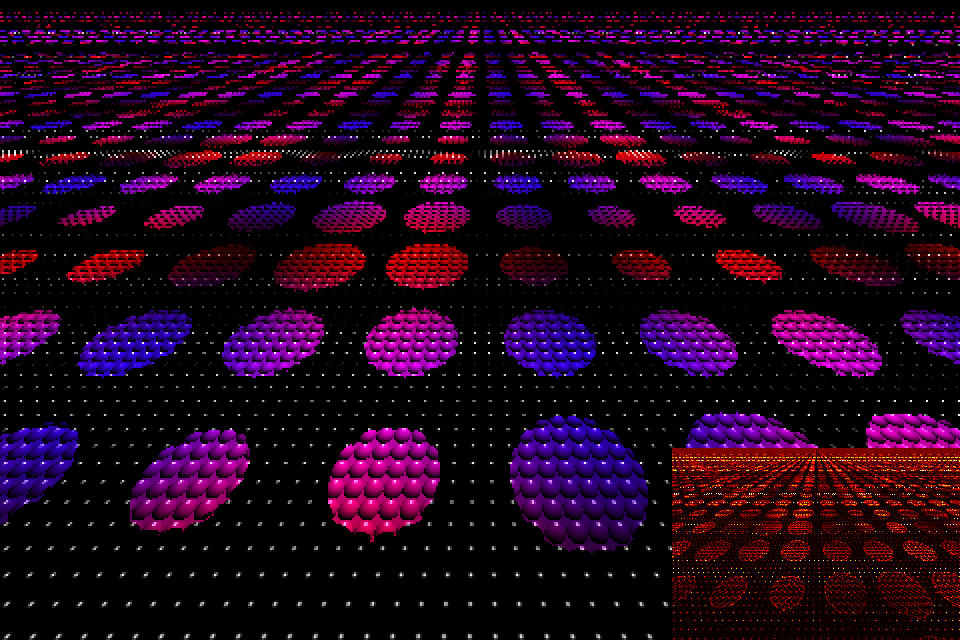} & \includegraphics[width=\w]{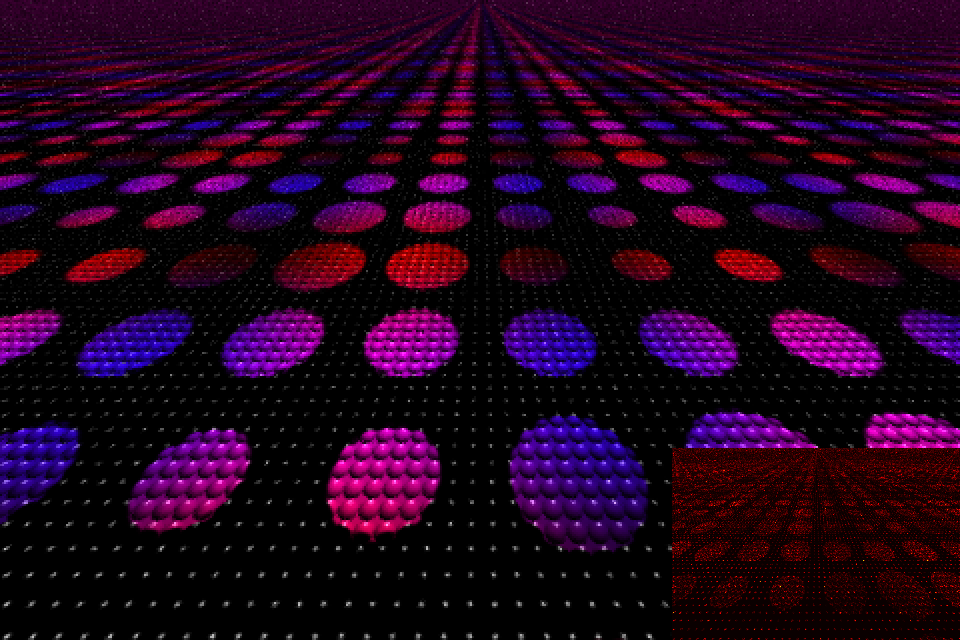} & \includegraphics[width=\w]{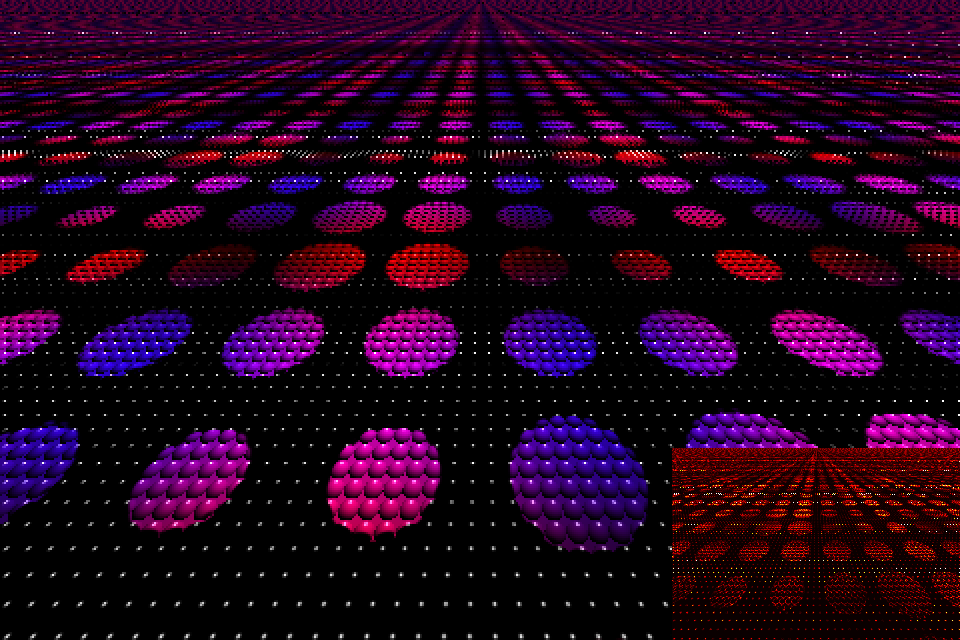} & \includegraphics[width=\w]{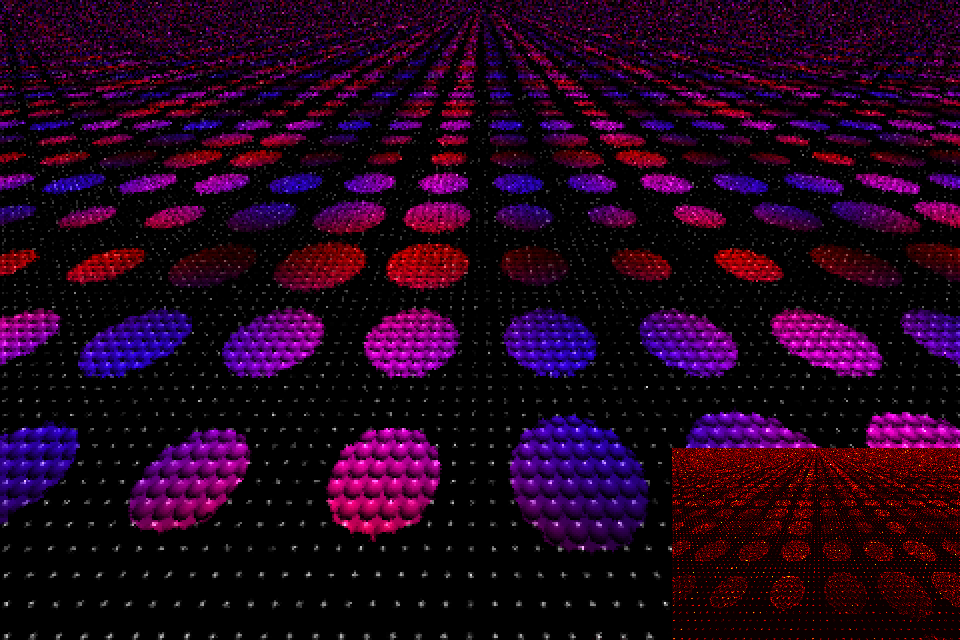}\tablegap

& & {\small 37 ms, $L^2$ error: 0.098} & {\small 149 ms (4x), $L^2$ error: 0.039} & {\small 56 ms (2x), $L^2$ error: 0.079} & {\small 112 ms (3x), $L^2$ error: 0.061} \\

\rottext{Fire}{with Bumps} &
\includegraphics[width=\w]{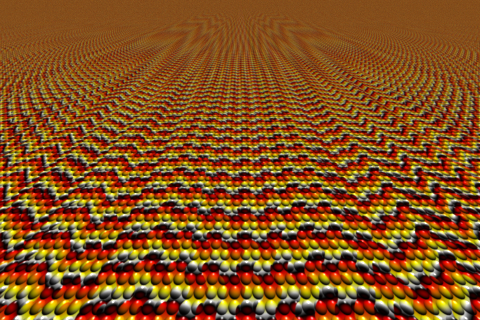} & \includegraphics[width=\w]{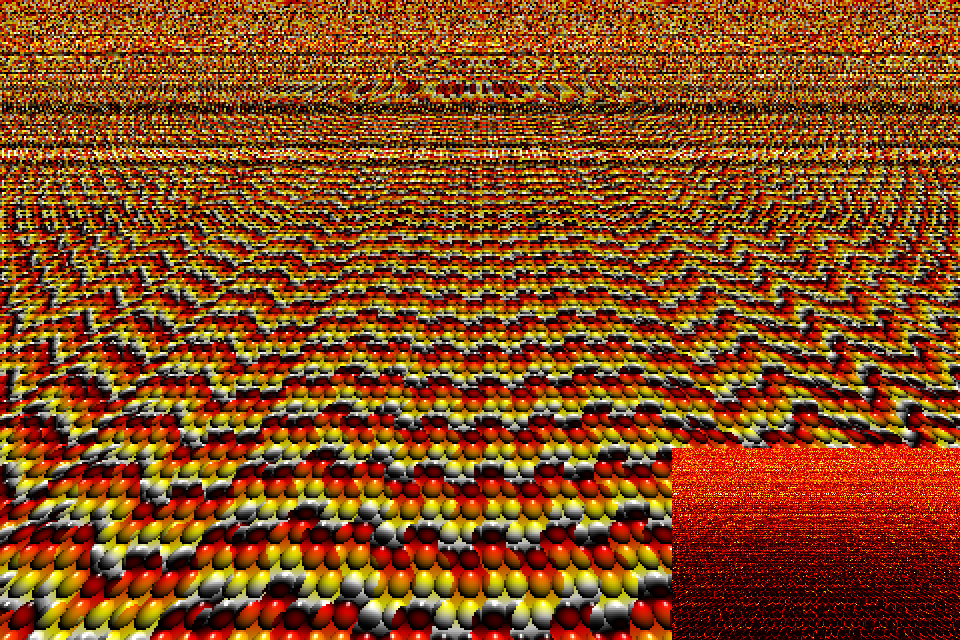} & \includegraphics[width=\w]{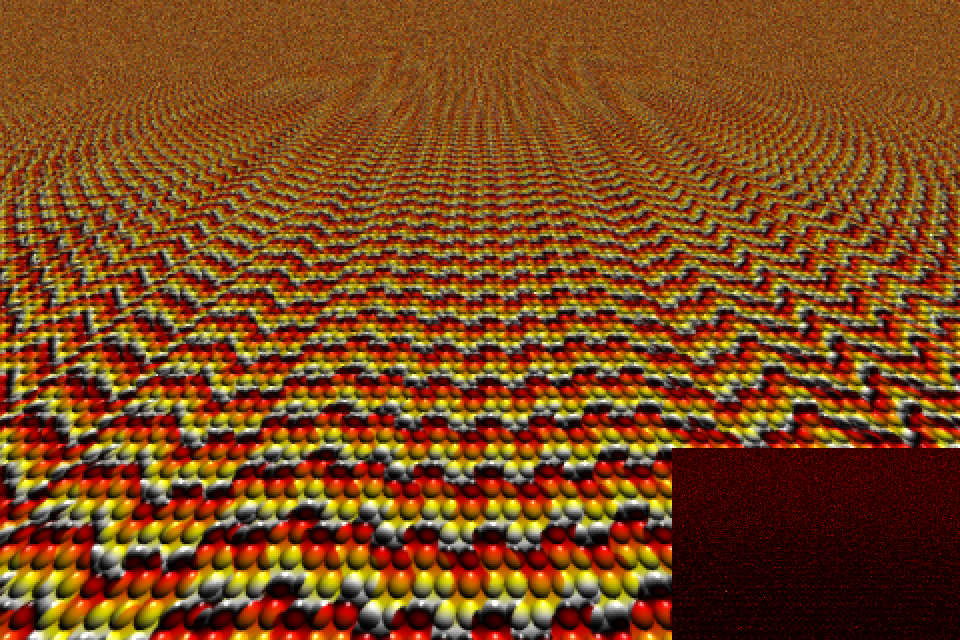} & \includegraphics[width=\w]{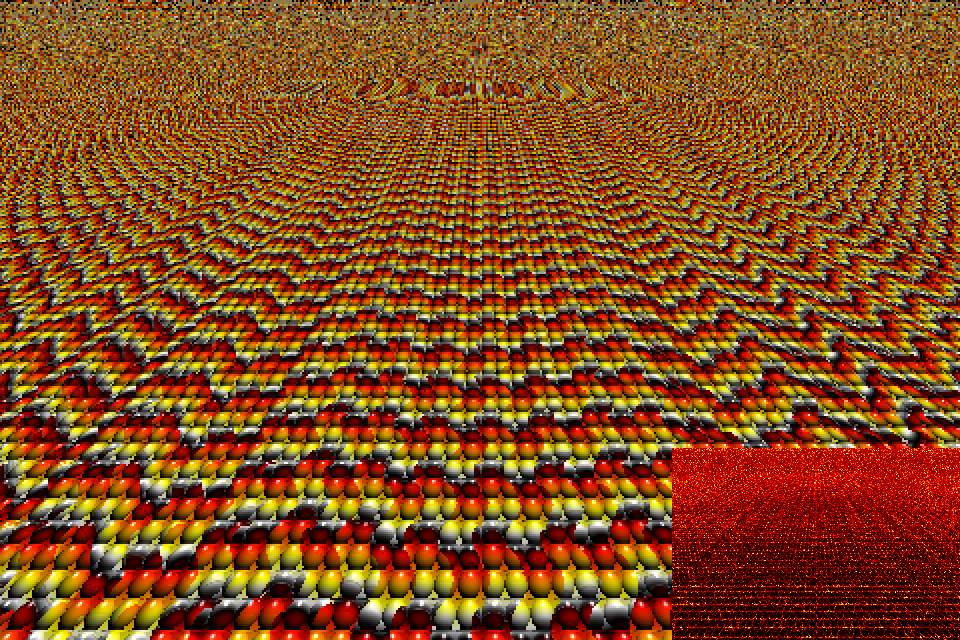} & \includegraphics[width=\w]{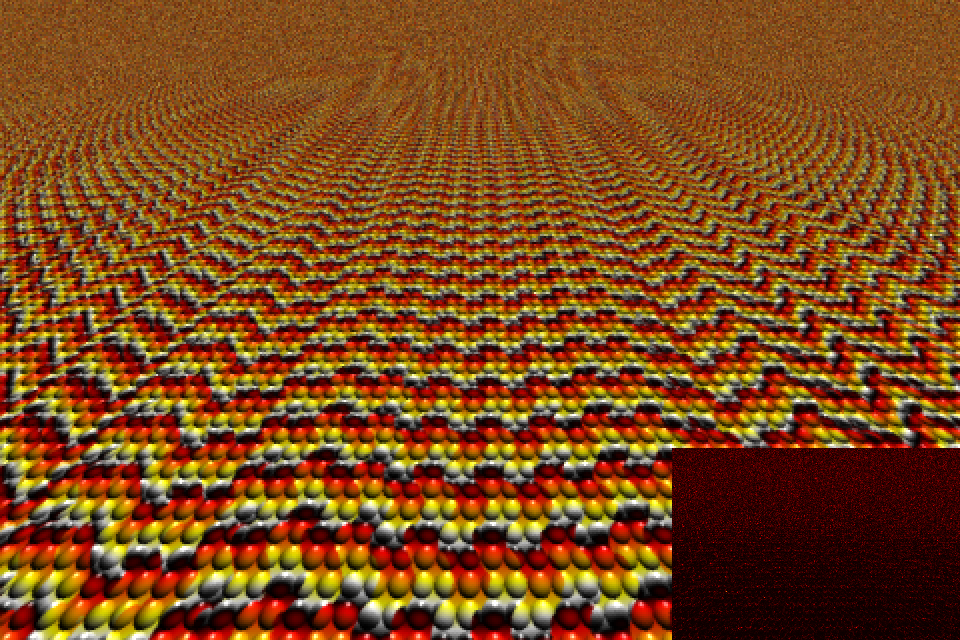}\tablegap

& & {\small 39 ms, $L^2$ error: 0.170} & {\small 698 ms (18x), $L^2$ error: 0.037} & {\small 67 ms (2x), $L^2$ error: 0.136} & {\small 705 ms (18x), $L^2$ error: 0.037} \\

\rottext{Quadratic Sine}{with Ripples} &
\includegraphics[width=\w]{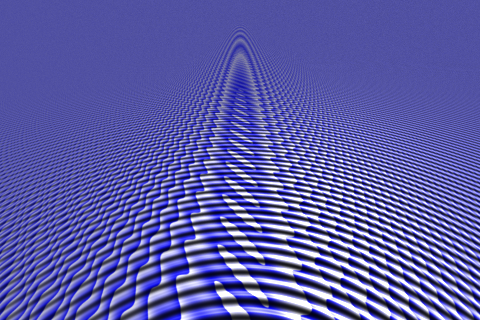} & \includegraphics[width=\w]{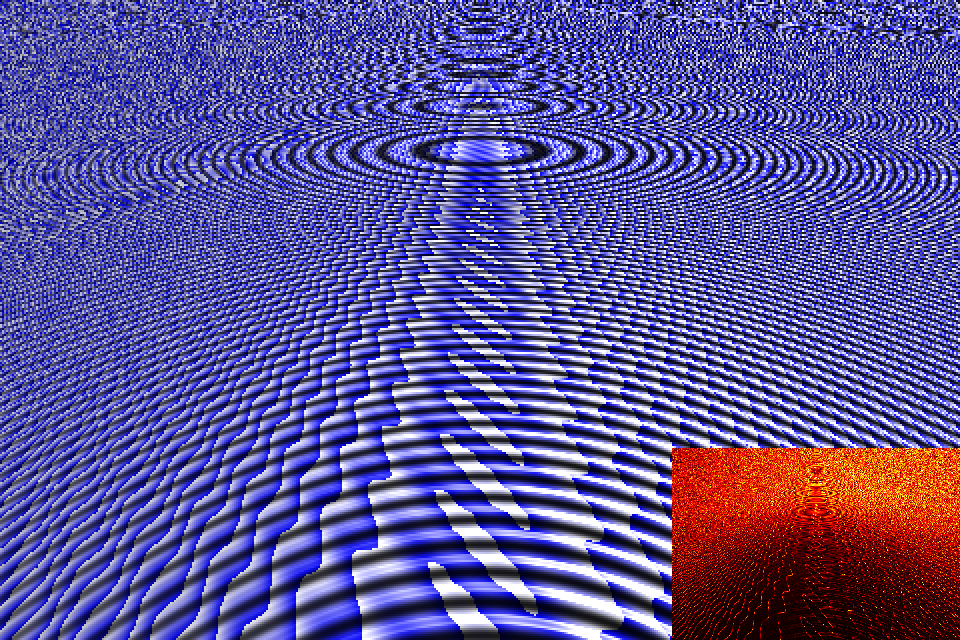} & \includegraphics[width=\w]{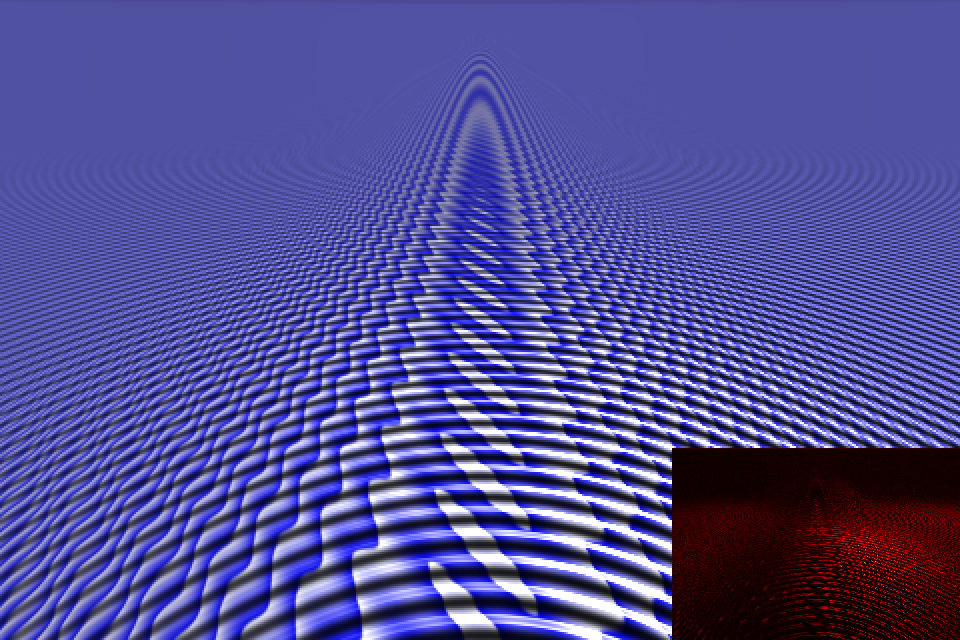} & \includegraphics[width=\w]{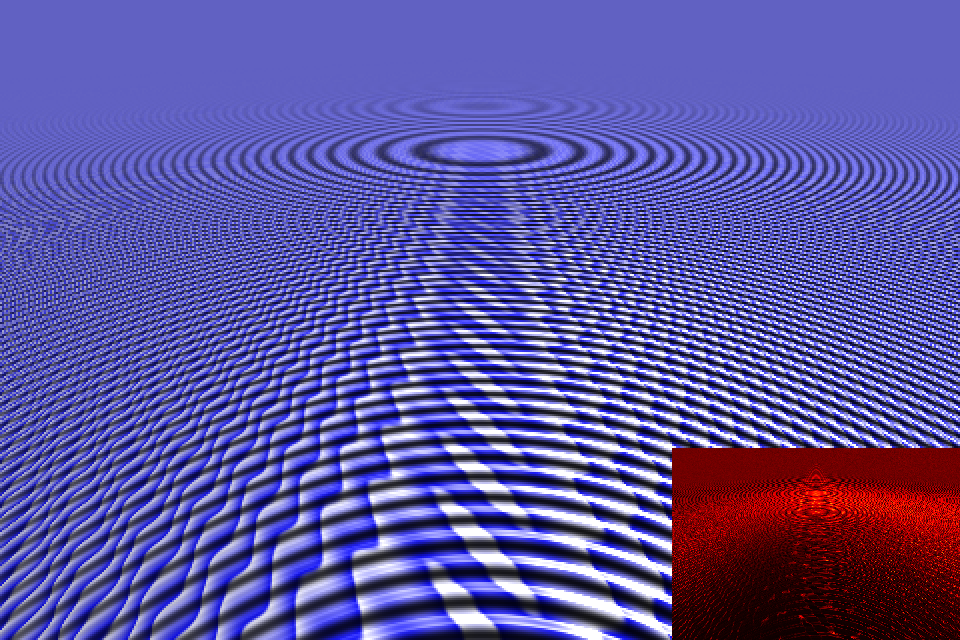} & \includegraphics[width=\w]{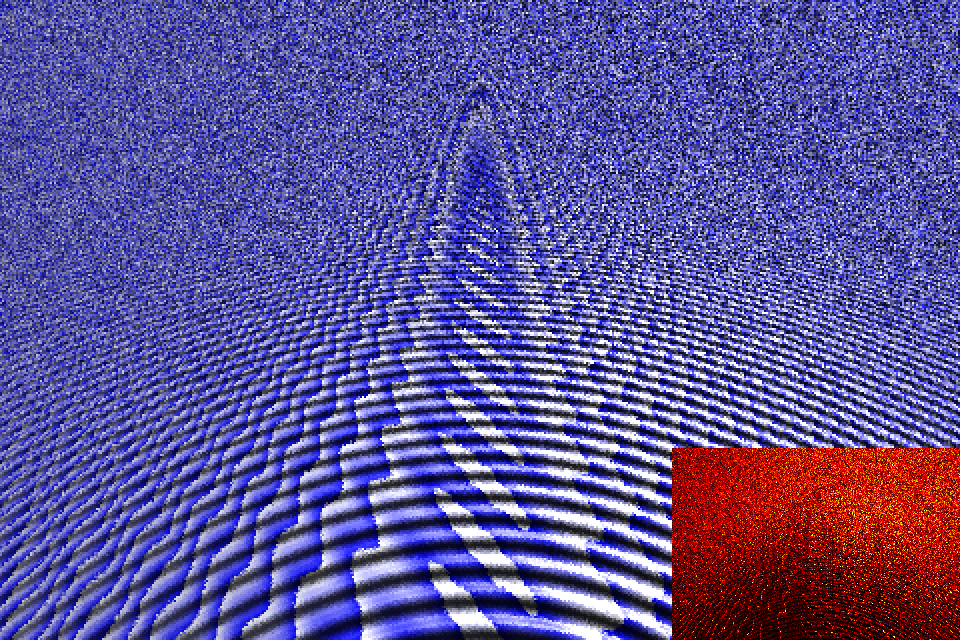}\tablegap

& & {\small 38 ms, $L^2$ error: 0.184} & {\small 81 ms (2x), $L^2$ error: 0.045} & {\small 59 ms (2x), $L^2$ error: 0.094} & {\small 99 ms (3x), $L^2$ error: 0.158} \\

\rottext{Zigzag}{with Ripples} &
\includegraphics[width=\w]{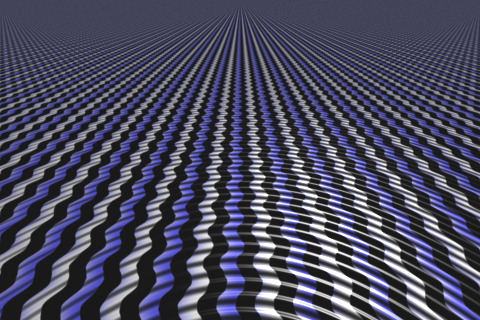} & \includegraphics[width=\w]{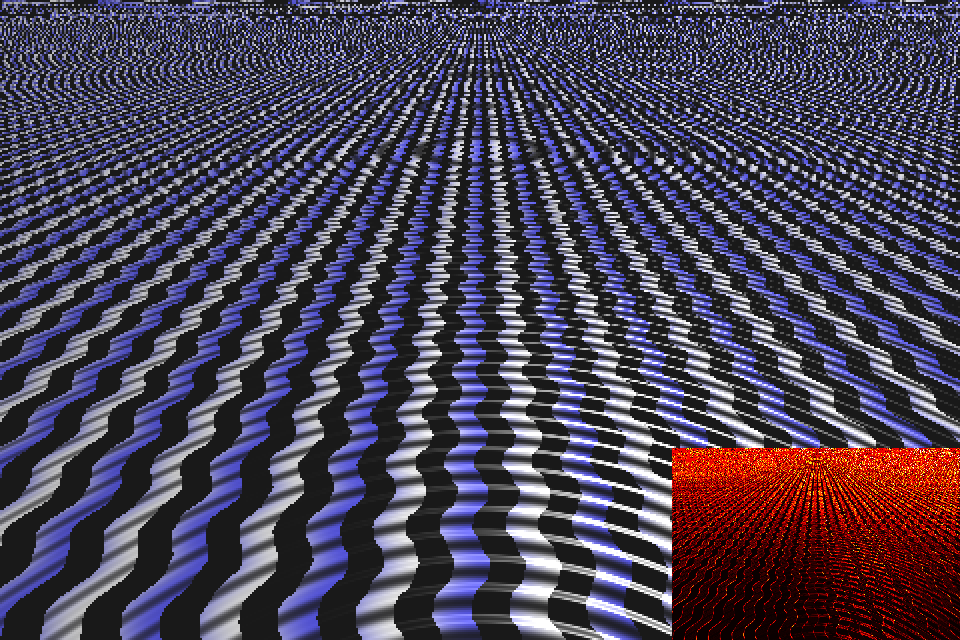} & \includegraphics[width=\w]{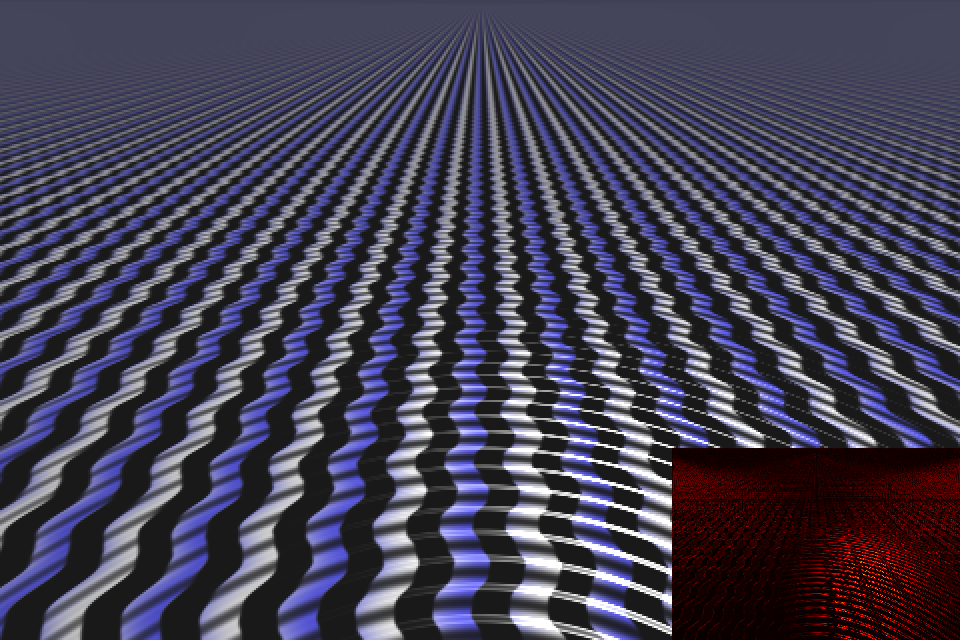} & \includegraphics[width=\w]{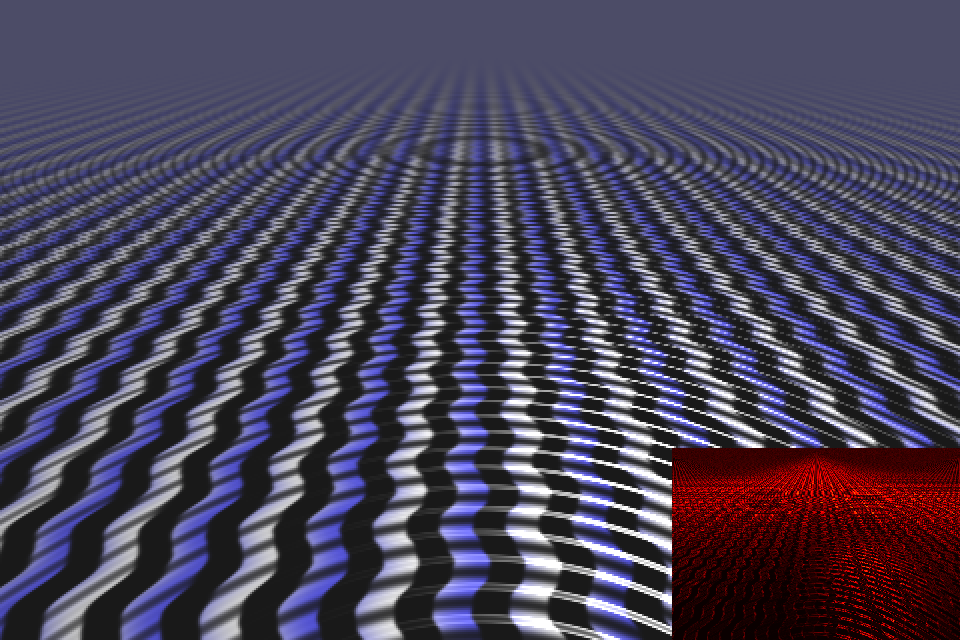} & \includegraphics[width=\w]{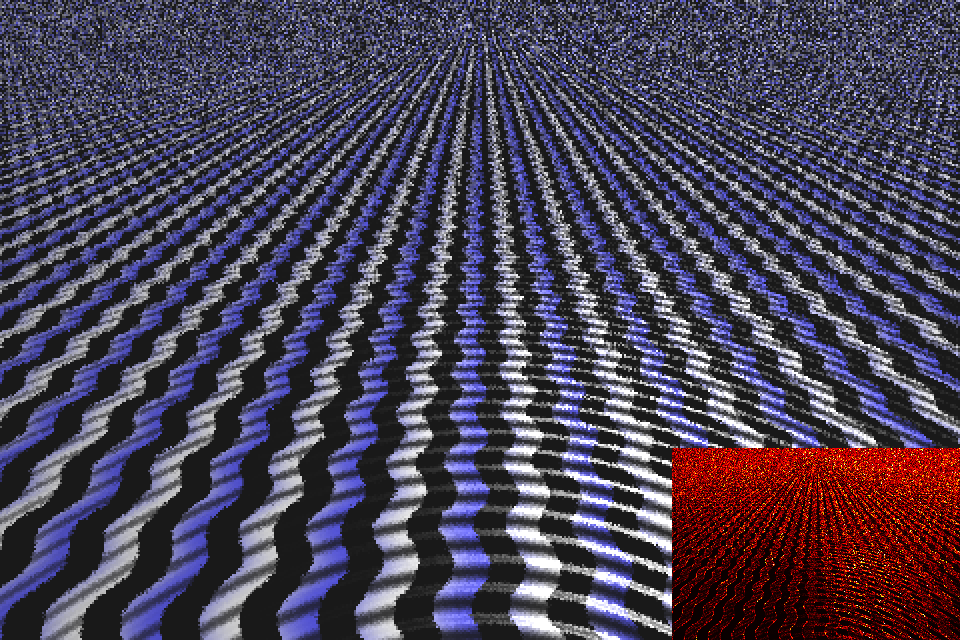}\tablegap

& & {\small 57 ms, $L^2$ error: 0.139} & {\small 77 ms (1x), $L^2$ error: 0.045} & {\small 59 ms (1x), $L^2$ error: 0.072} & {\small 83 ms (1x), $L^2$ error: 0.122} \\


\end{tabular}

\caption{Selected result images for 6 shaders. Please see the supplemental video for a comprehensive comparison of all shaders. Reported below each shader are the time to render a frame, time relative to no antialiasing, and $L^2$ error. Please zoom in to see aliasing and noise patterns in the different methods. Program variants with comparable time were selected: see \sect{sec:evaluation} for more details. Note that the amount of aliasing and error for our result is significantly less than Dorn~et~al.~\shortcite{dorn2015}. Note that we typically have significantly less error and noise than the comparable MSAA results. }
\label{fig:results}
\end{figure*}

\begin{figure*}
\setlength{\tabcolsep}{-2pt}
\setlength{\w}{1.07in}
\centering
\begin{tabular}{ccccccc}
 \includegraphics[width=\w]{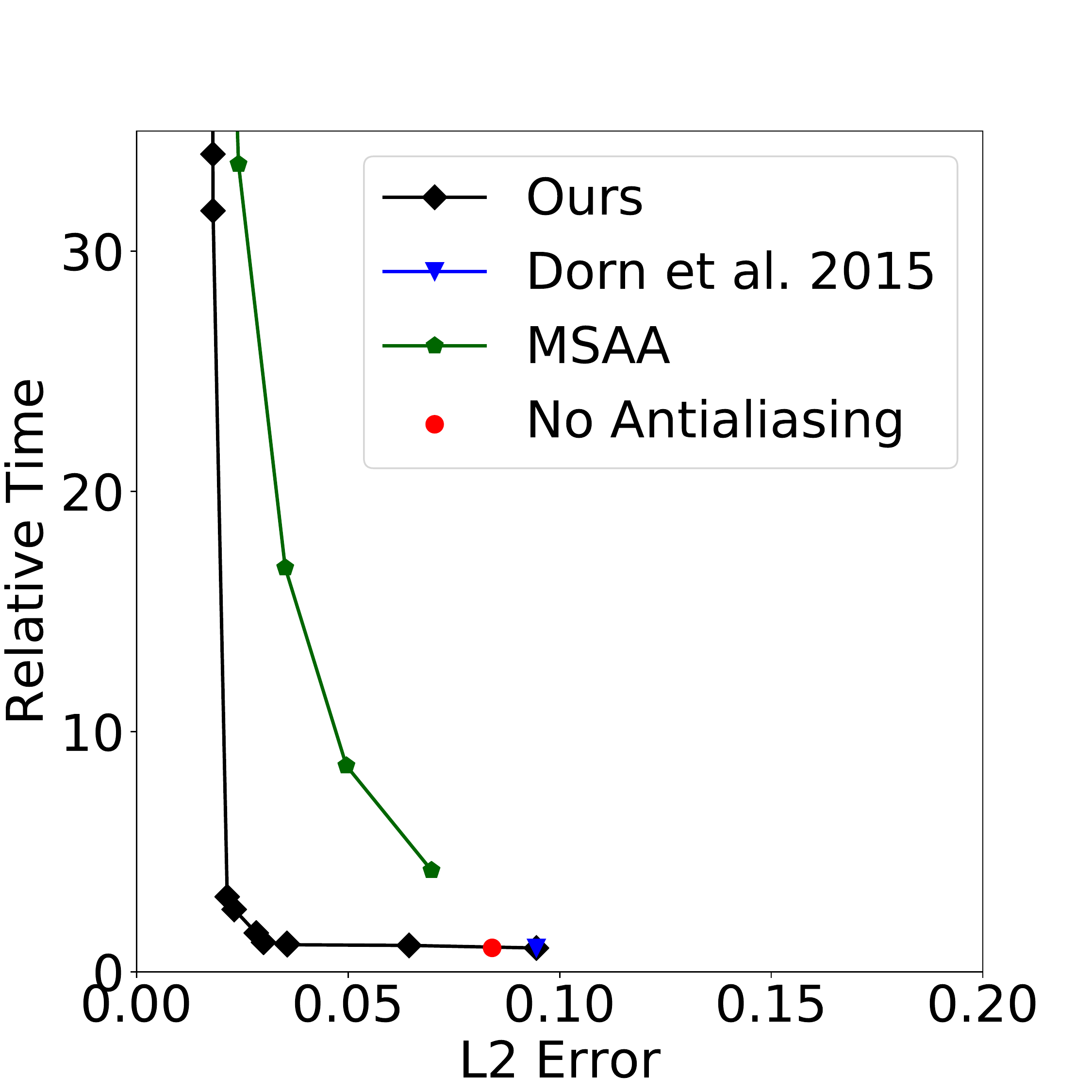} & \includegraphics[width=\w]{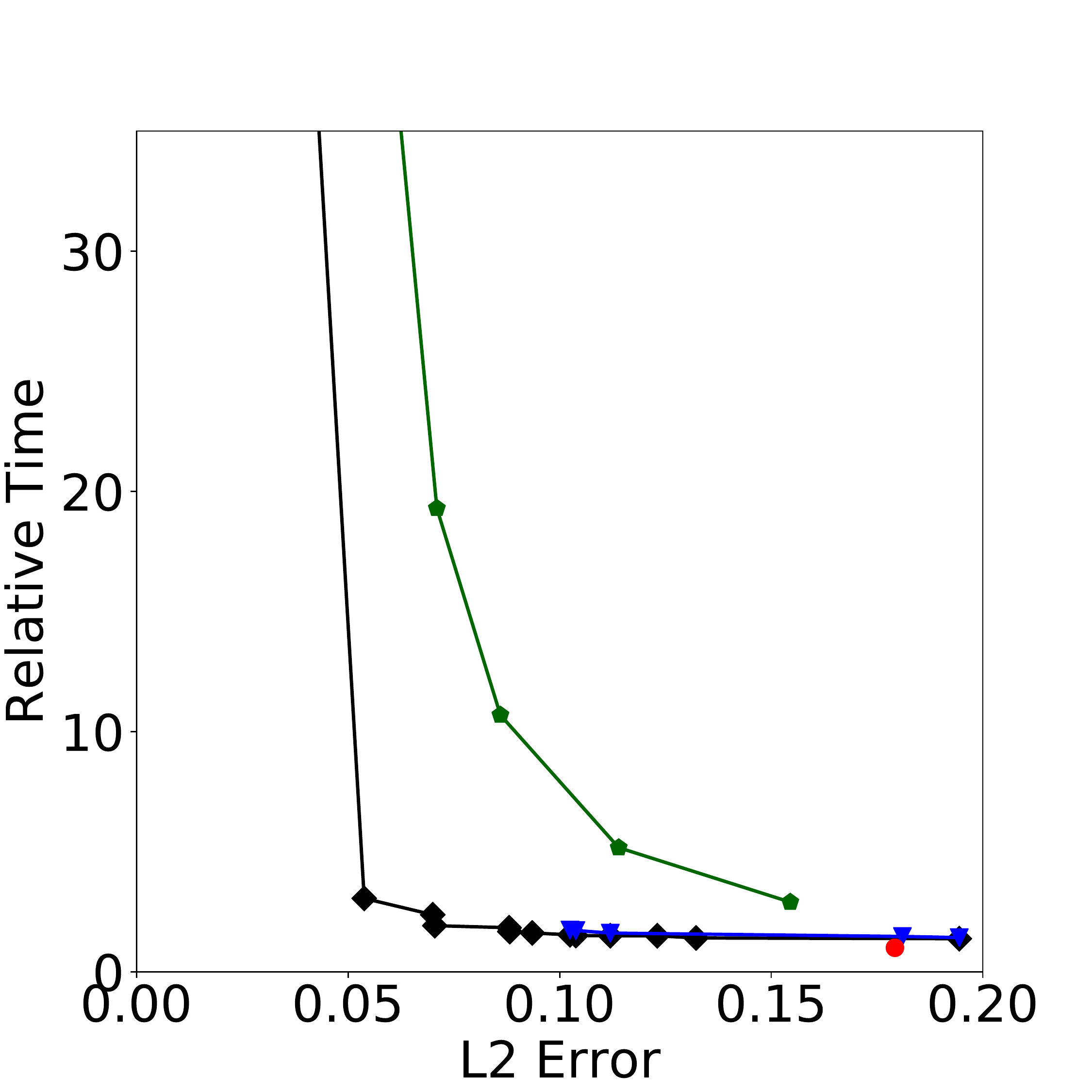} & \includegraphics[width=\w]{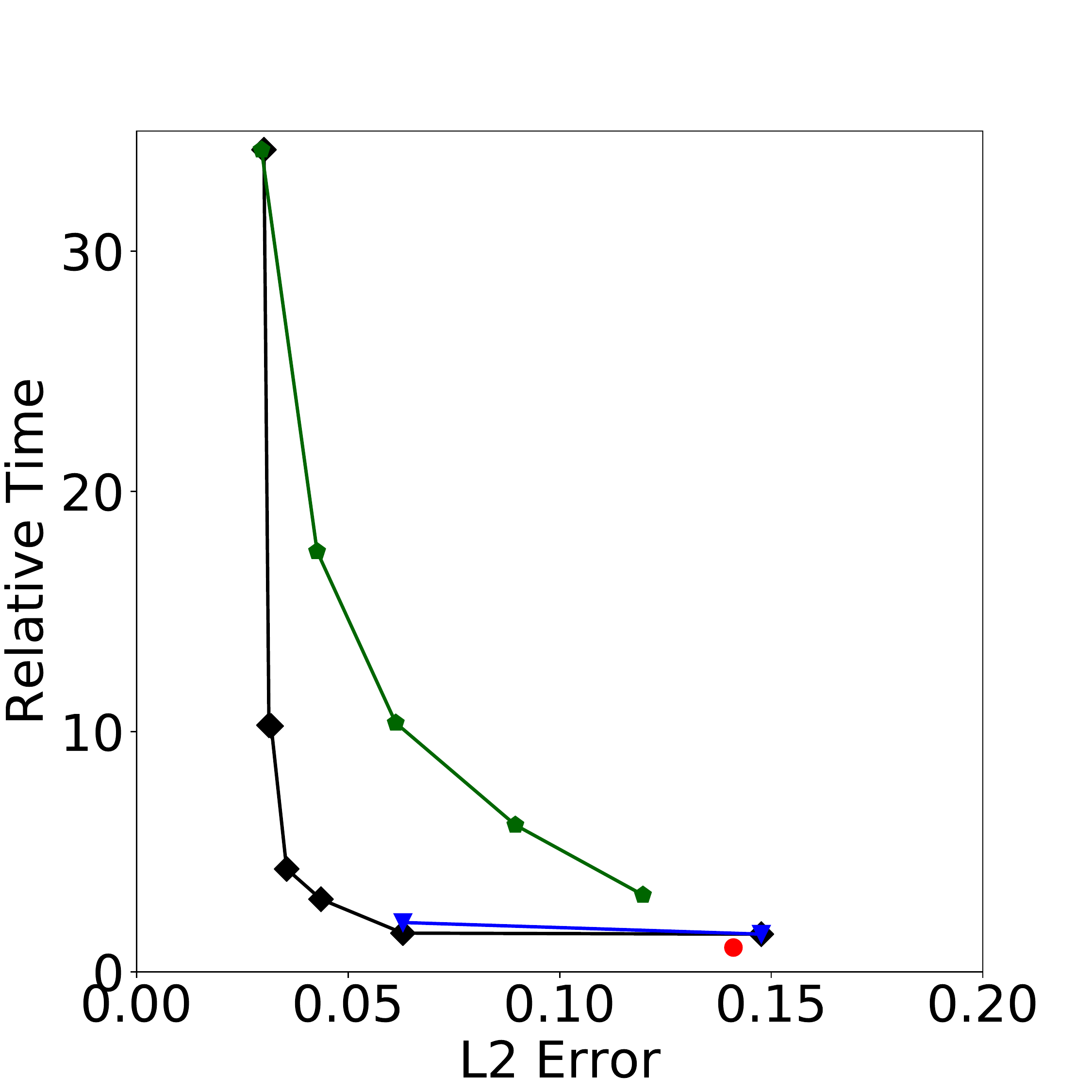} & \includegraphics[width=\w]{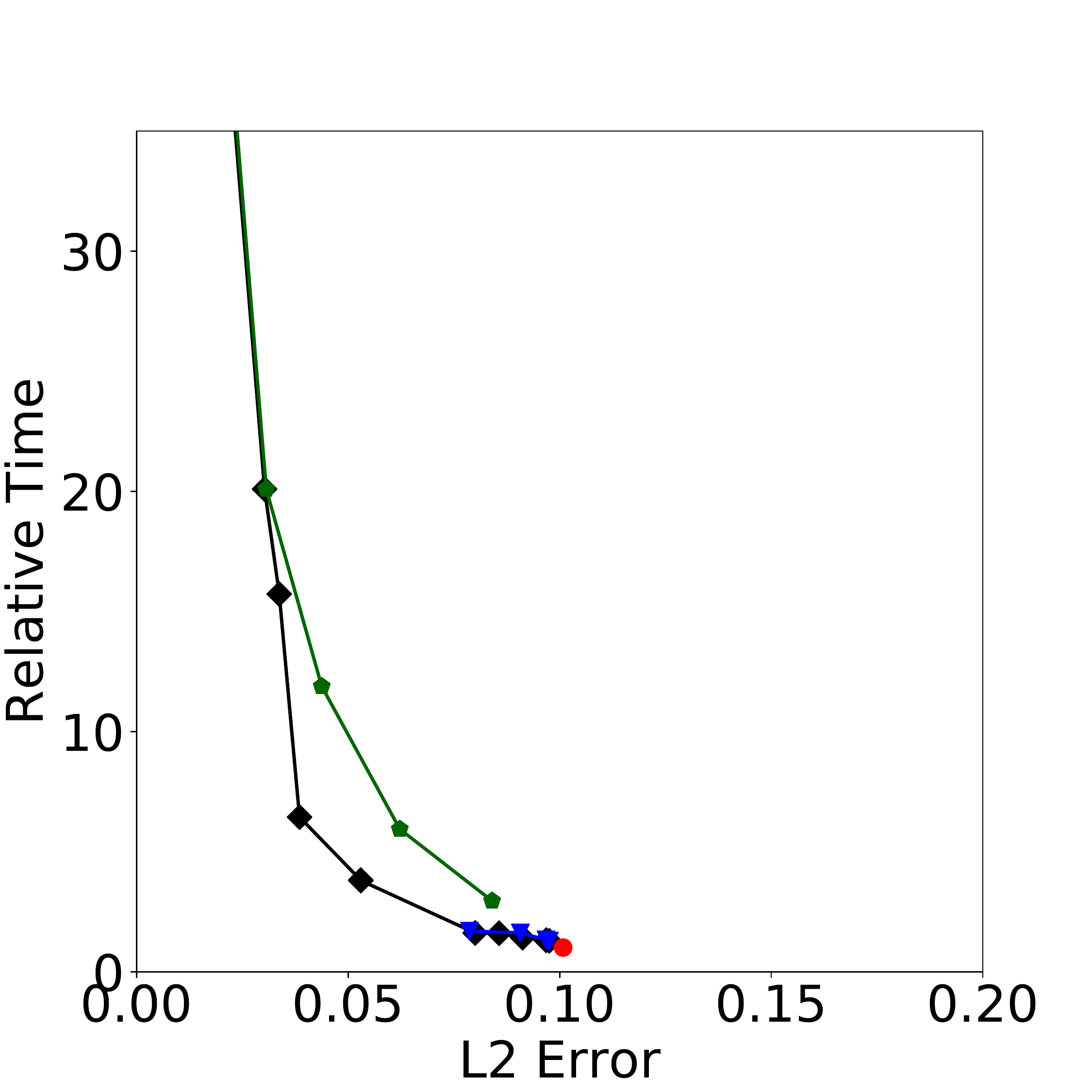} &
 \includegraphics[width=\w]{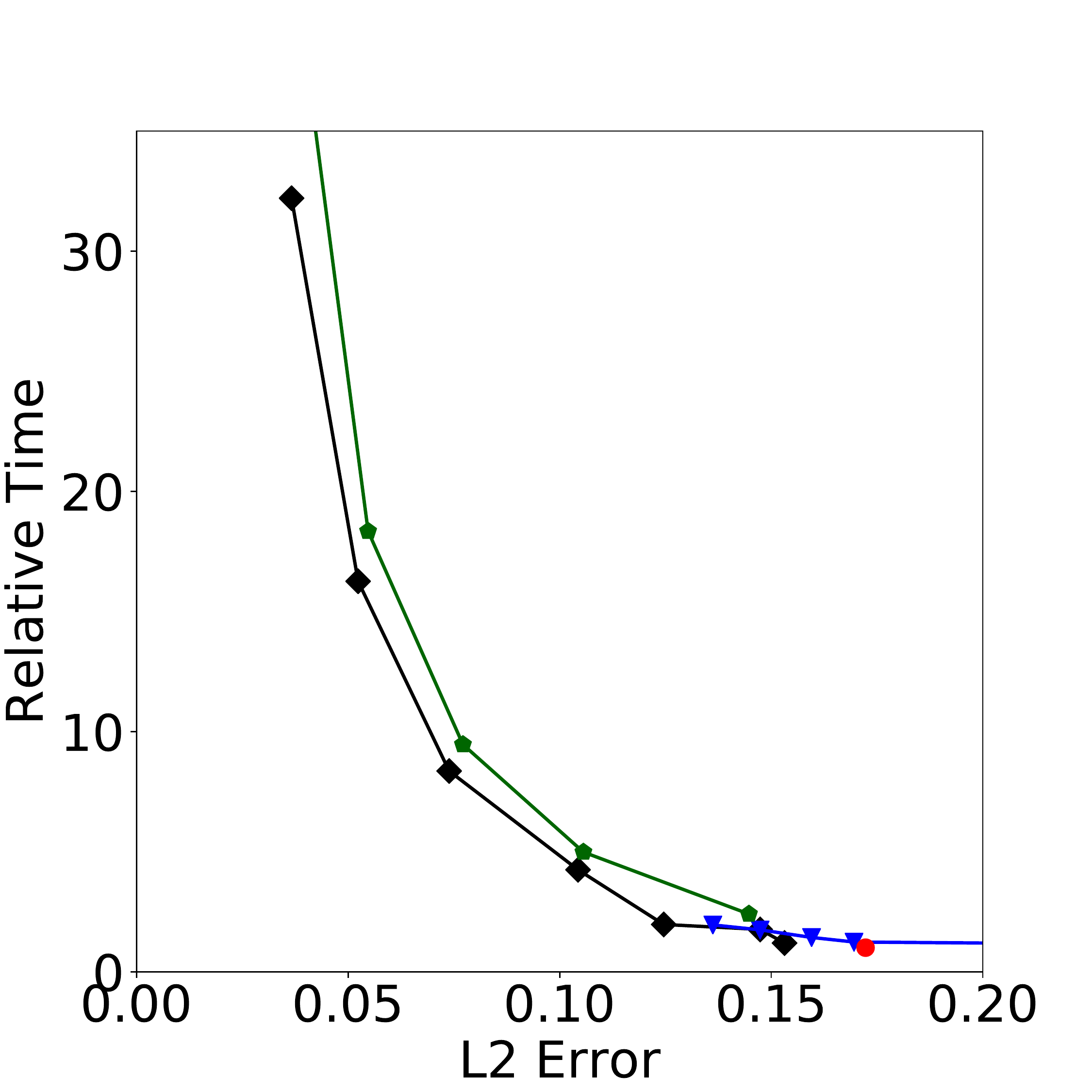} & \includegraphics[width=\w]{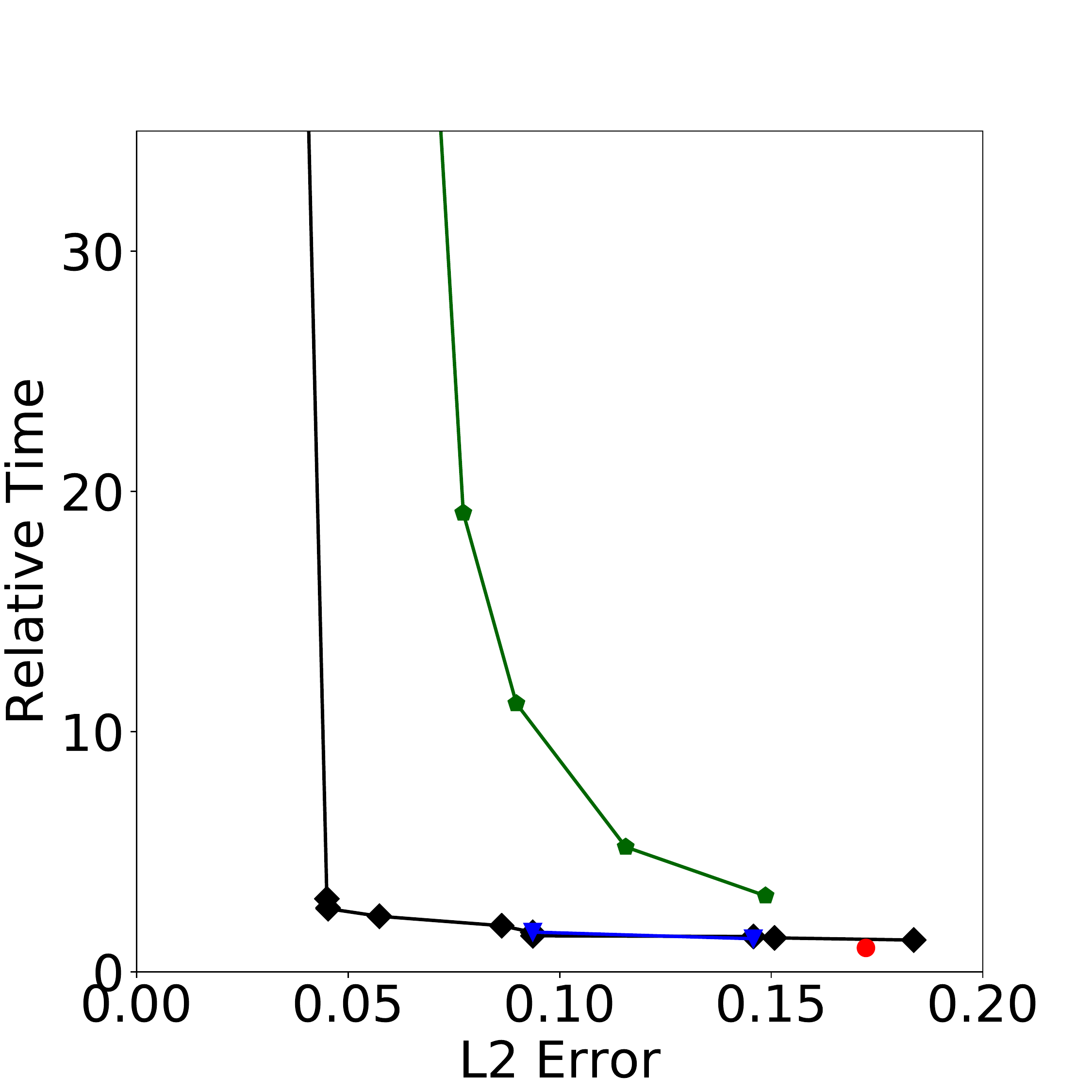} &
\includegraphics[width=\w]{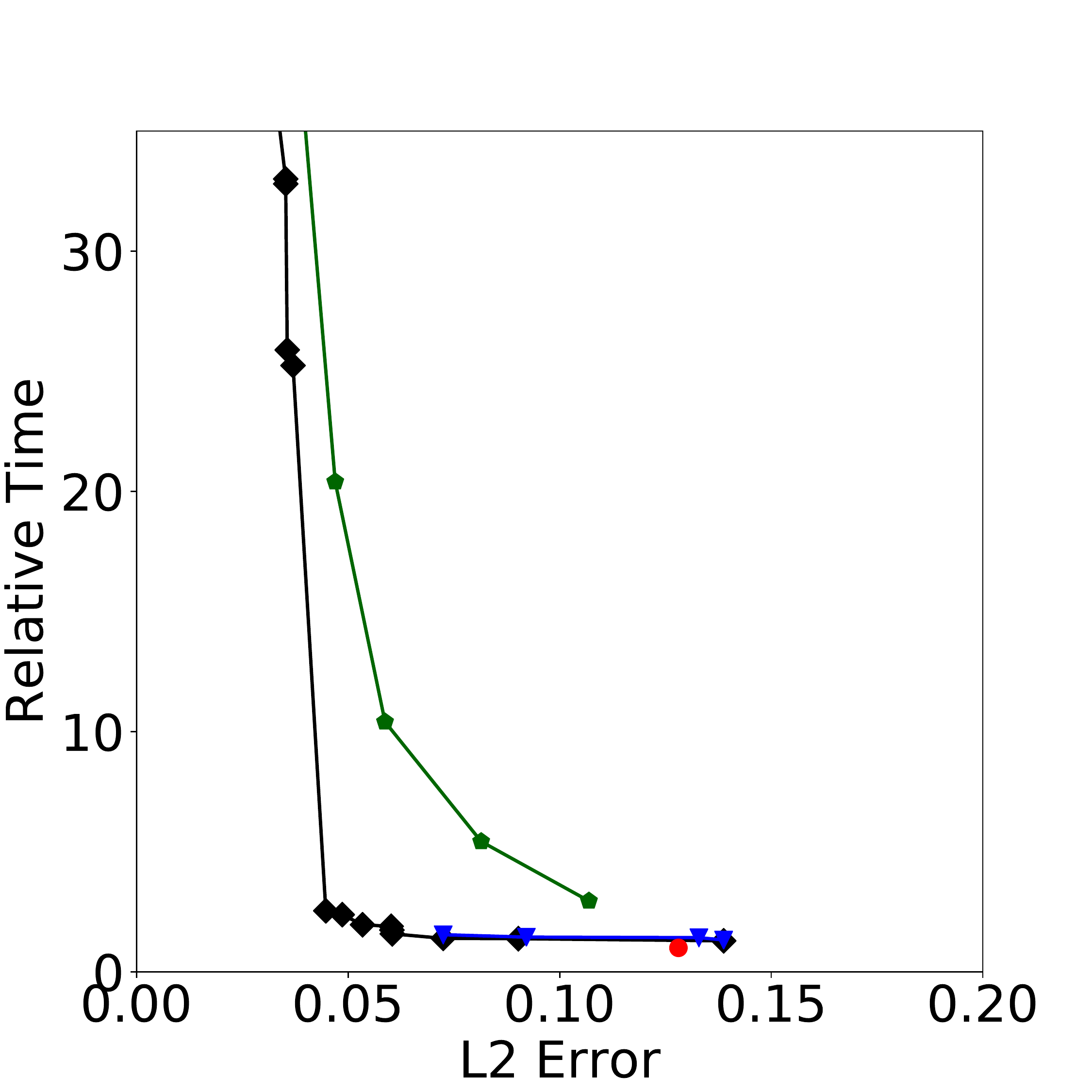} \\
 {\small Bricks} & {\small Checkerboard} &{\small  Circles} & {\small Color Circles} & {\small Fire} & {\small Quadratic Sine} & {\small Zigzag} \\
 {\small with None} &  {\small with Ripples} &  {\small with None} & {\small with Bumps} & {\small with Bumps} & {\small with Ripples} & {\small with Ripples} \\


\end{tabular}

\caption{Time versus error plots for the 7 shaders in \fig{fig:teaser} and \fig{fig:results}. Here we show the Pareto frontier of  program variants that optimally trade off running time and $L^2$ error. We show results for our method, Dorn~et~al~\shortcite{dorn2015}, MSAA with varying numbers of samples, and the input shader without antialiasing. Note that our approach typically has significantly less error than Dorn~et~al~\shortcite{dorn2015} and is frequently an order of magnitude faster than MSAA for comparable error.}
\label{fig:pareto_results}
\end{figure*}




%
\section{Discussion and Conclusion}

In this paper, we presented a novel compiler framework that smoothes an arbitrary program over the floats by convolving it with a Gaussian kernel. We explained several different approximations and discussed the accuracy of each. We then demonstrated that our framework allows shader programs to be automatically bandlimited. This shader bandlimiting application achieves state-of-the-art results: it often has substantially better error than Dorn~et~al.~\shortcite{dorn2015} even after our improvements, and is frequently an order of magnitude faster than multi-sample antialiasing (MSAA). Our framework is quite general, and we believe it could be useful for other problems in graphics and across the sciences. In order to facilitate reproducible research, we intend to release our source code under an open source license.

\section*{Acknowledgements}

We thank Zack Verham for authoring some shaders, Ning Yu for helping produce the supplementary video, and Francesco Di Plinio for providing references about the heat equation and its Taylor expansion. This project was partially funded by NSF grants HCC 1011444 and SHF 1619123.

\appendix
\section{Table of Smoothed Formulas}

In \tbl{tbl:bandlimited}, we show a table of functions and their corresponding convolutions with box and Gaussian kernels. These are needed for the approximations we developed in \sect{sec:approximation}. This table can be viewed as an extension of the table presented in Dorn~et~al.~\shortcite{dorn2015}. Note that in particular, for each function $f(x)$, we report not only the result of smoothing $f(x)$ but also smoothing $f^2(x)$ (e.g. if we report $\cos(x)$ then we also report $\cos^2(x)$). This is needed to determine the standard deviations output by a given compute stage for the adaptive Gaussian approximation rule of \sect{sec:gaussian}.

In \tbl{tbl:bandlimited}, we give that bandlimiting $x^n$ by a Gaussian is a generalized Hermite polynomial $He_n^{[\alpha]}(x)$. This can be derived trivially from a property of generalized Hermite polynomials: the $n$th noncentral moment of a Gaussian distribution $X$ with expected value $\mu$ and variance $\sigma$ is a generalized Hermite polynomial~\cite{wiki:hermite}. The ordinary Hermite polynomial is defined as:
	\begin{equation} \label{eq_hermite}
		He_n(x) = \sum_{k=0}^{\lfloor{\frac{n}{2}}\rfloor} \frac{n!}{(n - 2k)! k!} (-2)^{-k} x^{n-2k}
	\end{equation}
	
Then $He_n^{[\alpha]}(x)$ is the generalized Hermite polynomial with parameter $\alpha$, defined by:
	\begin{equation} \label{eq_g_hermite}
	\begin{split}
		He_n^{[\alpha]}(x) & = \alpha^{\frac{n}{2}} He_n\left(\frac{x}{\sqrt{\alpha}}\right) \\
                                               & = \sum_{k=0}^{\lfloor{\frac{n}{2}}\rfloor} \frac{n!}{(n-2k)!k!} (-2)^{-k} x^{n-2k} \alpha^{k} \\
	\end{split}
	\end{equation}


\begin{table*}[t]
\caption {A table of univariate functions, and their corresponding bandlimited result, using a box kernel $B$ and a Gaussian $G$. The box kernel is the PDF of the uniform random variable $U[-\sqrt{3}\sigma, \sqrt{3}\sigma]$. The Gaussian kernel is the PDF of the random variable $\mathcal{N}(0,\sigma^2)$. Each random variable has standard deviation $\sigma$. We define $\sinc(x) = \sin(x)/x$, and the Heaviside step function $H(x)$ is $0$ for $x \leq 0$ and $1$ for $x$ positive. Note that functions with undefined regions, such as $x^p$ for negative or fractional $p$ have $\sigma$ limited as described in \sect{sec:undef}.}
\begin{tabular}{|l|l|l|}
	\hline
Function $f(x)$ & Bandlimited with box kernel: $\hat{f}^{B}(x)$ & Bandlimited with Gaussian kernel: $\hat{f}^{G}(x)$\\
	\hline
	$x^p, p \neq -1$ & $\frac{1}{\sqrt{12}\sigma(p+1)}\left[(x+\sqrt{3}\sigma)^{p+1} - (x-\sqrt{3}\sigma)^{p+1}\right]$ & $He_p^{[-\sigma^2]}(x)$ \\
	$x^{-2}$ & $(x^2 - 3\sigma^2)^{-1}$ & \missing \\
	$x^{-1}$ & $\frac{1}{\sqrt{12}\sigma}\log\left|\frac{x + \sqrt{3}\sigma}{x - \sqrt{3}\sigma}\right|$ & \missing \\
	$x$ & $x$ & $x$ \\
	$x^2$ & $x^2+\sigma^2$ & $x^2+\sigma^2$ \\
	$x^3$ & $x^3+3x\sigma^2$ & $x^3 + 3x\sigma^2$ \\
	$x^4$ & $x^4 + 6x^2\sigma^2 + \frac{9}{5}\sigma^4$ & $x^4 + 6x^2\sigma^2 + 3\sigma^4$ \\
	$x^5$ & $x^5 + 10x^3\sigma^2 + 9x\sigma^4$ & $x^5 + 10x^3\sigma^2 + 15x\sigma^4$ \\
	$x^6$ & $x^6 + 15x^4\sigma^2 + 27x^2\sigma^4 + \frac{27}{7}\sigma^6$ & $x^6 + 15x^4\sigma^2 + 45x^2\sigma^4 + 15\sigma^6$ \\
	$x^7$ & $x^7 + 21x^5\sigma^2 + 63x^3\sigma^4 + 27x\sigma^6$ & $x^7 + 21x^5\sigma^2 + 105x^3\sigma^4 + 105x\sigma^6$\\
	$x^8$ & $x^8 + 28x^6\sigma^2 + 126x^4\sigma^4 + 108x^2\sigma^6 + 9\sigma^8$ & $x^8 + 28x^6\sigma^2 + 210x^4\sigma^4 + 420x^2\sigma^6 + 105\sigma^8$\\
	$\sin(x)$ & $\sin(x)\sinc(\sqrt{3}\sigma)$ & $\sin(x)e^{-\frac{\sigma^2}{2}}$ \\
	$\cos(x)$ & $\cos(x)\sinc(\sqrt{3}\sigma)$ & $\cos(x)e^{-\frac{\sigma^2}{2}}$ \\
	$\tan(x)$ & $\frac{-1}{\sqrt{12}\sigma}\log\left|\frac{\cos(x+\sqrt{3}\sigma)}{\cos(x-\sqrt{3}\sigma)}\right|$ & \missing \\
	$\sinh(x)$ &$\frac{1}{\sqrt{12}\sigma}(\cosh(x+\sqrt{3}\sigma) - \cosh(x-\sqrt{3}\sigma))$ & $\frac{1}{2}(e^{x + \frac{1}{2}\sigma^2} - e^{-x + \frac{1}{2}\sigma^2})$ \\
	$\cosh(x)$ & $\frac{1}{\sqrt{12}\sigma}(\sinh(x+\sqrt{3}\sigma) - \sinh(x-\sqrt{3}\sigma))$ & $\frac{1}{2}(e^{x + \frac{1}{2}\sigma^2} + e^{-x + \frac{1}{2}\sigma^2})$ \\
	$\tanh(x)$ & $\frac{1}{\sqrt{12}\sigma}(\log(\cosh(x+\sqrt{3}\sigma)) - \log(\cosh(x-\sqrt{3}\sigma)))$ & \missing \\
	$\sinh^2(x)$ &$\frac{1}{8\sqrt{3}\sigma}(-4\sqrt{3}\sigma + \sinh(2\sqrt{3}\sigma - 2x) + \sinh(2\sqrt{3}\sigma + 2x))$ & \\
	$\cosh^2(x)$ &$\frac{1}{8\sqrt{3}\sigma}(4\sqrt{3}\sigma + \sinh(2\sqrt{3}\sigma - 2x) + \sinh(2\sqrt{3}\sigma + 2x))$ & \\
	$\tanh^2(x)$ & $\frac{1}{2\sqrt{3}\sigma}(2\sqrt{3}\sigma - \tanh(\sqrt{3}\sigma - x) - \tanh(\sqrt{3}\sigma + x))$ & \missing \\
	$e^x$ & $\frac{1}{\sqrt{12}\sigma}\left(e^{x+\sqrt{3}\sigma} - e^{x-\sqrt{3}\sigma}\right)$ & $e^{x + \frac{1}{2}\sigma^2}$ \\
	$\sin^2(x)$ & $\frac{1}{2} - \frac{1}{2}\cos(2x)\sinc(\sqrt{12}\sigma)$ & $\frac{1}{2} - \frac{1}{2}\cos(2x)e^{-2\sigma^2}$ \\
	$\cos^2(x)$ & $\frac{1}{2} + \frac{1}{2}\cos(2x)\sinc(\sqrt{12}\sigma)$ & $\frac{1}{2} + \frac{1}{2}\cos(2x)e^{-2\sigma^2}$ \\
	$\tan^2(x)$ & $\frac{1}{\sqrt{12}\sigma}\left(\tan(x+\sqrt{3}\sigma) - \tan(x-\sqrt{3}\sigma)\right) - 1$ & \missing \\
	$H(x)$ & $\begin{cases} 0 & x\le -\sqrt{3}\sigma \\ \frac{x}{2\sqrt{3}\sigma} + \frac{1}{2} & -\sqrt{3}\sigma\le x\le\sqrt{3}\sigma \\ 1 & x\ge\sqrt{3}\sigma \end{cases}$ & $\frac{1}{2}(1+\erf\frac{x}{\sqrt{2}\sigma})$ \\
	$\fract(x)$ & $\frac{1}{\sqrt{48}\sigma}(\fract^2(x+\sqrt{3}\sigma) + \floor{x+\sqrt{3}\sigma} - $& \\
	& \hspace*{0.7cm}$\fract^2(x-\sqrt{3}\sigma) - \floor{x-\sqrt{3}\sigma})$ & \\
	$\fract^2(x)$ & $\frac{1}{\sqrt{108}\sigma}(\fract^3(x+\sqrt{3}\sigma) + \floor{x+\sqrt{3}\sigma} - $& \\
	& \hspace*{0.7cm}$\fract^3(x-\sqrt{3}\sigma) - \floor{x-\sqrt{3}\sigma})$ & \\
	$\floor{x}$ & $x - \widehat{\mathrm{fract}}(x)$ & \\
	$\floor{x}^2$ & $\widehat{x^2} + \widehat{\mathrm{fract}^2}(x) - F(x + \sqrt{3}\sigma) + F(x - \sqrt{3}\sigma)$& \\
	& where $F(x) = 2(\frac{\floor{x}}{3} + \frac{\floor{x}(\floor{x}-1)}{4} + \frac{\floor{x} \widehat{\mathrm{fract}}^2(x)}{2} + \frac{\widehat{\mathrm{fract}}^3(x)}{3})$& \\
	$\ceil{x}$ & $x + \widehat{\mathrm{fract}}(-x)$ & \\
	$\ceil{x}^2$ & $\widehat{\floor{-x}^2}$ & \\
    \hline
\end{tabular}
\label{tbl:bandlimited}
\end{table*}

\section{Correlation Coefficients for Multivariate Functions}
\label{sec:rho}
In this appendix, we describe rules to compute the correlation coefficient $\rho$, which is briefly discussed in \sect{sec:gaussian}. Specifically, we are given a binary function $f(a, b)$, which receives two inputs $a$ and $b$, with associated random variables $A$ and $B$, respectively. We discuss the following two rules: a) assume $\rho$ is constant and estimate by sampling and b) compute $\rho$ under the assumption that computations are affine.

\textbf{Estimate $\rho$ by sampling}. In a training stage, we use $n$ samples to approximate $\rho$ of two random variables $A$ and $B$. The samples drawn from these two distributions are represented as $a_i$ and $b_i$ with corresponding sample mean $\overline{a}$ and $\overline{b}$. Thus, $\rho$ can be estimated by \eqn{eq_rho_sample} \cite{wiki:correlation}.
\begin{equation} \label{eq_rho_sample}
    \rho = \frac{\sum_{i=1}^n (a_i - \overline{a}) (b_i - \overline{b})}{\sqrt{\sum_{i=1}^n(a_i - \overline{a})^2} \sqrt{\sum_{i=1}^n(b_i - \overline{b})^2}}
\end{equation}

\textbf{Estimate $\rho$ by an affine assumption.} When we calculate $\rho$ under this rule, we assume the variables $a$ and $b$ input to $f$ are affine transformations of the variables $x_1$, ..., $x_n$ which are input to the program. Here, $x_i$s are inputs to the function that is being smoothed, which for a shader program could be the $(x, y)$ spatial coordinate. Under this assumption, $a$ and $b$ can be expressed as follows.

\begin{equation} \label{eq_affine}
    \begin{split}
        a = a_c + \sum_{i=1}^n a_i x_i \\
        b = b_c + \sum_{i=1}^n b_i x_i
    \end{split}
\end{equation}

In \eqn{eq_affine}, $a_i$ and $b_i$ are coefficients of the affine transformation, and $a_c$ and $b_c$ end up not mattering for the $\rho$ computation, so we ignore these constants. In our implementation, we find $a_i$ and $b_i$ by taking the gradient, via the automatic differentiation of the expression nodes $a$ and $b$ with respect to the inputs $x_i$. Here $\rho$ is computed as:

\begin{equation} \label{eq_rho_gradient}
    \rho = \frac{\sum_{i=1}^n a_i b_i}{\sqrt{\sum_{i=1}^n a_{i}^2} \sqrt{\sum_{i=1}^n b_{i}^2}}
\end{equation}

\section{Smoothing Result for Periodic Functions}

In this section, we derive a convenient formula that gives the bandlimited result for any periodic function if its integral within a single period is known. We extend the analysis of fract() made by \dorn{} to any periodic function. We use Heckbert's technique of repeated integration \cite{heckbert1986filtering} to derive the convolution of a periodic function with a box kernel.

Specifically, we assume the periodic function $f(x)$ has period T and its first and second integrals within one period are also known. These are denoted as $F_p(x)$ and $F_{p2}(x)$, respectively.

\begin{equation} \label{eq_periodic_first}
    \begin{split}
        F_p(x) & = \int_{0}^{x} f(u) du \\
        F_{p2}(x) & = \int_{0}^{x} F_p(u) du \\
        x & \in [0, T)
    \end{split}
\end{equation}

Using \eqn{eq_periodic_first}, we derive the first and second integral of $f(x)$ as follows.

\begin{equation} \label{eq_periodic_inf}
    \begin{split}
        F(x) = & \int_{0}^x f(u) du \\ 
        = & \left(\floor{\frac{x}{T}} + 1\right) \cdot F_p(T) - \int_{x-T\cdot \floor{\frac{x}{T}}}^T f(u) du \\
        = & \left(\floor{\frac{x}{T}} + 1\right) \cdot F_p(T) - F_p(T) + F_p\left(x - T \cdot \floor{\frac{x}{T}}\right) \\
        = & \floor{\frac{x}{T}} \cdot F_p(T) + F_p\left(x - T \cdot \floor{\frac{x}{T}}\right)
    \end{split}
\end{equation}

\begin{equation} \label{eq_periodic_inf2}
    \begin{split}
        F_2(x) = & \int_{0}^x F(u) du \\
        = & \int_{0}^x \floor{\frac{u}{T}} \cdot F_p(T) du + \int_{0}^x F_p(u - T \cdot \floor{\frac{u}{T}}) du \\
        = & F_p(T) \cdot T \sum_{i=0}^{\floor{\frac{x}{T}} - 1} i + \left(x - T \floor{\frac{x}{T}}\right) \cdot \floor{\frac{x}{T}} \cdot F_p(T) + \\
        & \floor{\frac{x}{T}}\cdot  F_{p2}(T) + F_{p2}\left(x - T\cdot \floor{\frac{x}{T}}\right) \\
        = & F_p(T)\cdot \left(\frac{T \cdot (q-1) \cdot q}{2} + (x-T\cdot q) \cdot q\right) + \\
        & F_{p2}(T) \cdot q + F_{p2}(x - T \cdot q) \\
        \text{Here, } & q = \floor{\frac{x}{T}}.
    \end{split}
\end{equation}

Using Heckbert's result, the convolution of the periodic function $f(x)$ with a box kernel that has support [$-\sqrt{3} \sigma, \sqrt{3}\sigma$] (corresponding to a uniform kernel with standard deviation $\sigma$) can be expressed as follows.

\begin{equation} \label{eq_periodic_box}
    \hat{f}(x, \sigma) = \frac{F(x + \sqrt{3} \sigma) - F(x + \sqrt{3} \sigma)}{2 \sqrt{3} \sigma}
\end{equation}

And the convolution of the periodic function $f(x)$ with a tent kernel that has support [$-\sqrt{6} \sigma, \sqrt{6} \sigma$] (corresponding to a uniform kernel with standard deviation $\sigma$) can be expressed as follows.

\begin{equation} \label{eq_periodic_tent}
    \hat{f}(x, \sigma) = \frac{F_2(x + \sqrt{6} \sigma) - 2 \cdot F_2(x) + F_2(x - \sqrt{6} \sigma)}{6\sigma^2}
\end{equation}

\section{Proof of Second Order Approximation for a Single Composition}
\label{sec:secondorder}

Here we show for a univariate function, applying function composition using our adaptive Gaussian approximation from \sect{sec:gaussian} is accurate up to the second order in standard deviation $\sigma$. Suppose we wish to approximate the composition of two functions: $f(x) = f_2(f_1(x))$, where $f_1, f_2: \mathbb{R} \to \mathbb{R}$. Assume the input random variable is $X_0 \sim \mathcal{N}(x,\sigma^2)$: the Gaussian kernel centered at $x$. The output from $f_1$ is an intermediate value in the computation: we can represent this with another random variable $X_1 = f_1(X_0)$. Similarly, the output random variable $X_2 = f_2(X_1)$. We conclude that $f(X_0) = f_2(f_1(X_0)) = f_2(X_1) = X_2$.

We apply \eqn{eq_taylor} and \eqn{eq_unary_var} to $f_1$, and obtain the following mean and standard deviation.
\begin{equation} \label{eq_hat_f1}
    \begin{split}
        \mu_{X_1} & = \hat{f_1}(x, \sigma^2) = f_1(x) + \frac{1}{2}\sigma^2 f_{1}^{''}(x) + \mathcal{O}(\sigma^4) \\
        \widehat{f_{1}^{2}}(x, \sigma^2) & = f_{1}^{2}(x) + \frac{1}{2} \sigma^2 \frac{\partial^2}{\partial x^2} (f_{1}^{2}(x)) + \mathcal{O}(\sigma^4) \\
        & = f_{1}^{2}(x) + \frac{1}{2}\sigma^2 (2 f_1 f_{1}^{''} + 2 (f_{1}^{'})^2)(x) + \mathcal{O}(\sigma^4) \\
        \sigma_{X_1}^2 & = \widehat{f_{1}^{2}}(x, \sigma^2) - \hat{f_1}(x, \sigma^2)^2 \\
        & = \sigma^2 (f_{1}^{'})^2(x) + \mathcal{O}(\sigma^4)
    \end{split}
\end{equation}

Using our composition rule, $X_1$ is approximated as a normal distribution using the mean and standard deviation calculated from \eqn{eq_hat_f1}. That is, we approximate $X_1$ as being distributed as $\mathcal{N}(\mu_{X_1},\sigma_{X_1}^2)$. Similarly, $\mu_{X_2}$, which is the output we care about, can be computed based on \eqn{eq_taylor}, \eqn{eq_hat_f1}, and repeated Taylor expansion in $\sigma$ around $\sigma = 0$.

\begin{equation} \label{eq_hat_f2}
    \begin{split}
        \mu_{X_2} = & \hat{f_2}(\hat{f_1}(x, \sigma^2), \sigma_{X_1}^2) \\
        = & f_2(f_1(x) + \frac{1}{2} \sigma^2 f_{1}^{''}(x) + \mathcal{O}(\sigma^4)) + \\
        & \frac{1}{2} \sigma_{X_1}^2 f_{2}^{''}(\hat{f_1}(x, \sigma^2)) + \mathcal{O}(\sigma_{X_1}^4) \\
        = & f(x) + \frac{1}{2} \sigma^2 f_{2}^{'}(f_1(x))f_{1}^{''}(x) + \\
        & \frac{1}{2} \sigma^2 f_{2}^{''}(f_1(x))(f_{1}^{'})^2(x) + \mathcal{O}(\sigma^4) \\
        = & f(x) + \frac{1}{2} \sigma^2 f^{''}(x) + \mathcal{O}(\sigma^4)
    \end{split}
\end{equation}

Comparing \eqn{eq_hat_f2} with \eqn{eq_taylor}, the function composition in our framework agrees up to the second order term in the Taylor expansion.

We conclude that our approximation is accurate up to the second order in standard deviation for a single composition of univariate functions. The same property for additional compositions of univariate functions can be shown by induction.


\bibliography{paper}{}


\begin{thebibliography}{00}


\ifx \showCODEN    \undefined \def \showCODEN     #1{\unskip}     \fi
\ifx \showDOI      \undefined \def \showDOI       #1{{\tt DOI:}\penalty0{#1}\ }
  \fi
\ifx \showISBNx    \undefined \def \showISBNx     #1{\unskip}     \fi
\ifx \showISBNxiii \undefined \def \showISBNxiii  #1{\unskip}     \fi
\ifx \showISSN     \undefined \def \showISSN      #1{\unskip}     \fi
\ifx \showLCCN     \undefined \def \showLCCN      #1{\unskip}     \fi
\ifx \shownote     \undefined \def \shownote      #1{#1}          \fi
\ifx \showarticletitle \undefined \def \showarticletitle #1{#1}   \fi
\ifx \showURL      \undefined \def \showURL       #1{#1}          \fi
\providecommand\bibfield[2]{#2}
\providecommand\bibinfo[2]{#2}
\providecommand\natexlab[1]{#1}
\providecommand\showeprint[2][]{arXiv:#2}

\bibitem[\protect\citeauthoryear{Akenine-M{\"o}ller, Haines, and
  Hoffman}{Akenine-M{\"o}ller et~al\mbox{.}}{2008}]%
        {akenine2008real}
\bibfield{author}{\bibinfo{person}{Tomas Akenine-M{\"o}ller},
  \bibinfo{person}{Eric Haines}, {and} \bibinfo{person}{Naty Hoffman}.}
  \bibinfo{year}{2008}\natexlab{}.
\newblock \bibinfo{booktitle}{{\em Real-time rendering}}.
\newblock \bibinfo{publisher}{CRC Press}.
\newblock


\bibitem[\protect\citeauthoryear{Apodaca, Gritz, and Barzel}{Apodaca
  et~al\mbox{.}}{2000}]%
        {apodaca2000advanced}
\bibfield{author}{\bibinfo{person}{Anthony~A Apodaca}, \bibinfo{person}{Larry
  Gritz}, {and} \bibinfo{person}{Ronen Barzel}.}
  \bibinfo{year}{2000}\natexlab{}.
\newblock \bibinfo{booktitle}{{\em Advanced RenderMan: Creating CGI for motion
  pictures}}.
\newblock \bibinfo{publisher}{Morgan Kaufmann}.
\newblock


\bibitem[\protect\citeauthoryear{Baker and Sutlief}{Baker and Sutlief}{2003}]%
        {baker2003green}
\bibfield{author}{\bibinfo{person}{M Baker} {and} \bibinfo{person}{S Sutlief}.}
  \bibinfo{year}{2003}\natexlab{}.
\newblock \showarticletitle{Green’s Functions in Physics Version 1}.
\newblock  (\bibinfo{year}{2003}).
\newblock


\bibitem[\protect\citeauthoryear{Bako, Vogels, McWilliams, Meyer, Nov\'ak,
  Harvill, Sen, DeRose, and Rousselle}{Bako et~al\mbox{.}}{2017}]%
        {Bako17}
\bibfield{author}{\bibinfo{person}{Steve Bako}, \bibinfo{person}{Thijs Vogels},
  \bibinfo{person}{Brian McWilliams}, \bibinfo{person}{Mark Meyer},
  \bibinfo{person}{Jan Nov\'ak}, \bibinfo{person}{Alex Harvill},
  \bibinfo{person}{Pradeep Sen}, \bibinfo{person}{Tony DeRose}, {and}
  \bibinfo{person}{Fabrice Rousselle}.} \bibinfo{year}{2017}\natexlab{}.
\newblock \showarticletitle{Kernel-Predicting Convolutional Networks for
  Denoising Monte Carlo Renderings}.
\newblock \bibinfo{journal}{{\em ACM Transactions on Graphics (TOG)
  (Proceedings of SIGGRAPH 2017)\/}} \bibinfo{volume}{36}, \bibinfo{number}{4}
  (\bibinfo{date}{July} \bibinfo{year}{2017}).
\newblock


\bibitem[\protect\citeauthoryear{Brady, Lawrence, Peers, and Weimer}{Brady
  et~al\mbox{.}}{2014}]%
        {brady2014genbrdf}
\bibfield{author}{\bibinfo{person}{Adam Brady}, \bibinfo{person}{Jason
  Lawrence}, \bibinfo{person}{Pieter Peers}, {and} \bibinfo{person}{Westley
  Weimer}.} \bibinfo{year}{2014}\natexlab{}.
\newblock \showarticletitle{genBRDF: Discovering new analytic BRDFs with
  genetic programming}.
\newblock \bibinfo{journal}{{\em ACM Transactions on Graphics (TOG)\/}}
  \bibinfo{volume}{33}, \bibinfo{number}{4} (\bibinfo{year}{2014}),
  \bibinfo{pages}{114}.
\newblock


\bibitem[\protect\citeauthoryear{Buades, Coll, and Morel}{Buades
  et~al\mbox{.}}{2005}]%
        {buades2005non}
\bibfield{author}{\bibinfo{person}{Antoni Buades}, \bibinfo{person}{Bartomeu
  Coll}, {and} \bibinfo{person}{J-M Morel}.} \bibinfo{year}{2005}\natexlab{}.
\newblock \showarticletitle{A non-local algorithm for image denoising}. In
  \bibinfo{booktitle}{{\em Computer Vision and Pattern Recognition, 2005. CVPR
  2005. IEEE Computer Society Conference on}}, Vol.~\bibinfo{volume}{2}. IEEE,
  \bibinfo{pages}{60--65}.
\newblock


\bibitem[\protect\citeauthoryear{Buades, Coll, and Morel}{Buades
  et~al\mbox{.}}{2011}]%
        {buades2011non}
\bibfield{author}{\bibinfo{person}{Antoni Buades}, \bibinfo{person}{Bartomeu
  Coll}, {and} \bibinfo{person}{Jean-Michel Morel}.}
  \bibinfo{year}{2011}\natexlab{}.
\newblock \showarticletitle{Non-local means denoising}.
\newblock \bibinfo{journal}{{\em Image Processing On Line\/}}
  \bibinfo{volume}{1} (\bibinfo{year}{2011}), \bibinfo{pages}{208--212}.
\newblock


\bibitem[\protect\citeauthoryear{Chaudhuri and Solar-Lezama}{Chaudhuri and
  Solar-Lezama}{2011}]%
        {chaudhuri2011smoothing}
\bibfield{author}{\bibinfo{person}{Swarat Chaudhuri} {and}
  \bibinfo{person}{Armando Solar-Lezama}.} \bibinfo{year}{2011}\natexlab{}.
\newblock \showarticletitle{Smoothing a program soundly and robustly}. In
  \bibinfo{booktitle}{{\em International Conference on Computer Aided
  Verification}}. Springer, \bibinfo{pages}{277--292}.
\newblock


\bibitem[\protect\citeauthoryear{Chen and Chen}{Chen and Chen}{1999}]%
        {chen1999global}
\bibfield{author}{\bibinfo{person}{Bintong Chen} {and} \bibinfo{person}{Xiaojun
  Chen}.} \bibinfo{year}{1999}\natexlab{}.
\newblock \showarticletitle{A global and local superlinear
  continuation-smoothing method for P 0 and R 0 NCP or monotone NCP}.
\newblock \bibinfo{journal}{{\em SIAM Journal on Optimization\/}}
  \bibinfo{volume}{9}, \bibinfo{number}{3} (\bibinfo{year}{1999}),
  \bibinfo{pages}{624--645}.
\newblock


\bibitem[\protect\citeauthoryear{Chen and Xiu}{Chen and Xiu}{1999}]%
        {chen1999global2}
\bibfield{author}{\bibinfo{person}{Bintong Chen} {and} \bibinfo{person}{Naihua
  Xiu}.} \bibinfo{year}{1999}\natexlab{}.
\newblock \showarticletitle{A Global Linear and Local Quadratic Noninterior
  Continuation Method for Nonlinear Complementarity Problems Based on
  Chen--Mangasarian Smoothing Functions}.
\newblock \bibinfo{journal}{{\em SIAM Journal on Optimization\/}}
  \bibinfo{volume}{9}, \bibinfo{number}{3} (\bibinfo{year}{1999}),
  \bibinfo{pages}{605--623}.
\newblock


\bibitem[\protect\citeauthoryear{Cook}{Cook}{1986}]%
        {Cook:1986:SSC:7529.8927}
\bibfield{author}{\bibinfo{person}{Robert~L. Cook}.}
  \bibinfo{year}{1986}\natexlab{}.
\newblock \showarticletitle{Stochastic Sampling in Computer Graphics}.
\newblock \bibinfo{journal}{{\em ACM Trans. Graph.\/}} \bibinfo{volume}{5},
  \bibinfo{number}{1} (\bibinfo{date}{Jan.} \bibinfo{year}{1986}),
  \bibinfo{pages}{51--72}.
\newblock
\showISSN{0730-0301}
\showDOI{%
\url{http://dx.doi.org/10.1145/7529.8927}}


\bibitem[\protect\citeauthoryear{Crow}{Crow}{1977}]%
        {crow1977aliasing}
\bibfield{author}{\bibinfo{person}{Franklin~C Crow}.}
  \bibinfo{year}{1977}\natexlab{}.
\newblock \showarticletitle{The aliasing problem in computer-generated shaded
  images}.
\newblock \bibinfo{journal}{{\it Commun. ACM}} \bibinfo{volume}{20},
  \bibinfo{number}{11} (\bibinfo{year}{1977}), \bibinfo{pages}{799--805}.
\newblock


\bibitem[\protect\citeauthoryear{Crow}{Crow}{1984}]%
        {crow1984summed}
\bibfield{author}{\bibinfo{person}{Franklin~C Crow}.}
  \bibinfo{year}{1984}\natexlab{}.
\newblock \showarticletitle{Summed-area tables for texture mapping}.
\newblock \bibinfo{journal}{{\em ACM SIGGRAPH computer graphics\/}}
  \bibinfo{volume}{18}, \bibinfo{number}{3} (\bibinfo{year}{1984}),
  \bibinfo{pages}{207--212}.
\newblock


\bibitem[\protect\citeauthoryear{Dipp{\'e} and Wold}{Dipp{\'e} and
  Wold}{1985}]%
        {dippe1985antialiasing}
\bibfield{author}{\bibinfo{person}{Mark~AZ Dipp{\'e}} {and}
  \bibinfo{person}{Erling~Henry Wold}.} \bibinfo{year}{1985}\natexlab{}.
\newblock \showarticletitle{Antialiasing through stochastic sampling}.
\newblock \bibinfo{journal}{{\em ACM Siggraph Computer Graphics\/}}
  \bibinfo{volume}{19}, \bibinfo{number}{3} (\bibinfo{year}{1985}),
  \bibinfo{pages}{69--78}.
\newblock


\bibitem[\protect\citeauthoryear{Dorn, Barnes, Lawrence, and Weimer}{Dorn
  et~al\mbox{.}}{2015}]%
        {dorn2015}
\bibfield{author}{\bibinfo{person}{Jonathan Dorn}, \bibinfo{person}{Connelly
  Barnes}, \bibinfo{person}{Jason Lawrence}, {and} \bibinfo{person}{Westley
  Weimer}.} \bibinfo{year}{2015}\natexlab{}.
\newblock \showarticletitle{Towards Automatic Band-Limited Procedural Shaders}.
  In \bibinfo{booktitle}{{\em Computer Graphics Forum}},
  Vol.~\bibinfo{volume}{34}. Wiley Online Library.
\newblock


\bibitem[\protect\citeauthoryear{Ebert}{Ebert}{2003}]%
        {ebert2003texturing}
\bibfield{author}{\bibinfo{person}{David~S Ebert}.}
  \bibinfo{year}{2003}\natexlab{}.
\newblock \bibinfo{booktitle}{{\em Texturing \& modeling: a procedural
  approach}}.
\newblock \bibinfo{publisher}{Morgan Kaufmann}.
\newblock


\bibitem[\protect\citeauthoryear{Ermoliev and Norkin}{Ermoliev and
  Norkin}{1997}]%
        {ermoliev1997nonsmooth}
\bibfield{author}{\bibinfo{person}{Yuri~M Ermoliev} {and}
  \bibinfo{person}{Vladimir~I Norkin}.} \bibinfo{year}{1997}\natexlab{}.
\newblock \showarticletitle{On nonsmooth and discontinuous problems of
  stochastic systems optimization}.
\newblock \bibinfo{journal}{{\em European Journal of Operational Research\/}}
  \bibinfo{volume}{101}, \bibinfo{number}{2} (\bibinfo{year}{1997}),
  \bibinfo{pages}{230--244}.
\newblock


\bibitem[\protect\citeauthoryear{Ermoliev, Norkin, and Wets}{Ermoliev
  et~al\mbox{.}}{1995}]%
        {ermoliev1995minimization}
\bibfield{author}{\bibinfo{person}{Yuri~M Ermoliev},
  \bibinfo{person}{Vladimir~I Norkin}, {and} \bibinfo{person}{Roger~JB Wets}.}
  \bibinfo{year}{1995}\natexlab{}.
\newblock \showarticletitle{The minimization of semicontinuous functions:
  mollifier subgradients}.
\newblock \bibinfo{journal}{{\em SIAM Journal on Control and Optimization\/}}
  \bibinfo{volume}{33}, \bibinfo{number}{1} (\bibinfo{year}{1995}),
  \bibinfo{pages}{149--167}.
\newblock


\bibitem[\protect\citeauthoryear{Feiguin}{Feiguin}{2011}]%
        {mc_error}
\bibfield{author}{\bibinfo{person}{AE Feiguin}.}
  \bibinfo{year}{2011}\natexlab{}.
\newblock \bibinfo{title}{Monte Carlo error analysis}.
\newblock   (\bibinfo{year}{2011}).
\newblock
\showURL{%
\url{https://www.northeastern.edu/afeiguin/phys5870/phys5870/node71.html}}
\newblock
\shownote{[Online; accessed 22-May-2017].}


\bibitem[\protect\citeauthoryear{Hachisuka, Jarosz, Weistroffer, Dale,
  Humphreys, Zwicker, and Jensen}{Hachisuka et~al\mbox{.}}{2008}]%
        {hachisuka2008multidimensional}
\bibfield{author}{\bibinfo{person}{Toshiya Hachisuka},
  \bibinfo{person}{Wojciech Jarosz}, \bibinfo{person}{Richard~Peter
  Weistroffer}, \bibinfo{person}{Kevin Dale}, \bibinfo{person}{Greg Humphreys},
  \bibinfo{person}{Matthias Zwicker}, {and} \bibinfo{person}{Henrik~Wann
  Jensen}.} \bibinfo{year}{2008}\natexlab{}.
\newblock \showarticletitle{Multidimensional adaptive sampling and
  reconstruction for ray tracing}. In \bibinfo{booktitle}{{\em ACM Transactions
  on Graphics (TOG)}}, Vol.~\bibinfo{volume}{27}. ACM, \bibinfo{pages}{33}.
\newblock


\bibitem[\protect\citeauthoryear{Heckbert}{Heckbert}{1986}]%
        {heckbert1986filtering}
\bibfield{author}{\bibinfo{person}{Paul~S Heckbert}.}
  \bibinfo{year}{1986}\natexlab{}.
\newblock \showarticletitle{Filtering by repeated integration}. In
  \bibinfo{booktitle}{{\em ACM SIGGRAPH Computer Graphics}},
  Vol.~\bibinfo{volume}{20}. ACM, \bibinfo{pages}{315--321}.
\newblock


\bibitem[\protect\citeauthoryear{Kalantari, Bako, and Sen}{Kalantari
  et~al\mbox{.}}{2015}]%
        {kalantari2015machine}
\bibfield{author}{\bibinfo{person}{Nima~Khademi Kalantari},
  \bibinfo{person}{Steve Bako}, {and} \bibinfo{person}{Pradeep Sen}.}
  \bibinfo{year}{2015}\natexlab{}.
\newblock \showarticletitle{A machine learning approach for filtering Monte
  Carlo noise.}
\newblock \bibinfo{journal}{{\em ACM Trans. Graph.\/}} \bibinfo{volume}{34},
  \bibinfo{number}{4} (\bibinfo{year}{2015}), \bibinfo{pages}{122}.
\newblock


\bibitem[\protect\citeauthoryear{Koza}{Koza}{1992}]%
        {koza1992genetic}
\bibfield{author}{\bibinfo{person}{John~R Koza}.}
  \bibinfo{year}{1992}\natexlab{}.
\newblock \bibinfo{booktitle}{{\em Genetic programming: on the programming of
  computers by means of natural selection}}. Vol.~\bibinfo{volume}{1}.
\newblock \bibinfo{publisher}{MIT press}.
\newblock


\bibitem[\protect\citeauthoryear{Lagae, Lefebvre, Drettakis, and
  Dutr{\'e}}{Lagae et~al\mbox{.}}{2009}]%
        {lagae2009procedural}
\bibfield{author}{\bibinfo{person}{Ares Lagae}, \bibinfo{person}{Sylvain
  Lefebvre}, \bibinfo{person}{George Drettakis}, {and} \bibinfo{person}{Philip
  Dutr{\'e}}.} \bibinfo{year}{2009}\natexlab{}.
\newblock \showarticletitle{Procedural noise using sparse Gabor convolution}.
  In \bibinfo{booktitle}{{\em ACM Transactions on Graphics (TOG)}},
  Vol.~\bibinfo{volume}{28}. ACM, \bibinfo{pages}{54}.
\newblock


\bibitem[\protect\citeauthoryear{Liu, Wang, Chen, Guo, and Peng}{Liu
  et~al\mbox{.}}{2008}]%
        {liu2008robust}
\bibfield{author}{\bibinfo{person}{Yan-Li Liu}, \bibinfo{person}{Jin Wang},
  \bibinfo{person}{Xi Chen}, \bibinfo{person}{Yan-Wen Guo}, {and}
  \bibinfo{person}{Qun-Sheng Peng}.} \bibinfo{year}{2008}\natexlab{}.
\newblock \showarticletitle{A robust and fast non-local means algorithm for
  image denoising}.
\newblock \bibinfo{journal}{{\em Journal of computer science and technology\/}}
  \bibinfo{volume}{23}, \bibinfo{number}{2} (\bibinfo{year}{2008}),
  \bibinfo{pages}{270--279}.
\newblock


\bibitem[\protect\citeauthoryear{{\L}ysik}{{\L}ysik}{2012}]%
        {lysik2012mean}
\bibfield{author}{\bibinfo{person}{Grzegorz {\L}ysik}.}
  \bibinfo{year}{2012}\natexlab{}.
\newblock \showarticletitle{Mean-value properties of real analytic functions}.
\newblock \bibinfo{journal}{{\em Archiv der Mathematik\/}}
  \bibinfo{volume}{98}, \bibinfo{number}{1} (\bibinfo{year}{2012}),
  \bibinfo{pages}{61--70}.
\newblock


\bibitem[\protect\citeauthoryear{Mitchell}{Mitchell}{1987}]%
        {mitchell1987generating}
\bibfield{author}{\bibinfo{person}{Don~P Mitchell}.}
  \bibinfo{year}{1987}\natexlab{}.
\newblock \showarticletitle{Generating antialiased images at low sampling
  densities}. In \bibinfo{booktitle}{{\em ACM SIGGRAPH Computer Graphics}},
  Vol.~\bibinfo{volume}{21}. ACM, \bibinfo{pages}{65--72}.
\newblock


\bibitem[\protect\citeauthoryear{Mitchell}{Mitchell}{1991}]%
        {mitchell1991spectrally}
\bibfield{author}{\bibinfo{person}{Don~P Mitchell}.}
  \bibinfo{year}{1991}\natexlab{}.
\newblock \showarticletitle{Spectrally optimal sampling for distribution ray
  tracing}. In \bibinfo{booktitle}{{\em ACM SIGGRAPH Computer Graphics}},
  Vol.~\bibinfo{volume}{25}. ACM, \bibinfo{pages}{157--164}.
\newblock


\bibitem[\protect\citeauthoryear{Moore}{Moore}{1979}]%
        {moore1979methods}
\bibfield{author}{\bibinfo{person}{Ramon~E Moore}.}
  \bibinfo{year}{1979}\natexlab{}.
\newblock \bibinfo{booktitle}{{\em Methods and applications of interval
  analysis}}.
\newblock \bibinfo{publisher}{SIAM}.
\newblock


\bibitem[\protect\citeauthoryear{Nelder and Mead}{Nelder and Mead}{1965}]%
        {nelder1965simplex}
\bibfield{author}{\bibinfo{person}{John~A Nelder} {and} \bibinfo{person}{Roger
  Mead}.} \bibinfo{year}{1965}\natexlab{}.
\newblock \showarticletitle{A simplex method for function minimization}.
\newblock \bibinfo{journal}{{\em The computer journal\/}} \bibinfo{volume}{7},
  \bibinfo{number}{4} (\bibinfo{year}{1965}), \bibinfo{pages}{308--313}.
\newblock


\bibitem[\protect\citeauthoryear{Nesterov}{Nesterov}{2005}]%
        {nesterov2005smooth}
\bibfield{author}{\bibinfo{person}{Yu Nesterov}.}
  \bibinfo{year}{2005}\natexlab{}.
\newblock \showarticletitle{Smooth minimization of non-smooth functions}.
\newblock \bibinfo{journal}{{\em Mathematical programming\/}}
  \bibinfo{volume}{103}, \bibinfo{number}{1} (\bibinfo{year}{2005}),
  \bibinfo{pages}{127--152}.
\newblock


\bibitem[\protect\citeauthoryear{Norton, Rockwood, and Skolmoski}{Norton
  et~al\mbox{.}}{1982}]%
        {norton1982clamping}
\bibfield{author}{\bibinfo{person}{Alan Norton}, \bibinfo{person}{Alyn~P
  Rockwood}, {and} \bibinfo{person}{Philip~T Skolmoski}.}
  \bibinfo{year}{1982}\natexlab{}.
\newblock \showarticletitle{Clamping: A method of antialiasing textured
  surfaces by bandwidth limiting in object space}. In \bibinfo{booktitle}{{\em
  ACM SIGGRAPH Computer Graphics}}, Vol.~\bibinfo{volume}{16}. ACM,
  \bibinfo{pages}{1--8}.
\newblock


\bibitem[\protect\citeauthoryear{Petersen, Pedersen, et~al\mbox{.}}{Petersen
  et~al\mbox{.}}{2008}]%
        {petersen2008matrix}
\bibfield{author}{\bibinfo{person}{Kaare~Brandt Petersen},
  \bibinfo{person}{Michael~Syskind Pedersen}, {and} \bibinfo{person}{others}.}
  \bibinfo{year}{2008}\natexlab{}.
\newblock \showarticletitle{The matrix cookbook}.
\newblock \bibinfo{journal}{{\em Technical University of Denmark\/}}
  \bibinfo{volume}{7} (\bibinfo{year}{2008}), \bibinfo{pages}{15}.
\newblock


\bibitem[\protect\citeauthoryear{Rousselle, Knaus, and Zwicker}{Rousselle
  et~al\mbox{.}}{2012}]%
        {rousselle2012adaptive}
\bibfield{author}{\bibinfo{person}{Fabrice Rousselle}, \bibinfo{person}{Claude
  Knaus}, {and} \bibinfo{person}{Matthias Zwicker}.}
  \bibinfo{year}{2012}\natexlab{}.
\newblock \showarticletitle{Adaptive rendering with non-local means filtering}.
\newblock \bibinfo{journal}{{\em ACM Transactions on Graphics (TOG)\/}}
  \bibinfo{volume}{31}, \bibinfo{number}{6} (\bibinfo{year}{2012}),
  \bibinfo{pages}{195}.
\newblock


\bibitem[\protect\citeauthoryear{Sitthi-Amorn, Modly, Weimer, and
  Lawrence}{Sitthi-Amorn et~al\mbox{.}}{2011}]%
        {sitthi2011genetic}
\bibfield{author}{\bibinfo{person}{Pitchaya Sitthi-Amorn},
  \bibinfo{person}{Nicholas Modly}, \bibinfo{person}{Westley Weimer}, {and}
  \bibinfo{person}{Jason Lawrence}.} \bibinfo{year}{2011}\natexlab{}.
\newblock \showarticletitle{Genetic programming for shader simplification}.
\newblock \bibinfo{journal}{{\em ACM Transactions on Graphics (TOG)\/}}
  \bibinfo{volume}{30}, \bibinfo{number}{6} (\bibinfo{year}{2011}),
  \bibinfo{pages}{152}.
\newblock


\bibitem[\protect\citeauthoryear{Szirmay-Kalos and Umenhoffer}{Szirmay-Kalos
  and Umenhoffer}{2008}]%
        {szirmay2008displacement}
\bibfield{author}{\bibinfo{person}{L{\'a}szl{\'o} Szirmay-Kalos} {and}
  \bibinfo{person}{Tam{\'a}s Umenhoffer}.} \bibinfo{year}{2008}\natexlab{}.
\newblock \showarticletitle{Displacement Mapping on the GPU—State of the
  Art}. In \bibinfo{booktitle}{{\em Computer Graphics Forum}},
  Vol.~\bibinfo{volume}{27}. Wiley Online Library, \bibinfo{pages}{1567--1592}.
\newblock


\bibitem[\protect\citeauthoryear{Wells}{Wells}{1986}]%
        {wells1986efficient}
\bibfield{author}{\bibinfo{person}{William~M Wells}.}
  \bibinfo{year}{1986}\natexlab{}.
\newblock \showarticletitle{Efficient synthesis of Gaussian filters by cascaded
  uniform filters}.
\newblock \bibinfo{journal}{{\em IEEE Transactions on Pattern Analysis and
  Machine Intelligence\/}} \bibinfo{number}{2} (\bibinfo{year}{1986}),
  \bibinfo{pages}{234--239}.
\newblock


\bibitem[\protect\citeauthoryear{Wikipedia}{Wikipedia}{2017a}]%
        {wiki:bessel}
\bibfield{author}{\bibinfo{person}{Wikipedia}.}
  \bibinfo{year}{2017}\natexlab{a}.
\newblock \bibinfo{title}{Bessel's correction --- Wikipedia{,} The Free
  Encyclopedia}.
\newblock   (\bibinfo{year}{2017}).
\newblock
\showURL{%
\url{https://en.wikipedia.org/w/index.php?title=Bessel\%27s_correction&oldid=764629526}}
\newblock
\shownote{[Online; accessed 23-May-2017].}


\bibitem[\protect\citeauthoryear{Wikipedia}{Wikipedia}{2017b}]%
        {wiki:correlation}
\bibfield{author}{\bibinfo{person}{Wikipedia}.}
  \bibinfo{year}{2017}\natexlab{b}.
\newblock \bibinfo{title}{Correlation and dependence --- Wikipedia{,} The Free
  Encyclopedia}.
\newblock   (\bibinfo{year}{2017}).
\newblock
\showURL{%
\url{https://en.wikipedia.org/w/index.php?title=Correlation_and_dependence&oldid=778221524}}
\newblock
\shownote{[Online; accessed 23-May-2017].}


\bibitem[\protect\citeauthoryear{Wikipedia}{Wikipedia}{2017c}]%
        {wiki:entire_function}
\bibfield{author}{\bibinfo{person}{Wikipedia}.}
  \bibinfo{year}{2017}\natexlab{c}.
\newblock \bibinfo{title}{Entire function --- Wikipedia{,} The Free
  Encyclopedia}.
\newblock   (\bibinfo{year}{2017}).
\newblock
\showURL{%
\url{https://en.wikipedia.org/w/index.php?title=Entire_function&oldid=778079847}}
\newblock
\shownote{[Online; accessed 20-May-2017].}


\bibitem[\protect\citeauthoryear{Wikipedia}{Wikipedia}{2017d}]%
        {wiki:hermite}
\bibfield{author}{\bibinfo{person}{Wikipedia}.}
  \bibinfo{year}{2017}\natexlab{d}.
\newblock \bibinfo{title}{Hermite polynomials --- Wikipedia{,} The Free
  Encyclopedia}.
\newblock   (\bibinfo{year}{2017}).
\newblock
\showURL{%
\url{https://en.wikipedia.org/w/index.php?title=Hermite_polynomials&oldid=778044979}}
\newblock
\shownote{[Online; accessed 20-May-2017].}


\bibitem[\protect\citeauthoryear{Williams}{Williams}{1983}]%
        {williams1983pyramidal}
\bibfield{author}{\bibinfo{person}{Lance Williams}.}
  \bibinfo{year}{1983}\natexlab{}.
\newblock \showarticletitle{Pyramidal parametrics}. In \bibinfo{booktitle}{{\em
  Acm siggraph computer graphics}}, Vol.~\bibinfo{volume}{17}. ACM,
  \bibinfo{pages}{1--11}.
\newblock


\bibitem[\protect\citeauthoryear{Wu}{Wu}{1996}]%
        {wu1996effective}
\bibfield{author}{\bibinfo{person}{Zhijun Wu}.}
  \bibinfo{year}{1996}\natexlab{}.
\newblock \showarticletitle{The effective energy transformation scheme as a
  special continuation approach to global optimization with application to
  molecular conformation}.
\newblock \bibinfo{journal}{{\em SIAM Journal on Optimization\/}}
  \bibinfo{volume}{6}, \bibinfo{number}{3} (\bibinfo{year}{1996}),
  \bibinfo{pages}{748--768}.
\newblock


\bibitem[\protect\citeauthoryear{Yang, Nehab, Sander, Sitthi-amorn, Lawrence,
  and Hoppe}{Yang et~al\mbox{.}}{2009}]%
        {yang2009amortized}
\bibfield{author}{\bibinfo{person}{Lei Yang}, \bibinfo{person}{Diego Nehab},
  \bibinfo{person}{Pedro~V Sander}, \bibinfo{person}{Pitchaya Sitthi-amorn},
  \bibinfo{person}{Jason Lawrence}, {and} \bibinfo{person}{Hugues Hoppe}.}
  \bibinfo{year}{2009}\natexlab{}.
\newblock \showarticletitle{Amortized supersampling}. In
  \bibinfo{booktitle}{{\em ACM Transactions on Graphics (TOG)}},
  Vol.~\bibinfo{volume}{28}. ACM, \bibinfo{pages}{135}.
\newblock


\end{thebibliography}
\bibliographystyle{ACM-Reference-Format}

\end{document}